\newif\iffinal
\finalfalse	
\finaltrue 

\newif\ifmarek
\marektrue

\documentclass[reqno,twoside,11pt]{amsart}
\iffinal\else\usepackage[notref,notcite]{showkeys}\fi
\usepackage{cite}
\usepackage[toc,title,page]{appendix}
\usepackage{subcaption}
\usepackage{amsmath}
\usepackage{amsmath,lipsum}
\usepackage{amsfonts}
\usepackage{amssymb}
\usepackage{graphicx}
\usepackage{booktabs}
\usepackage{enumitem}
\usepackage{booktabs}
\usepackage{multirow}
\usepackage{tikz}
\usetikzlibrary{arrows,automata,positioning}
\graphicspath{{/Users/aubain2015/Documents/york/}}
\usepackage{verbatim}
\usepackage{color}
\usepackage{float}
\usepackage{dirtytalk}
\IfFileExists{epsf.def}{\input epsf.def}{\usepackage{epsf}}
\ifmarek
\IfFileExists{myowntimes.sty}{\usepackage{myowntimes}\usepackage{mathrsfs}}
	{\usepackage{times}\newcommand{\mathscr}{\mathcal}}

\usepackage{mathptmx}
\fi

\DeclareFontFamily{OT1}{eusb}{} \DeclareFontShape{OT1}{eusb}{m}{n}
{<5> <6> <7> <8> <9> <10> <11> <12> <14.4> eusb10}{}
\DeclareMathAlphabet{\eusb}{OT1}{eusb}{m}{n}

\DeclareFontFamily{OT1}{eusm}{} \DeclareFontShape{OT1}{eusm}{m}{n}
{<5> <6> <7> <8> <9> <10> <11> <12> <14.4> eusm10}{}
\DeclareMathAlphabet{\eusm}{OT1}{eusm}{m}{n}

\DeclareFontFamily{OT1}{eufm}{} \DeclareFontShape{OT1}{eufm}{m}{n}
{<5> <6> <7> <8> <9> <10> <11> <12> <14.4> eufm10}{}
\DeclareMathAlphabet{\mathfrak}{OT1}{eufm}{m}{n}

\DeclareFontFamily{OT1}{fraktura}{}
\DeclareFontShape{OT1}{fraktura}{m}{n} {<5> <6> <7> <8> <9> <10>
<11> <12> <13> <14.4> [1.1] eufm10}{}
\DeclareMathAlphabet{\fraktura}{OT1}{fraktura}{m}{n}

\DeclareFontFamily{OT1}{cmfi}{} \DeclareFontShape{OT1}{cmfi}{m}{n}
{<5> <6> <7> <8> <9> <10> <11> <12> <13> <14.4> [0.9] cmfi10}{}
\DeclareMathAlphabet{\cmfi}{OT1}{cmfi}{b}{n}

\DeclareFontFamily{OT1}{cmss}{} \DeclareFontShape{OT1}{cmss}{m}{n}
{<5> <6> <7> <8> <9> <10> <11> <12> <13> <14.4> cmss10}{}
\DeclareMathAlphabet{\cmss}{OT1}{cmss}{m}{n}

\newtheoremstyle{thm}{1.5ex}{1.5ex}{\itshape\rmfamily}{}
{\bfseries\rmfamily}{}{2ex}{}

\newtheoremstyle{def}{1.5ex}{1.5ex}{\slshape\rmfamily}{}
{\bfseries\rmfamily}{}{2ex}{}

\newtheoremstyle{rem}{1.3ex}{1.3ex}{\rmfamily}{}
{\itshape}
{} {1.5ex}{}

\theoremstyle{thm}

\theoremstyle{def}

\theoremstyle{rem}

\numberwithin{equation}{section}

\renewcommand{\subsection}{\secdef\subsct\sbsect}
\newcommand{\subsct}[2][default]{\refstepcounter{subsection}
\addcontentsline{toc}{subsection}
{{\tocsection{\!\!}{\hspace{1.2em}\thesubsection}{\!\!\!\!#1\dotfill}}{}}
\nopagebreak\vspace{0.45\baselineskip} {\flushleft\bf
\thesection.\arabic{subsection}~\bf #1.~}
\\*[3mm]\noindent
\nopagebreak}
\newcommand{\sbsect}[1]{\vspace{0.1cm}\noindent
\textbf{#1.~}\vspace{0.1cm}}

\renewcommand{\subsubsection}{%
\secdef \subsubsect\sbsbsect}
\newcommand{\subsubsect}[2][default]{%
\refstepcounter{subsubsection} 
\addcontentsline{toc}{subsubsection}{{\tocsection{\!\!}
{\hspace{3.05em}\thesubsubsection}{\!\!\!\!#1\dotfill}}{}}
\nopagebreak
\vspace{0.15\baselineskip} \nopagebreak {\flushleft\rmfamily
\itshape\arabic{section}.\arabic{subsection}.\arabic{subsubsection}
\ \rmfamily #1\/.}\ }
\newcommand{\sbsbsect}[1]{\vspace{0.1cm}\noindent
\rmfamily \itshape
\arabic{section}.\arabic{subsection}.\arabic{subsubsection} \
\sffamily #1\/.\ }
\iffinal

\else

\fi
   
\usepackage[toc,title,page]{appendix}

\newcommand{\scrF}{\mathscr{F}}

\title[]
{\large Fitting Infinitely divisible distribution: \\
Case of Gamma-Variance Model}

\author [] {A.H.Nzokem$^1$}
\thanks {$^1$ hilaire77@gmail.com}
\begin{document}

\maketitle
\begin{abstract}
The paper examines the Fractional Fourier Transform (FRFT) based technique as a tool for obtaining the probability density function and its derivatives; and mainly for fitting stochastic model with the fundamental probabilistic relationships of infinite divisibility. The probability density functions are computed and the distributional proprieties such as leptokurtosis, peakedness, and asymmetry are reviewed for Variance-Gamma (VG) model and Compound Poisson with Normal Compounding model. The first and second derivatives of probability density function of the VG model are also computed in order to build the Fisher information matrix for the Maximum likelihood method.\\
\noindent
The VG model has been increasingly used as an alternative to the Classical Lognormal Model (CLM) in modelling asset price. The VG model with fives parameters was estimated by the FRFT. The data comes from the daily SPY ETF price data. The Kolmogorov-Smirnov (KS) goodness-of-fit shows that the VG model fits better the empirical cumulative distribution than the CLM. The best VG model comes from the FRFT estimation.\\

  \noindent
\end{abstract}

 \section {Introduction}
\noindent
Empirical studies have shown that asset returns are often characterized by peakedness, leptokurtosis and asymmetry. Most of these empirical properties were reviewed in \cite{cont2001empirical} as a set of stylized empirical facts that emerge from price variations in various types of financial markets. These facts provide evidence suggesting the assumptions of the Classical Lognormal Model (CLM) are not consistent with the empirical observations. In order to reduce the theoretical-empirical gap, it has been used in literature two sophisticated stochastic processes with the fundamental probabilistic relationships of infinite divisibility. A natural generalization of the Brownian motion, which is the L\'evy process. As a stochastic process, the L\'evy process has independent and stationary increments; and as a function, the process is right continuous with left limit. However, the L\'evy process model is usually incomplete, as the associated risk-neutral equivalent martingale measure is usually not unique; and this situation leads to many different possible prices for European options\cite {schoutens2003levy, Mercuri2010OptionPI}. \\
Another way to generalize the Classical Lognormal Model (CLM) is the method of subordination \cite {clark1973subordinated, hurst1997subordinated, Mercuri2010OptionPI}.  The subordinated process is obtained by substituting the physical time in the CLM by any independent and stationary increments random process, called the subordinator. If we consider the random process to be a Gamma process, we have a Variance Gamma (VG) model, which is the model the paper will be investigating. The Variance Gamma (VG) model was proposed by \cite{madan1990variance} as an alternative to the Classical Lognormal Model (CLM).
In contrast to the CLM, the VG model does not have an explicit closed-form for probability density function and its derivatives. However, the fundamental probabilistic relationship of infinite indivisibility allows the characteristic function of the VG model and the Fourier of probability density function to have a closed-form function.  In the study, the VG model has five parameters: parameters of location ($\mu$), symmetric ($\delta$), volatility ($\sigma$), and the Gamma parameters of shape ($\alpha$) and scale ($\theta$). The VG model density function is proven to be (\ref {eq:l01}).
\begin{align}
f(y) &=\frac {1} {\sigma\Gamma(\alpha) \theta^{\alpha}}\int_{0}^{+\infty} \frac{1}{\sqrt{2\pi v}}e^{-\frac{(y-\mu-\delta v)^2}{2v\sigma^2}}v^{\alpha -1}e^{-\frac{v}{\theta}} \,dv \label{eq:l01}
 \end{align}\\
\noindent 
The integral in (\ref {eq:l01}) makes it difficult to utilize the density function and its derivatives, and to perform the Maximum likelihood method. However, in the literature, many studies have found a way to circumvent the lack of closed-form by decreasing the number of parameters and making use of approximation function or analytical expression with modified second kind Bessel function. In fact, \cite{madan1987chebyshev} developed a procedure to approximate  (\ref {eq:l01}) by Chebyshev Polynomials expansion. \cite{madan1998variance} and \cite{seneta2004fitting} got (\ref {eq:l01}) by analytical expression with modified Bessel function of second kind and third kind respectively. \cite {Mercuri2010OptionPI} got (\ref {eq:l01}) through Gauss-Laguerre quadrature approximation with Laguerre polynomial of degree $10$.\cite{hurst1997subordinated} used the Fast Fourier Transform (FFT).\\
The paper will implement the Fractional Fourier Transform (FRFT)-based technique on the Fourier Transform of the VG model function in order to obtain probability density function and its derivatives. Contrary to the Fast Fourier Transform (FFT), the advantage of using the FRFT mentioned in \cite {Bailey1994AFM} can be found at three levels:(1) both the input function values and the output transform function values are equally spaced; (2) a large fraction of the density function is either zero or smaller than the computer machine epsilon; and (3) only a limited range of the density function are required.\\
The paper is structured as follows, the next section presents the analytical framework, with an overview of the FRFT computing formulas, their applications on three infinitely divisible distributions, and a review of the Maximum likelihood method. The third section will present the Variance Gamma (VG) model and the sample data, before performing the parameter estimations of the VG model and the Kolmogorov-Smirnov (KS) goodness-of-fit test.

\section{Analytical Framework}
\noindent
The continuous Fourier transform (CFT) of function $f(t)$ and its inverse are defined by:
\begin{align}
 \scrF[f](x) = \int_{-\infty} ^{+\infty}\! f(y)e^{-ixy} \, \mathrm{d}y \label {eq:l1}\\
 f(x) = \frac{1} {2\pi}\int_{-\infty}^{+\infty}\! \scrF[f](y) e^{ixy}\, \mathrm{d}y \label {eq:l2}\end{align}
where i is the imaginary unit.\\
\noindent
 The Fast Fourier Transform (FFT) is commonly used to evaluate the integrals (\ref {eq:l1}) and (\ref {eq:l2}) . The fundamentally inflexible nature \cite {Bailey1994AFM} of FFT is the main weakness of the algorithm. The advantages of computing with the FRFT mentioned in \cite {Bailey1994AFM} can be found at three levels: (1) both the input function values $f(x_k) $ and the output transform values $\scrF[f](x_k) $ are equally spaced;  (2) a large fraction of $f(x_k) $ are either zero or smaller than the computer machine epsilon; and (3) only a limited range of $f(x_k) $ are required.\\ The FRFT is set up on $n$-long sequence ($x_1$, $x_ {2} $, \dots, $x_{n}$) by:
 \begin{equation}
 G_k(x,\delta)=\sum_{j=0}^{n-1}\! x_ie^{-2\pi i jk\delta} \hspace{5mm}
   \hbox{$0\leq k<M$}
   \label {eq:l3}
\end{equation}  
\noindent
For $\delta=\frac{1}{n} $, we have the Discrete Fourier Transform (DFT). In the general scheme, the parameter $\delta$ is not limited to fractions; it can take complex number. It is shown in \cite {bailey1991fractional} that $ G_k(x,\delta)$ is a composition of ${DFT}^ {-1} $ and ${DFT} $.
   \begin{equation}
 G_k(x, \delta)=e^{-\pi ik^2\delta}{DFT}_k^{-1}[{{DFT}_j(y){DFT}_j(z)}] 
   \label {eq: l3}
\end{equation} 
where $y$ and $z$ are two 2n-long sequences defined by:
  \begin{align}
y_j &= e^{-2\pi ij^2\delta}   &\quad z_j &= e^{-2\pi ij^2\delta}      &\quad  \hspace{5mm}  \hbox{$0\leq k<n$}  \label{eq:l4}\\
y_j&=0   &\quad z_j&=e^{-2\pi i(j-2m)^2\delta}  &\quad  \hspace{5mm}
   \hbox{$n\leq k<2n$}
   \label{eq:l5}
  \end{align}
\noindent
As shown in $(\ref {eq: l3}) $, the FRFT is efficiently computed through the FFT on 2n-point. We assume that $\scrF[f](t)$ is zero outside the interval $[-\frac{a}{2}, \frac{a}{2}]$, and $\beta=\frac{a}{n} $ is the step size of the $n$ input values $\scrF[f](t) $; we define $t_j=(j-\frac{n}{2}) \beta$ for $ 0 \leq j <n$. We have also $\gamma$ as the step size of the $n$ output values of $f(t)$ and $x_k=(k-\frac{n}{2}) \gamma$ for $ 0 \leq k <n$. By choosing the step size $\beta$ on the input side and the step size $\gamma$ in the output side, we fix the FRFT parameter $\delta=\frac{\beta\gamma}{2\pi}$.\\
 \noindent
The density function $f$ at $x_k$ can be written as (\ref{eq:l7}). The proof is provided in Appendix \ref{eq:an01}.
 \begin{align}
 \hat{f}(x_k) = \frac{\gamma} {2\pi}e^{-\pi i(k-\frac{n}{2})n\delta}G_{k}(\scrF[f](y_{j})e^{-\pi i jn\delta}),-\delta) 
 \label {eq:l7}
 \end{align}  
\noindent
In order to perform $f(t)$ function from the Fourier Transform (FT), we assume $a=20$, $n=2048$, $\beta=\gamma=\frac{a}{n} $. For more detail on FRFT, See \cite {Bailey1994AFM}, \cite {bailey1991fractional}.

\subsection{Infinitely Divisible Distributions}
The FRFT probability density functions are computed for each of the three infinitely divisible distributions: Normal distribution with two parameters, Variance Gamma (VG) Distribution with five parameters, and Compound Poisson with Normal compounding with three parameters. The distributional properties are reviewed and the results also show that the Kurtosis statistics tell very little about the peakedness\cite{westfall2014kurtosis}.\\ For detailed proof of infinite divisibility for each distribution. See \cite{klenke2008probability},\cite{feller2008introduction},\cite{mainardi2008origin}
\subsubsection{Normal distribution}
\begin{figure}[ht]
 \centering
  \begin{subfigure}[b]{0.4\linewidth}
    \includegraphics[width=\linewidth]{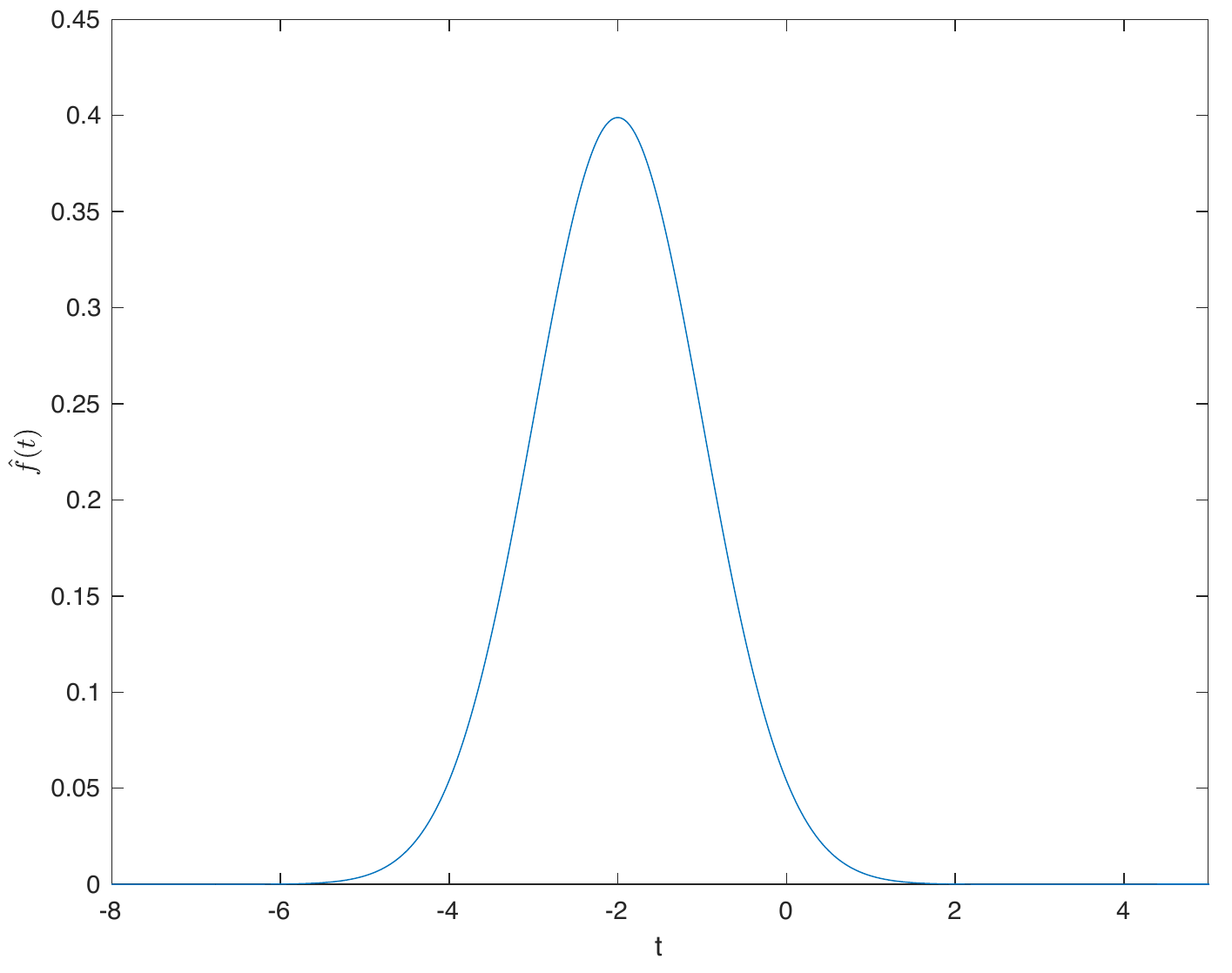}
     \caption{ $\hat{f}(t)$ with $\mu=-2$, $\sigma=1$}
         \label{fig11}
  \end{subfigure}
  \begin{subfigure}[b]{0.4\linewidth}
    \includegraphics[width=\linewidth]{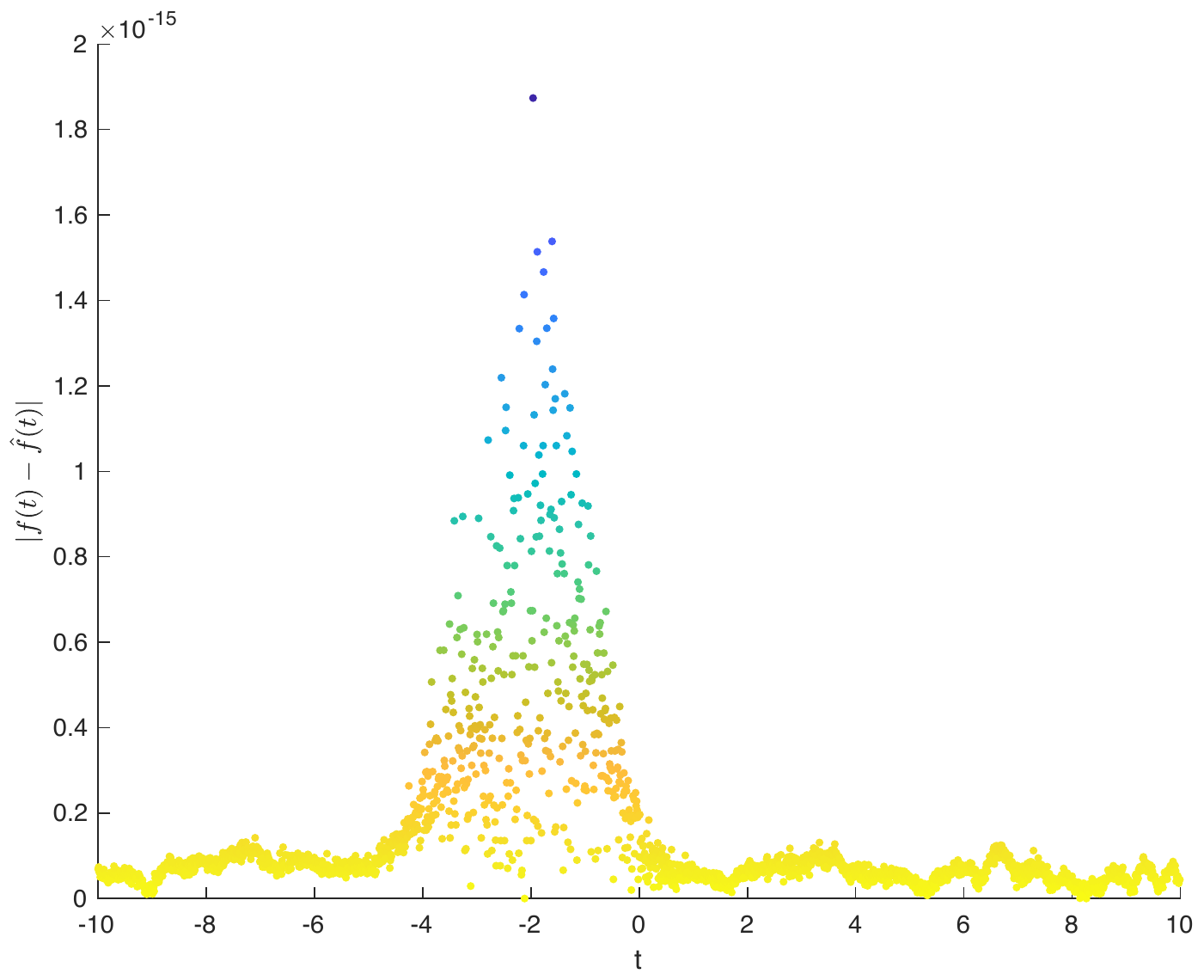}
      \caption{$\sup_{t}{|f(t)-\hat{f}(t)|}=1.8738e-15$}
         \label{fig12}
  \end{subfigure}
    \caption{FRFT estimation of the Normal density function}
  \label{fig1} 

\end{figure}

 \begin{align*}
X \sim \mu + N(0,\sigma^{2})
 \end{align*}
\noindent
The normal distribution is an infinitely divisible distribution. It was shown in \cite{klenke2008probability} that the normal distribution $N(\mu,\sigma)$ is the $n^{th}$ convolution power of a probability measure $N(\frac{\mu}{n},\frac{\sigma}{n})$, for every n. The density and the Fourier transform functions have an explicit closed form as shown below.
\begin{align}
 \scrF[f](x) &= e^{-uix -\frac{1}{2}\sigma^{2}x^2} \label {eq:l8}\\
f(x) &= \frac{1}{\sigma\sqrt{2\pi}}e^{-\frac{(y-\mu)^2}{2\sigma^2}} \label {eq:l9}
 \end{align}
For mean ($\mu=-2$) and standard deviation  ($\sigma=1$), Fig \ref{fig11} displays the FRFT estimation of the Normal density function   and  Fig \ref{fig12} shows the absolute value between the estimations and the exact values  of the normal density function.   

\noindent
The highest computation error is $1.8738e-15$, which is negligible. The probability density function is symmetric with the Kurtosis statistic $3$ for the normal distribution.

\subsubsection{Variance Gamma (VG) Distribution}
\begin{figure}[ht]
  \centering
  \begin{subfigure}[b]{0.4\linewidth}
\vspace{-0.5cm}
    \includegraphics[width=\linewidth]{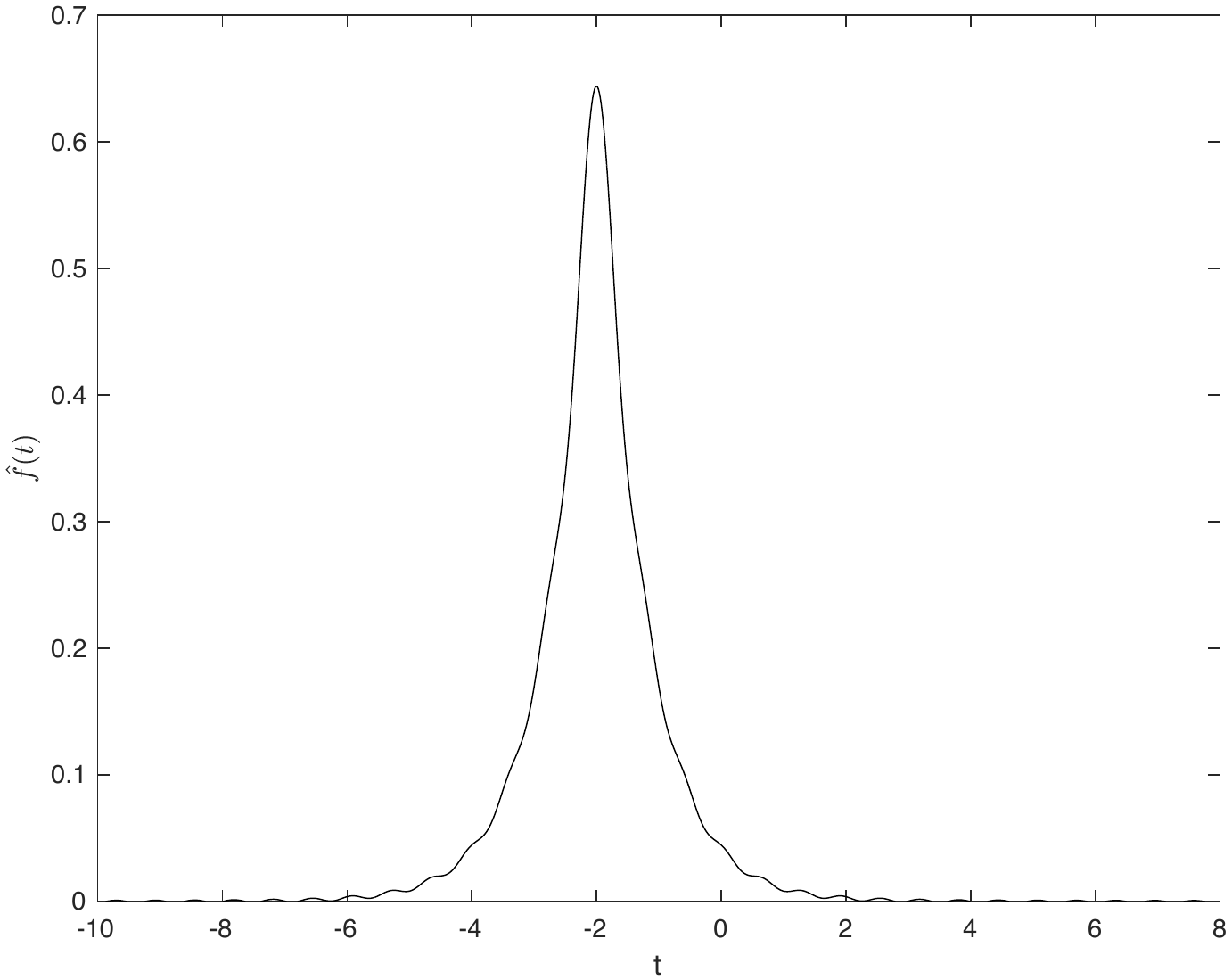}
    \vspace{-0.7cm}     
    \caption{$\hat{f}$:$\mu=-2$,$\delta=0$,$\sigma=1$,$\alpha=1$,$\theta=1$}
         \label{fig21}
  \end{subfigure}
  \begin{subfigure}[b]{0.4\linewidth}
    \vspace{-0.5cm}
    \includegraphics[width=\linewidth]{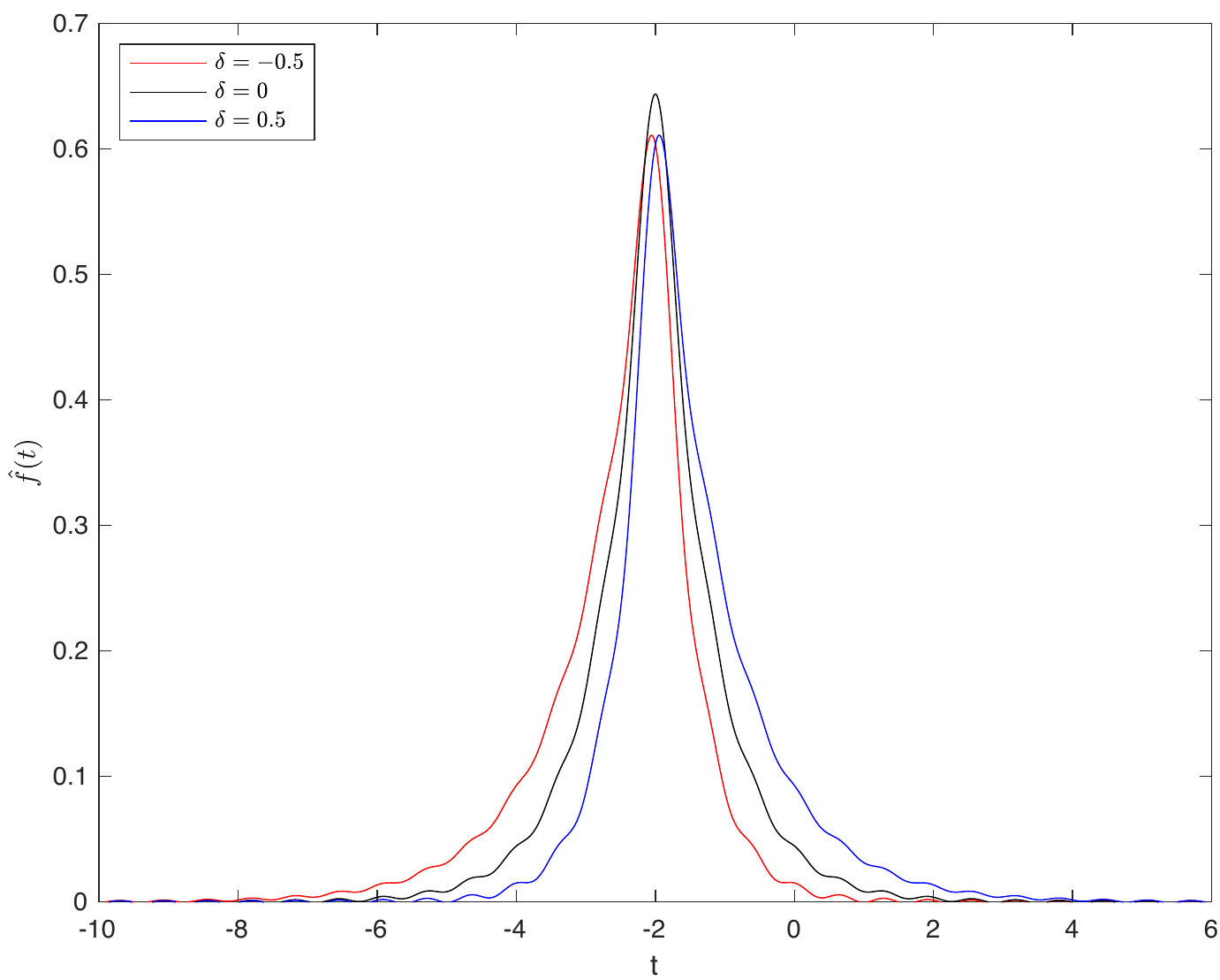}
    \vspace{-0.7cm}
      \caption{$\hat{f}(t)$ and symmetric parameter ($\delta$)}
         \label{fig22}
  \end{subfigure}\\
  \begin{subfigure}[b]{0.38\linewidth}
   \vspace{-0.2cm}
    \includegraphics[width=\linewidth]{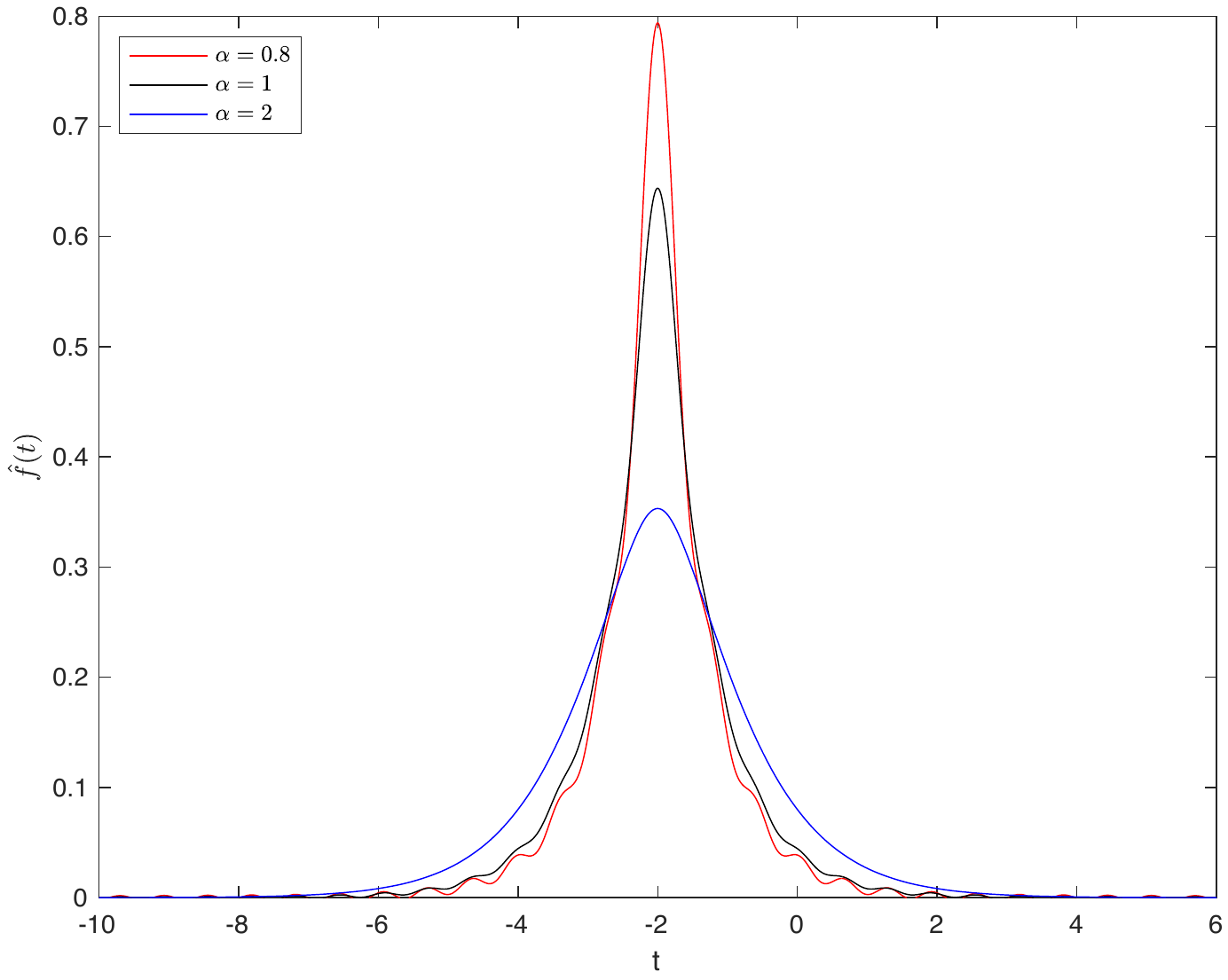}
     \vspace{-0.5cm}
      \caption{$\hat{f}(t)$ and shape parameter ($\alpha$)}
         \label{fig23}
  \end{subfigure}
  \begin{subfigure}[b]{0.38\linewidth} 
\vspace{-0.2cm}
    \includegraphics[width=\linewidth]{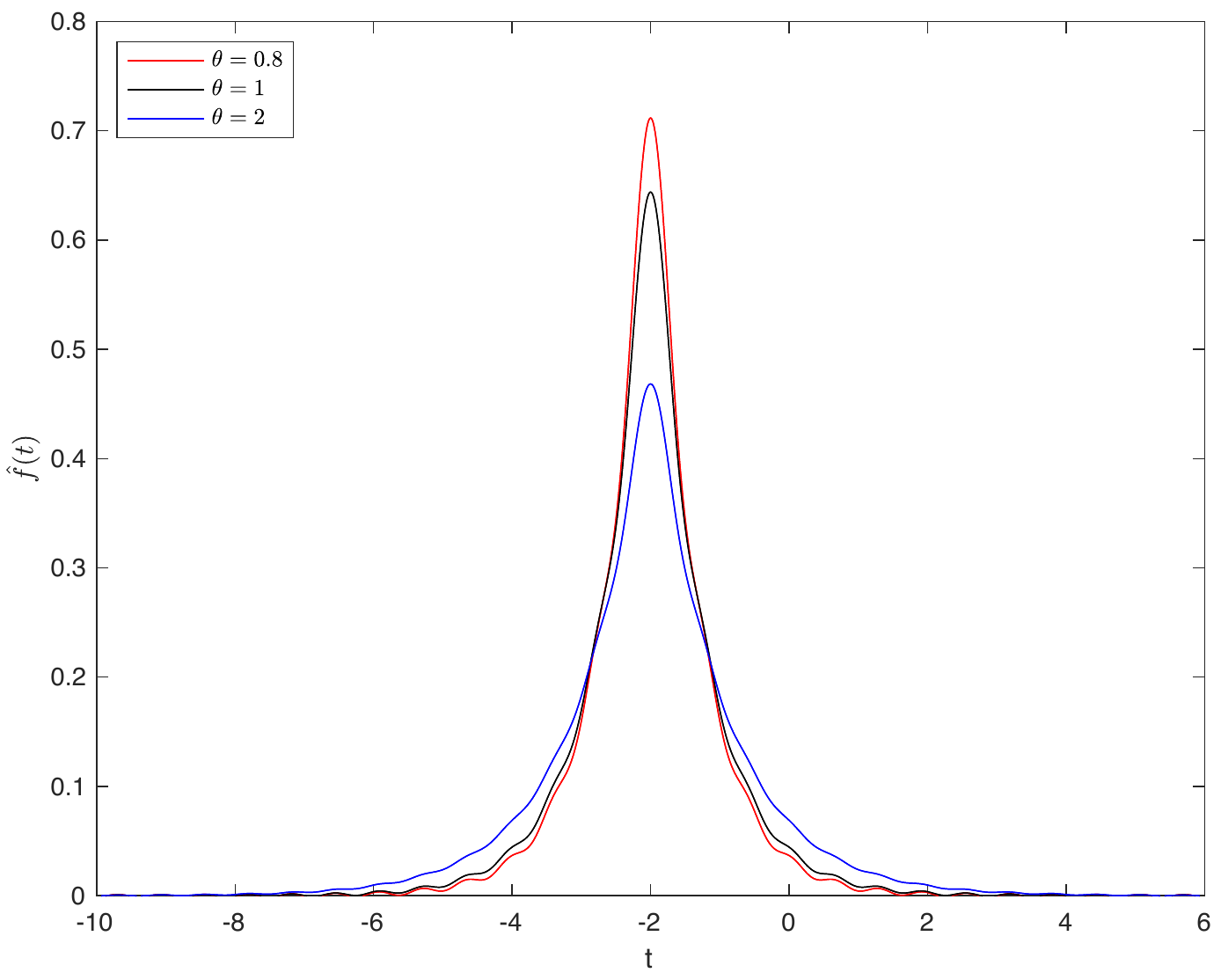}
   \vspace{-0.5cm}
     \caption{$\hat{f}(t)$ and scale parameter ($\theta$)}
         \label{fig24}
  \end{subfigure}
\vspace{-0.5cm}
  \caption{FRFT estimation of the Variance Gamma density function}
  \label{fig2}
\vspace{-0.8cm}
\end{figure}

\begin{align*}
X &=\mu + \gamma V +\sigma \sqrt{V}Z\\
Z &\sim N(0,1) \hspace{5mm}  \hbox{and } \hspace{5mm}   V \sim \Gamma(\alpha,\theta)
 \end{align*}
\noindent
The Variance Gamma distribution is an infinitely divisible distribution. The density function is (\ref {eq:l11}) and does not have a closed form. However, the Fourier transform function has explicit closed form (\ref {eq:l10}). See Appendix \ref{eq:an01} for proof
\begin{align}
 \scrF[f](x) &=\frac{e^{-i\mu x}}{\left(1+\frac{1}{2}\theta \sigma^{2}x^{2} + i\gamma\theta x\right)^{\alpha}}   \label {eq:l10}\\
 f(y) &=\frac {1} {\sigma\Gamma(\alpha) \theta^{\alpha}}\int_{0}^{+\infty} \frac{1}{\sqrt{2\pi v}}e^{-\frac{(y-\mu-\delta v)^2}{2v\sigma^2}}v^{\alpha -1}e^{-\frac{v}{\theta}} \,dv \label {eq:l11} 
  \end{align}
\noindent
When $\delta=0$, we have Symmetric Variance Gamma (SVG) Model, which is a special case of the Variance Gamma (VG) Model. It can be shown through Cumulant-generating function\cite{kendall1946advanced} that
\begin{align}
E(X)&=\mu &
Var(X)&=\alpha \theta \sigma^{2} &
Skew(X)&=0 & Kurt(X)&=3(1+\frac{1}{\alpha}) \label{eq:l011}
 \end{align} 
\noindent
For Parameter values: $\mu=-2$, $\delta=0$, $\sigma=1$, $\alpha=1$, $\theta=1$. Fig \ref{fig2} displays the FRFT estimation of the probability  density functions. As shown in Fig \ref{fig22}, the probability density is left asymmetric and right asymmetric when the parameter ($\delta$) is negative and positive respectively. For $\delta=0$, the density function is symmetric, as shown in $(\ref {eq:l011})$. The shape parameter ($\alpha$) impacts the peakedness and tails of the distribution, as illustrated in Fig \ref{fig23} and $(\ref {eq:l011})$; heavier is the tails, shorter is the peakedness. $\theta$ and $\sigma$ have the same impact on the distribution. As shown in $(\ref {eq:l011})$, both change only the variance.

\subsubsection{Compound Poisson with Normal Compounding}
\begin{figure}[ht]
  \centering
  \begin{subfigure}[b]{0.4\linewidth}
\vspace{-0.08cm}
    \includegraphics[width=\linewidth]{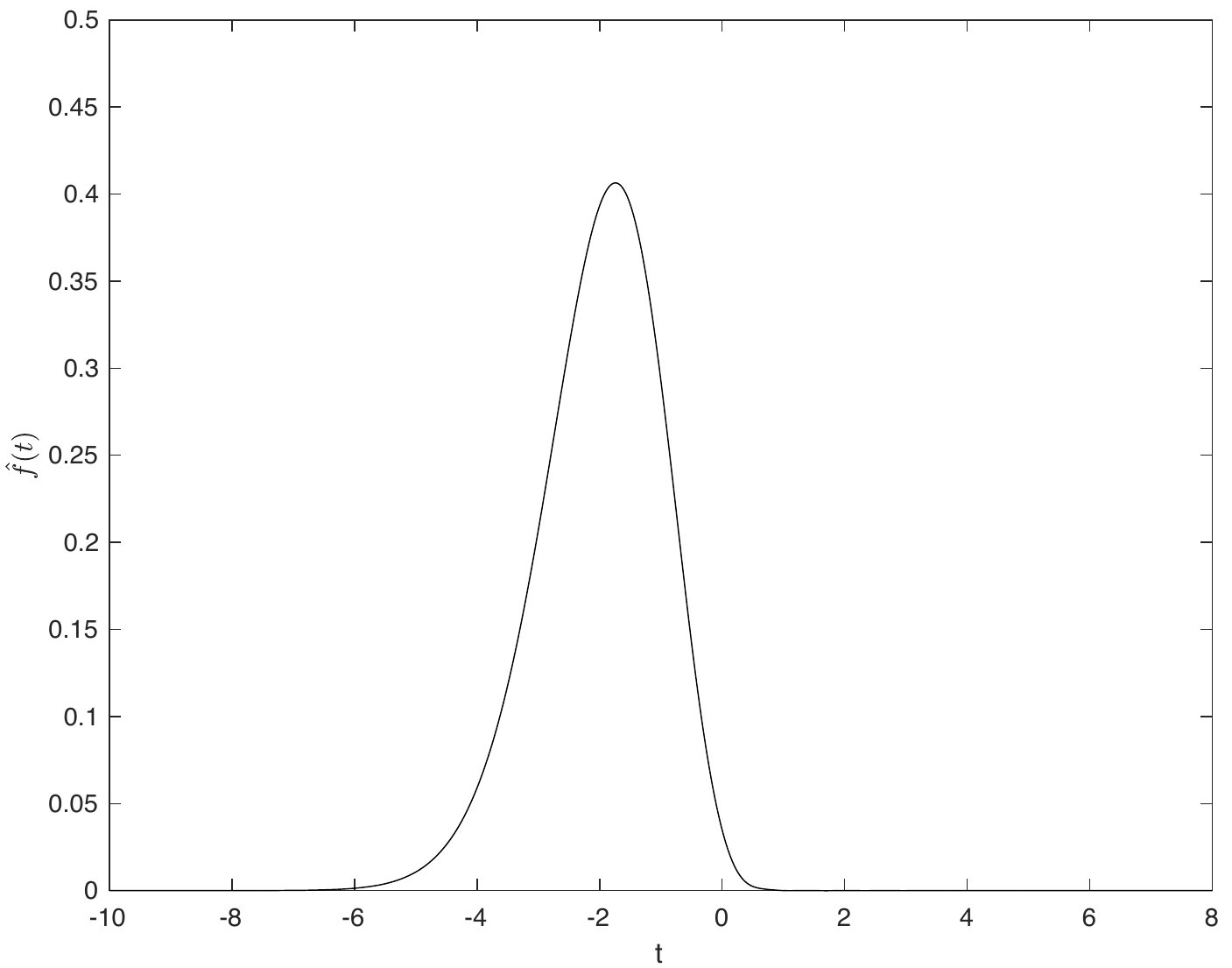}
    \vspace{-0.5cm}
     \caption{$\hat{f}(t)$, $\mu=-0.25$, $\sigma=0.25$, $\lambda=8$}
         \label{fig31}
  \end{subfigure}
  \begin{subfigure}[b]{0.4\linewidth}
\vspace{-0.1cm}
    \includegraphics[width=\linewidth]{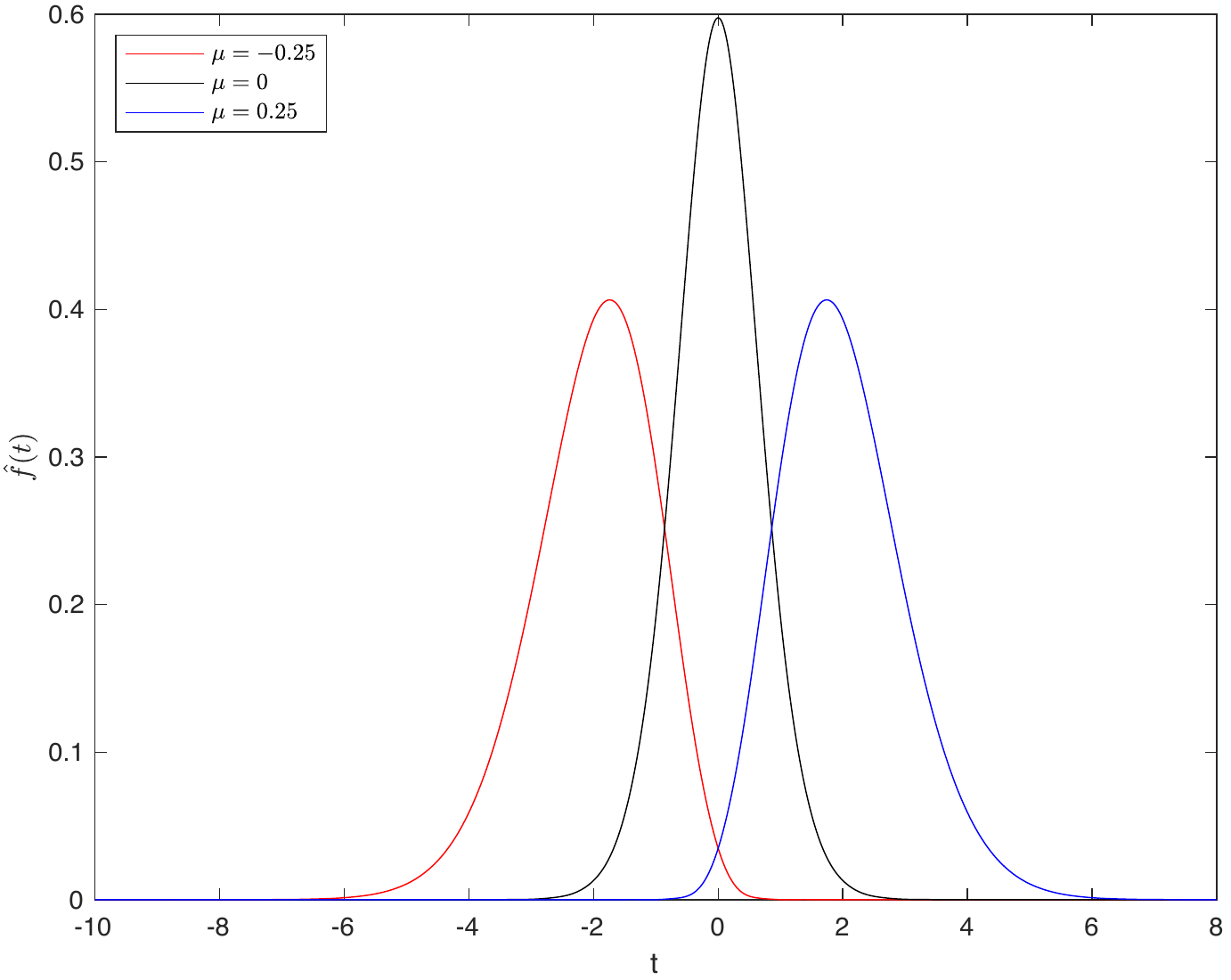} 
     \vspace{-0.5cm}
      \caption{$\hat{f}(t)$ and drift parameter ($\mu$)}
         \label{fig32}
  \end{subfigure}
  \begin{subfigure}[b]{0.4\linewidth}
\vspace{-0.2cm}
    \includegraphics[width=\linewidth]{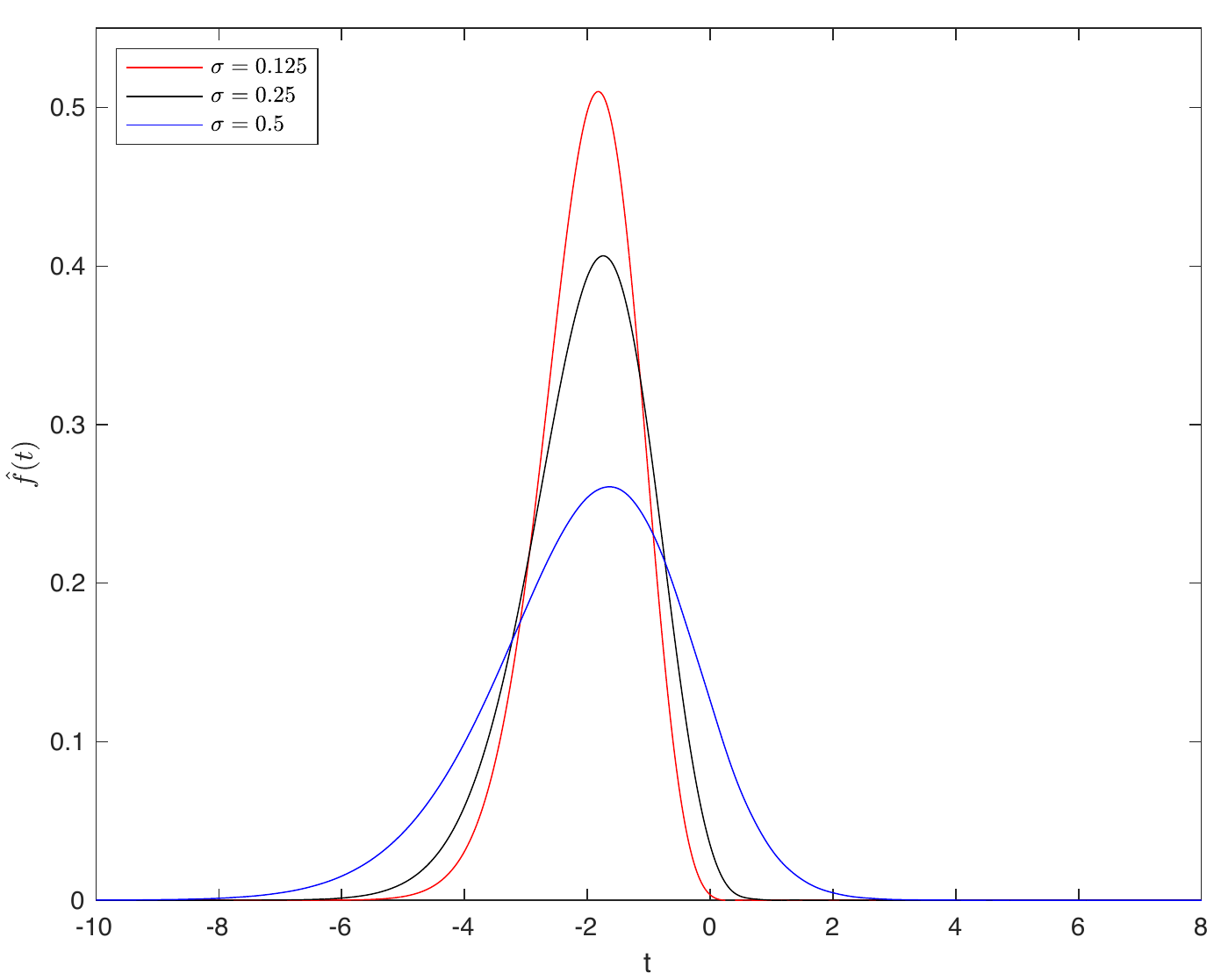}
\vspace{-0.5cm}
    \caption{$\hat{f}(t)$ and Volatility parameter ($\sigma$)}
         \label{fig33}
  \end{subfigure}
  \begin{subfigure}[b]{0.4\linewidth}
\vspace{-0.2cm}
    \includegraphics[width=\linewidth]{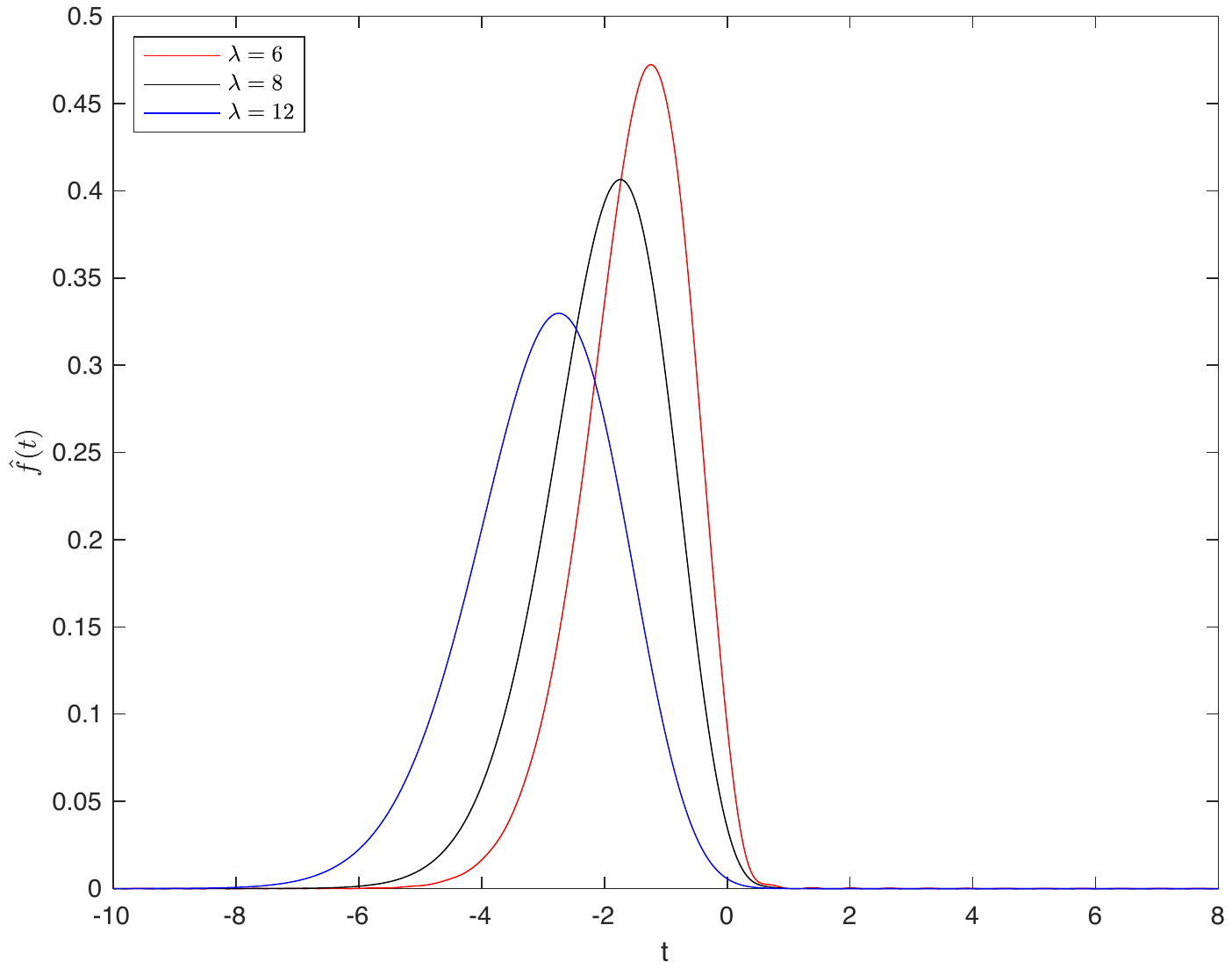}
\vspace{-0.5cm}
     \caption{$\hat{f}(t)$ and Poisson parameter ($\lambda$)}
         \label{fig34}
  \end{subfigure}
\vspace{-0.5cm}
  \caption{FRFT estimation of the Compound Poisson with Normal Compounding density function}
  \label{fig3}
\vspace{-0.5cm}
\end{figure}

\begin{align*}
X &=\sum_{i=1}^{N(\lambda)} X_{j} \\
N(\lambda) \sim Poisson(\lambda) \hspace{5mm}  & \hbox{and } \hspace{5mm}  X_{j} \sim \mu + N(0,\sigma^{2}) 
\end{align*}
\noindent
The density function was shown in \cite{press1968modified} to be (\ref{eq:l13}). However, the Fourier Transform density has explicit closed form $(\ref {eq:l12})$. See Appendix \ref{eq:an01} for proof.
    \begin{align}
 \scrF[f](x) &= e^{\lambda(e^{-uix -\frac{1}{2}\sigma^{2}x^2}-1)}  \label {eq:l12}\\
 f(x) &= \sum_{n=1}^{\infty}\frac{e^{-\lambda} \lambda^{n}}{n!}\frac{e^{-\frac{1}{2}\frac{(x-n\mu)^{2}}{n \sigma^2}}}{\sigma \sqrt{2n\pi}} \label {eq:l13}
 \end{align}
 \noindent
It can be shown through Cumulant-generating function\cite{kendall1946advanced} that
\begin{align}
E(X)&=\lambda\mu &Var(X)&=\lambda(\mu^{2} +\sigma^{2}) &Skew(X)&=\frac{\mu(\mu^{2} + 3\sigma^2)}{\sqrt{\lambda} (\mu^{2} + \sigma^{2})^{\frac{3}{2}}}&Kurt(X)&=3+\frac{\mu^{4} +6\sigma^{2} \mu^{2} + 3\sigma^{4}}{\lambda (\mu^{2}+\sigma^{2})^{2}}  \label{eq:l013}
 \end{align} 
\noindent
For Parameter values: $\mu=-0.1$, $\sigma=0.25$, $\lambda=16$. The FRFT produces the density functions in Fig \ref{fig3}. As shown in Fig \ref{fig32}, the probability density is asymmetric, shifts to the left and to the right when the parameter ($\mu$) is negative and positive respectively. For $\mu=0$, we have a symmetric density function. $\mu$ and $\sigma$ impact the symmetry, peakedness and tails of the distribution, as illustrated in Fig \ref{fig33} and $(\ref {eq:l013})$; heavier is the tails, shorter is the peakedness.  The Poisson parameter ($\lambda$) impacts all the statistics  in $(\ref {eq:l013})$. $\lambda$ affects  the asymmetry, peakedness and tails of the distribution simultaneously.
   
\subsection{Review of the Maximum Likelihood Method}
\begin{figure}[ht]
\vspace{-0.1cm}
  \centering
  \begin{subfigure}[b]{0.4\linewidth}
    \includegraphics[width=\linewidth]{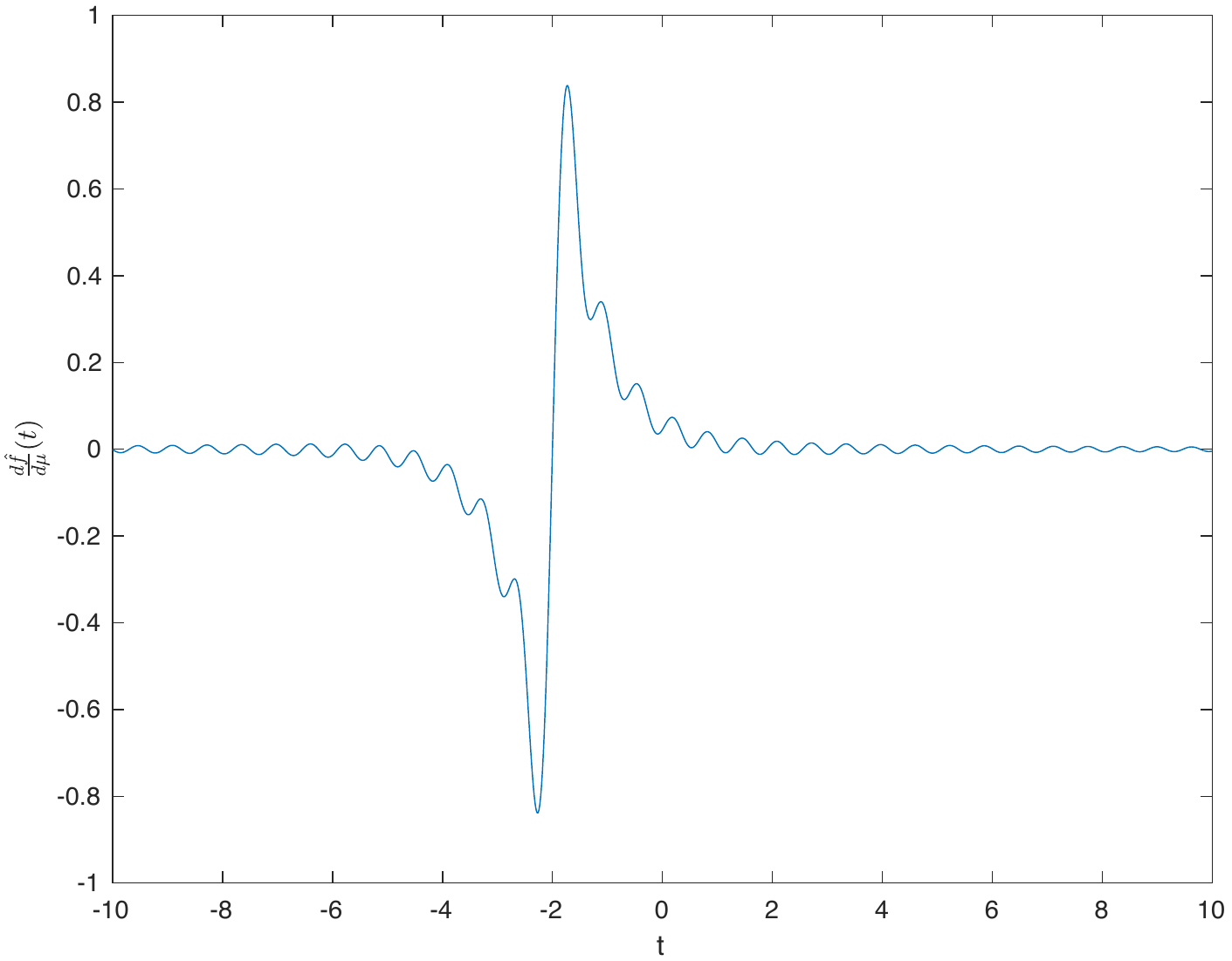}
\vspace{-0.7cm}
     \caption{$\frac{d\hat{f}}{du}(t)$ }
         \label{fig41}
  \end{subfigure}
  \begin{subfigure}[b]{0.4\linewidth}
    \includegraphics[width=\linewidth]{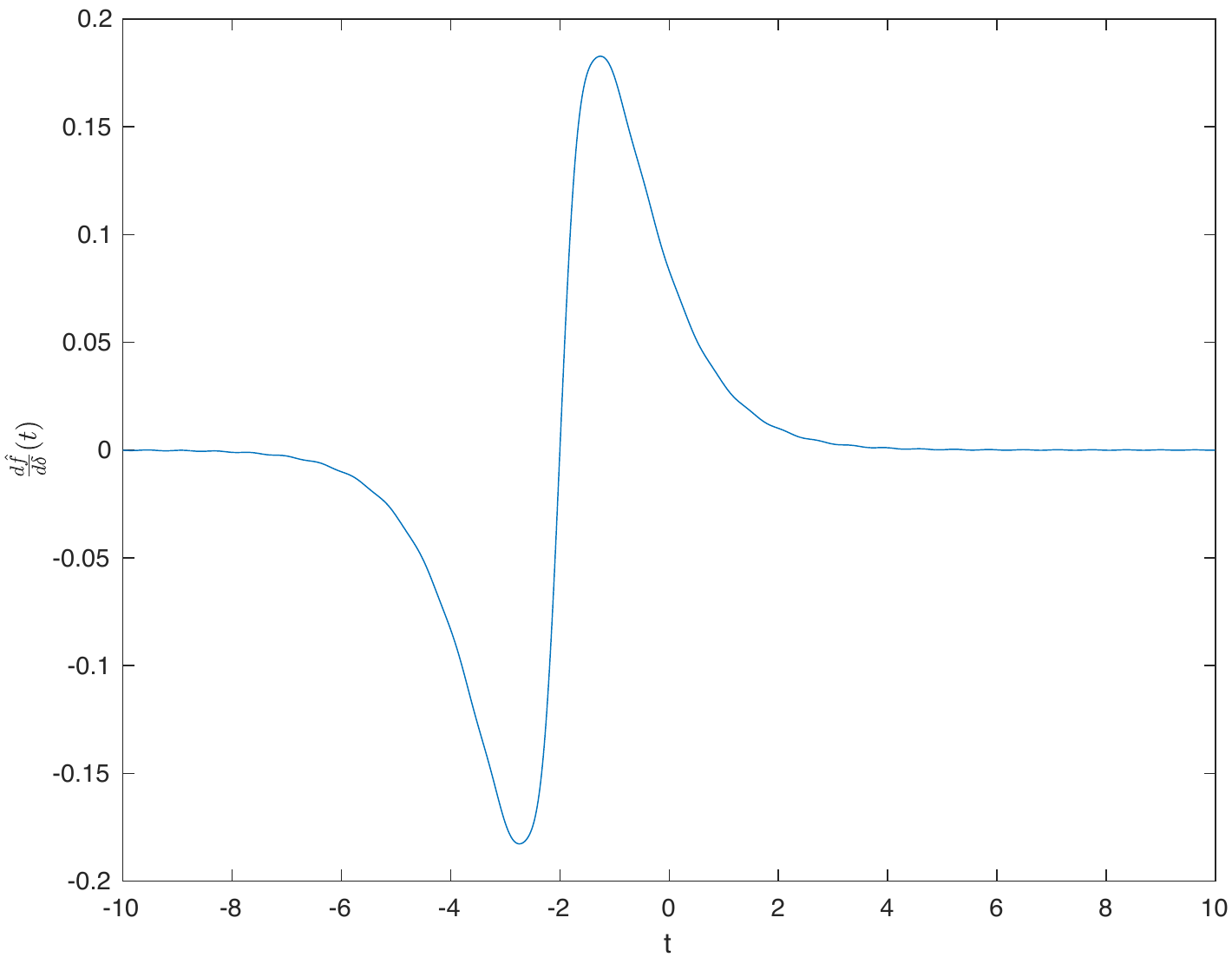}
\vspace{-0.7cm}
      \caption{$\frac{d\hat{f}}{d\delta}(t)$}
         \label{fig42}
  \end{subfigure}\\
  \begin{subfigure}[b]{0.4\linewidth}
    \vspace{-0.2cm}
    \includegraphics[width=\linewidth]{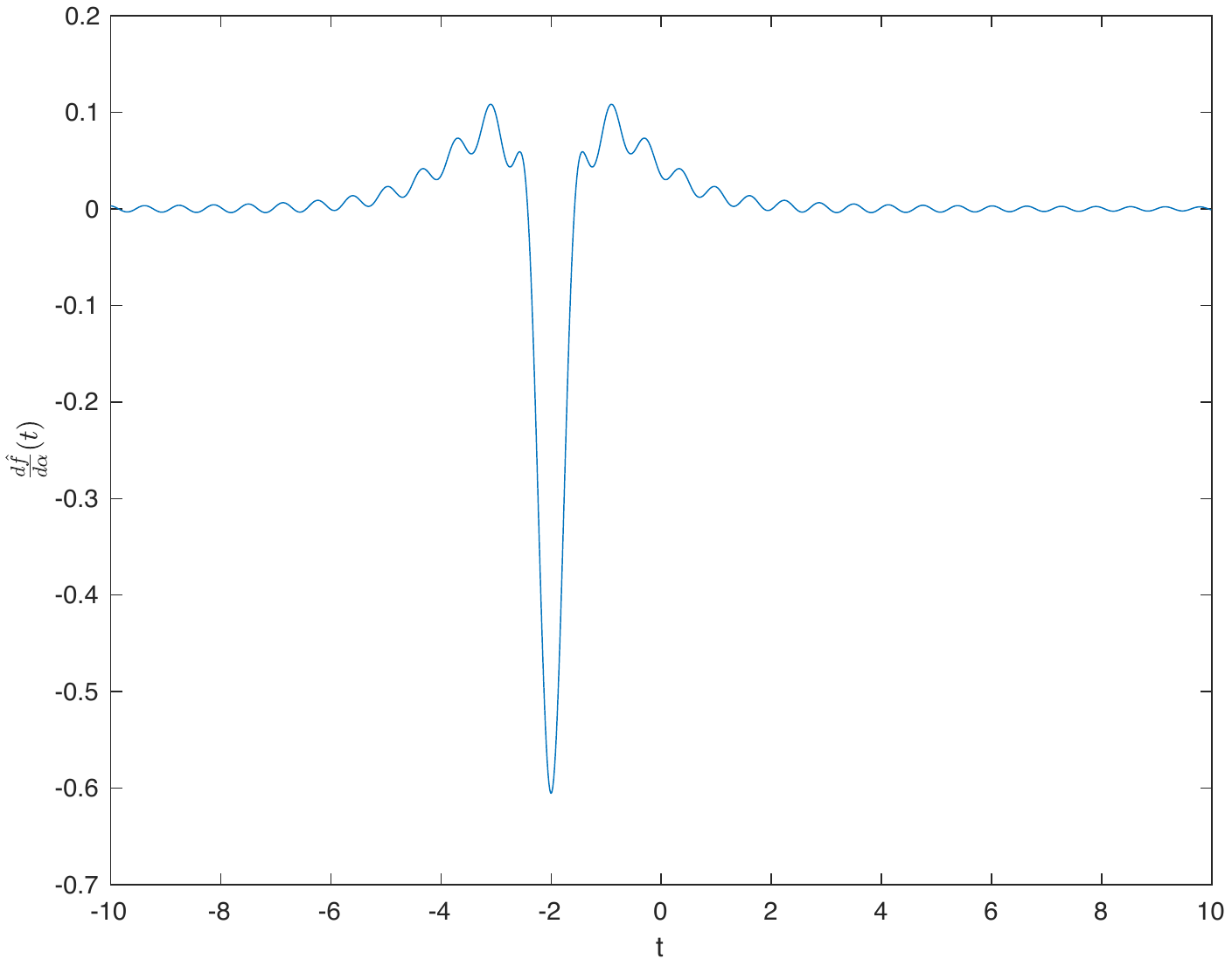}
    \vspace{-0.7cm}
    \caption{$\frac{d\hat{f}}{d\alpha}(t)$}
         \label{fig43}
  \end{subfigure}
  \begin{subfigure}[b]{0.4\linewidth}
     \vspace{-0.2cm}
    \includegraphics[width=\linewidth]{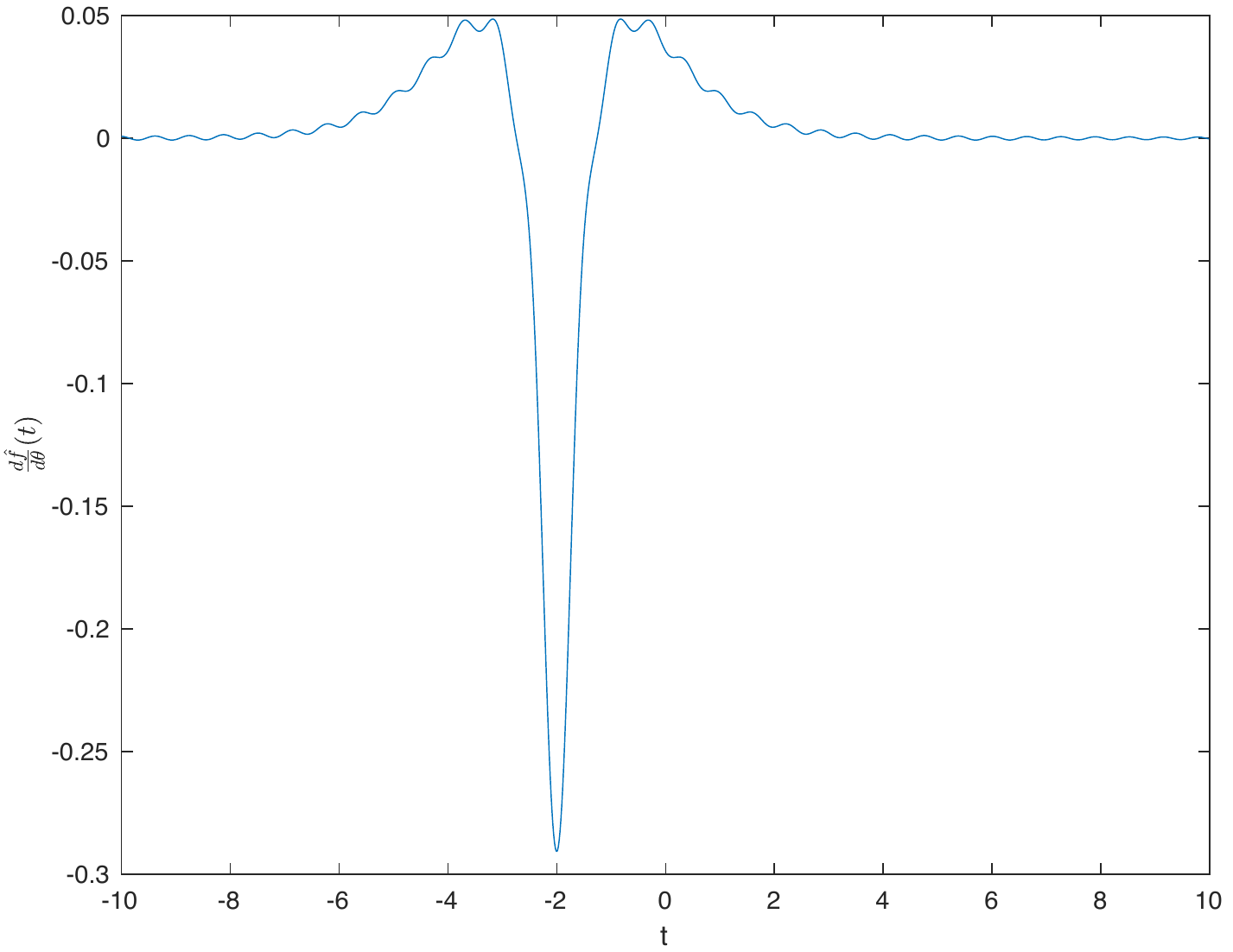}
   \vspace{-0.7cm}
     \caption{$\frac{d\hat{f}}{d\theta}(t)$}
         \label{fig44}
  \end{subfigure}
 \vspace{-0.5cm}
  \caption{ First order derivative of VG density function with $\mu=-2$, $\delta=0$, $\sigma=1$, $\alpha=1$, $\theta=1$}
  \label{fig4}
\vspace{-0.7cm}
\end{figure}

\noindent
From a probability density function $f(x,V)$ with parameter $V$ of size $p$ and the sample data $x$ of size $m$,  we definite  the Likelihood Function as shown in (\ref{eq:l14}).
  \begin{align}
 L(x,V) &= \prod_{j=1}^{m} f(x_{j},V)  \label {eq:l14}\\
 l(x,V) &= \sum_{j=1}^{m} log(f(x_{j},V))  \label {eq:l15}
 \end{align}

\begin{align}
 \frac{dl(x,V)}{dV_j} &= \sum_{i=1}^{m} \frac{\frac{df(x_{i},V)}{dV_j}}{f(x_{i},V)}\hspace{5mm}  \hbox{$1\leq j \leq p$} \label {eq:l16}\\
  \frac{d^{2}l(x,V)}{dV_{k}dV_{j}} &= \sum_{i=1}^{m} \left(\frac{\frac{d^{2}f(x_{i},V)}{dV_{k}dV_{j}}}{f(x_{i},V)}- \frac{\frac{df(x_{i},V)}{dV_{k}}}{f(x_{i},V)}\frac{\frac{df(x_{i},V)}{dV_j}}{f(x_{i},V)}\right)\hspace{5mm}  \hbox{$1\leq k\leq p \ \ and \ \ 1\leq j \leq p$} \label {eq:l17}
 \end{align}
\noindent
In order to perform the Maximum of the likelihood function  (\ref{eq:l14}), the quantities  $\frac{dl(x,V)}{dV_j}$ and $\frac{d^{2}l(x,V)}{dV_{k}dV_{j}}$ must be computed. We need  to compute  $\frac{df(x,V)}{dV_j}$ and $\frac{d^{2}f(x,V)}{dV_{k}dV_{j}}$, which are the first and second order derivative of the density function respectively. Variance Gamma (VG) distribution  has five parameters ($\mu$, $\delta$, $\sigma$, $\alpha$, $\theta$), using the closed form of Fourier transform in (\ref{eq:l10}) and the FRFT algorithm, we perform the first derivative functions:$\frac{df}{d\mu}$, $\frac{df}{d\delta}$, $\frac{df}{d\sigma}$, $\frac{df}{d\alpha}$ and  $\frac{df}{d\theta}$. The results of the computation are shown in Fig \ref{fig4}.

\noindent
As shown in Fig \ref{fig4}, we have two Odd functions in Fig \ref{fig41} and \ref{fig42} generated by the drift parameters $\mu$ and $\delta$; and the remainings are Even functions. The functions in Fig \ref{fig43}, and Fig \ref{fig44} are symmetric along the vertical line $x=-2$, whereas $\frac{df(x,V)}{d\mu}$ and $\frac{df(x,V)}{d\delta}$ have rotational symmetry at the point ($-2$,$0$). In addition, each graph has positive area values and negative area values. Having such properties is  a necessary condition to find $V(\mu, \delta, \sigma, \alpha, \theta)$ such as $(\frac{dl(x,v)}{d\mu},\frac{dl(x,v)}{d\delta},\frac{dl(x,v)}{d\sigma},\frac{dl(x,v)}{d\alpha},\frac{dl(x,v)}{d\theta})$ equals zero.\\
\noindent
The quantities $\frac{d^{2}l(x,V)}{dV_{k}dV_{j}}$ in (\ref{eq:l17}) are critical in computing the Hessian Matrix and the Fisher Information Matrix. For VG distribution, We need fifteen second derivative functions. The results of four functions are shown in Fig \ref{fig5}. The remaining functions are reported in Appendix \ref{eq:an02} and Appendix \ref{eq:an03}.
\begin{figure}[ht]
  \centering
  \begin{subfigure}[b]{0.4\linewidth}
    \vspace{-0.1cm}
    \includegraphics[width=\linewidth]{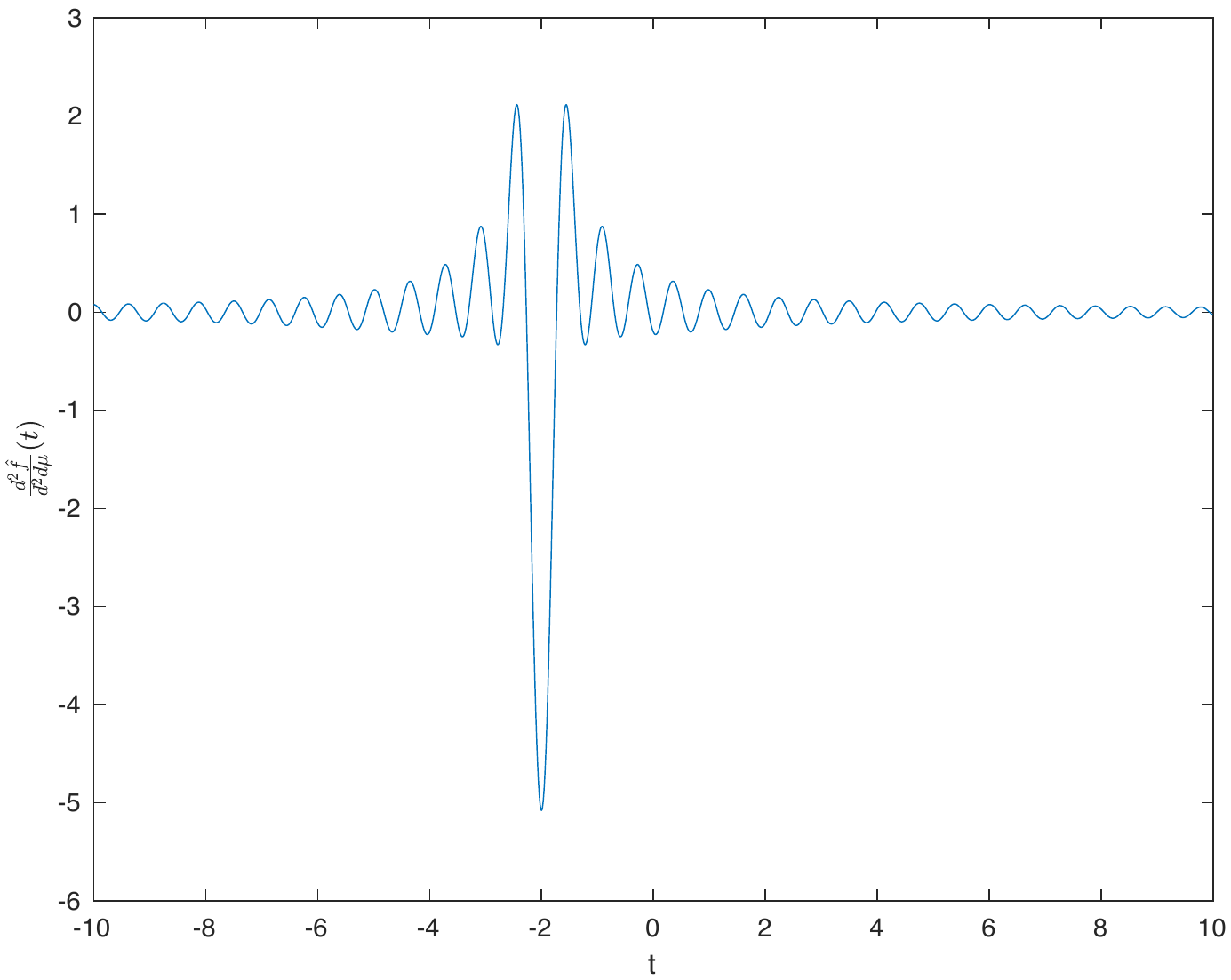}
     \vspace{-0.5cm}
     \caption{$\frac{d^{2}\hat{f}}{d^{2}\mu}(t)$ }
         \label{fig51}
  \end{subfigure}
  \begin{subfigure}[b]{0.4\linewidth}
\vspace{-0.1cm}
    \includegraphics[width=\linewidth]{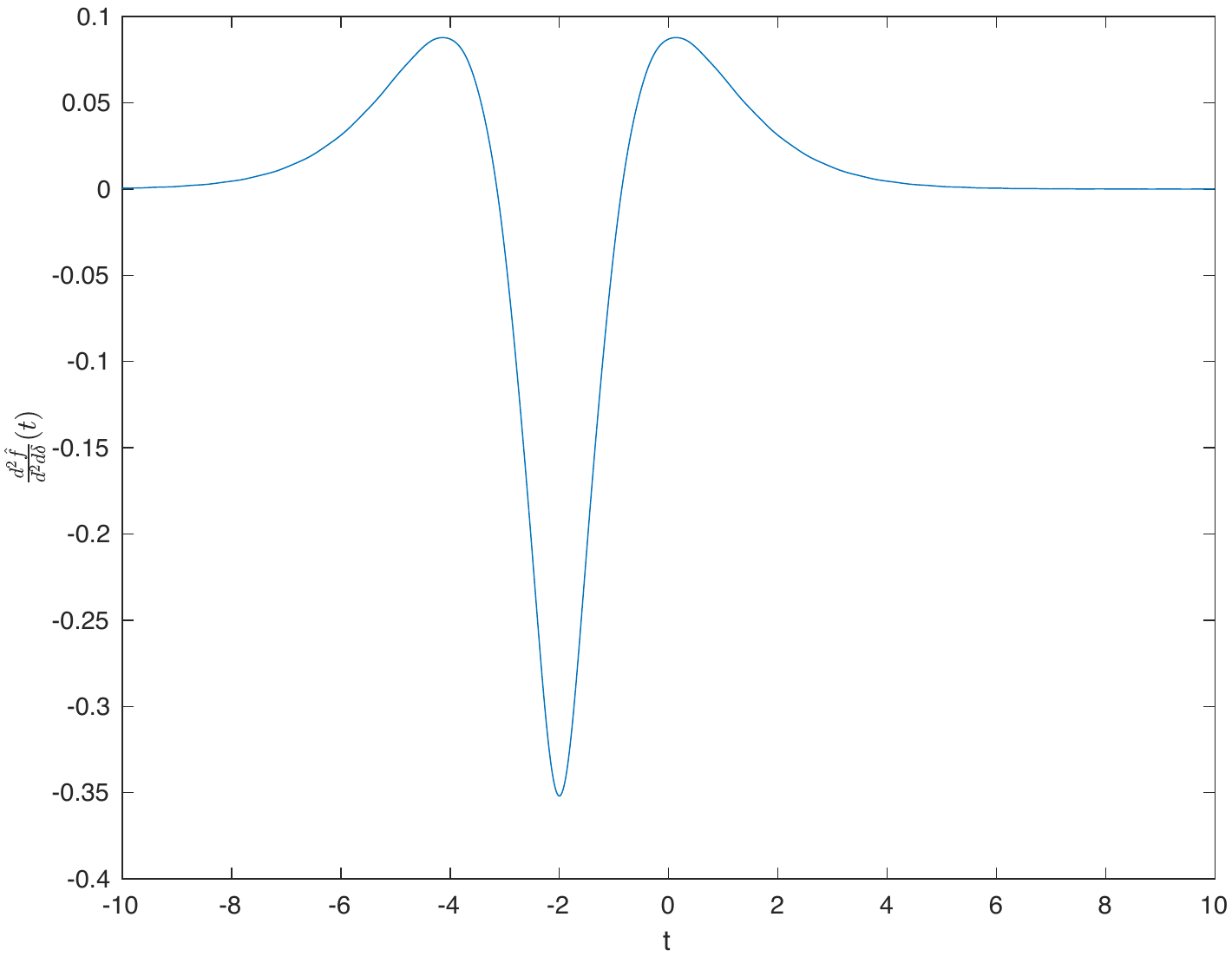} 
\vspace{-0.7cm}
      \caption{$\frac{d^{2}\hat{f}}{d^{2}\delta}(t)$}
         \label{fig52}
  \end{subfigure}\\
  \begin{subfigure}[b]{0.4\linewidth}
\vspace{-0.2cm}
    \includegraphics[width=\linewidth]{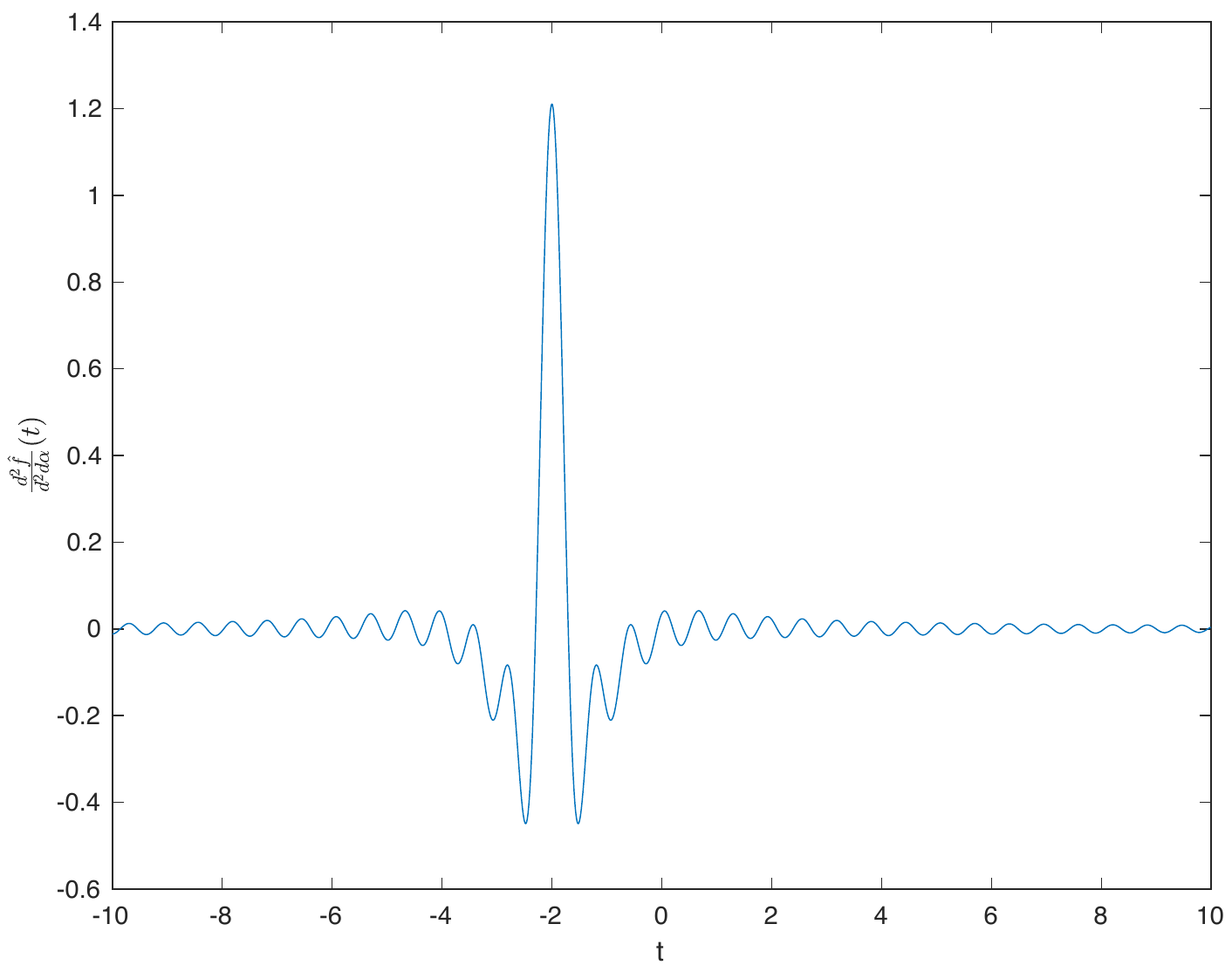}
\vspace{-0.7cm}
    \caption{$\frac{d^{2}\hat{f}}{d^{2}\alpha}(t)$}
         \label{fig53}
  \end{subfigure}
  \begin{subfigure}[b]{0.4\linewidth}
\vspace{-0.2cm}
    \includegraphics[width=\linewidth]{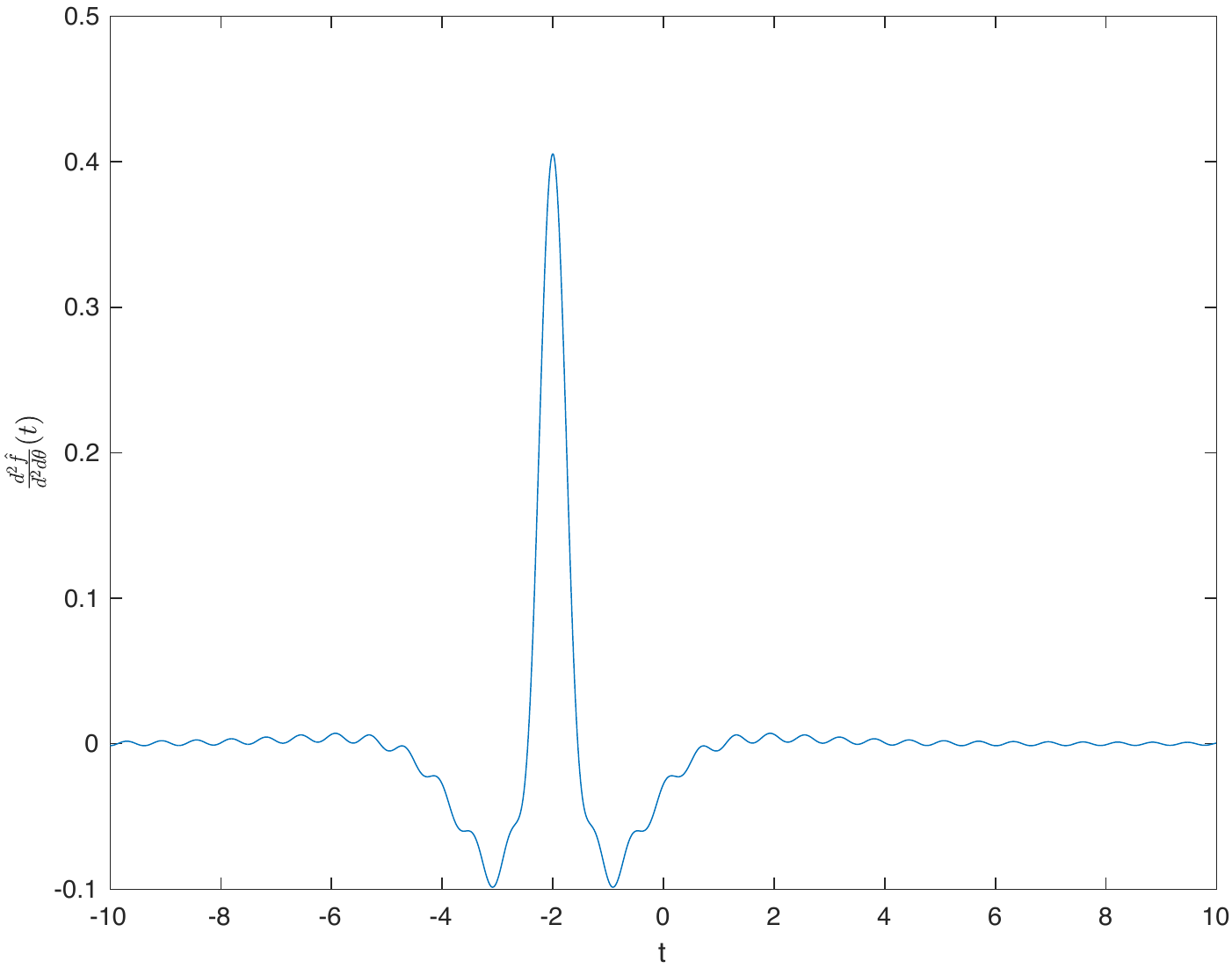}
\vspace{-0.7cm}
     \caption{$\frac{d^{2}\hat{f}}{d^{2}\theta}(t)$}
         \label{fig54}
  \end{subfigure}
\vspace{-0.3cm}
  \caption{Second order derivative of VG density function with $\mu=-2$, $\delta=0$, $\sigma=1$, $\alpha=1$, $\theta=1$}
  \label{fig5}
\vspace{-0.3cm}
\end{figure}

\noindent
As shown in Fig \ref{fig5}, all four functions are symmetric along the vertical line $x=-2$. We have the same characteristics for $\frac{d^{2}\hat{f}}{d\delta d\mu}$, $\frac{d^{2}\hat{f}}{d^{2}\sigma}$, $\frac{d^{2}\hat{f}}{d\alpha d\sigma}$, $\frac{d^{2}\hat{f}}{d\theta d\sigma}$, $\frac{d^{2}\hat{f}}{d\theta d\alpha}$ in Appendix \ref{eq:an02} and Appendix \ref{eq:an03}. The remaining derivatives have rotational symmetry at the point ($-2$,$0$). These features can be checked analytically through the Inverse Fourier Transform in (\ref {eq:l2}) and $\scrF[f]$ in (\ref {eq:l10}).\\
 \noindent
Given the parameters $V(\mu, \delta, \sigma, \alpha, \theta)$ and the sample data set $x$, from the previous development and computations,  we have the quantities:
  \begin{align}
 I'(x,V) =\left(\frac{dl(x,V)}{d\mu},\frac{dl(x,V)}{d\delta}, \frac{dl(x,V)}{d\sigma}, \frac{dl(x,V)}{d\alpha}, \frac{dl(x,V)}{d\theta}\right) \label {eq:l18}
 \end{align}
 
   \begin{align}
 I''(x,V)&= \begin{bmatrix}
\frac{d^{2}l}{d^{2}\mu} & \frac{d^{2}l}{d\delta d\mu} & \frac{d^{2}l}{d\sigma d\mu} & \frac{d^{2}l}{d\alpha d\mu} & \frac{d^{2}l}{d\theta d\mu}\\
\frac{d^{2}l}{d\delta d\mu} & \frac{d^{2}l}{d^{2}\delta} & \frac{d^{2}l}{d\sigma d\delta}   & \frac{d^{2}l}{d\delta d\alpha } & \frac{d^{2}l}{d\theta d\delta}\\
\frac{d^{2}l}{d\sigma d\mu}&  \frac{d^{2}l}{d\sigma d\delta}&\frac{d^{2}l}{d^{2}\sigma} & \frac{d^{2}l}{d\alpha d\sigma} & \frac{d^{2}l}{d\theta d\sigma}\\
\frac{d^{2}l}{d\alpha d\mu} & \frac{d^{2}l}{d\alpha d\delta} &\frac{d^{2}l}{d\alpha d\sigma} & \frac{d^{2}l}{d^{2}\alpha} & \frac{d^{2}l}{d\theta d\alpha}\\
\frac{d^{2}l}{d\theta d\mu} & \frac{d^{2}l}{d\theta d\delta}&\frac{d^{2}l}{d\theta d\sigma} & \frac{d^{2}l}{d\alpha d\theta} & \frac{d^{2}l}{d^{2}\theta}\\
\end{bmatrix}
\label {eq:l19}
 \end{align}

\noindent
$I'(x,V)$ is the score function with the property \cite {giudici2013wiley} that $E(I'(x,V))=0$.  $-I''(x,V)$ is the observed Fisher information and $I(V)=E(I''(x,V))$ is the Fisher information matrix.\\
 \noindent 
 Maximizing  L(x,V) in (\ref{eq:l14}) with respect to $V$ is equivalent to solving the equation system (\ref{eq:l20}).
  \begin{align}
 I'(x,V) =0 \label {eq:l20}
 \end{align}
 \noindent
  The solutions of (\ref{eq:l20}) are provided by the Newton--Raphson Iteration Algorithm (\ref{eq:l21}). In fact, the $Taylor$'s expansion can be applied to the score function to derive the Newton--Raphson Iteration Algorithm.
    \begin{align}
V^{n+1}=V^{n}+{\left(I''(x,V^{n})\right)^{-1}}I'(x,V^{n})\label {eq:l21}
 \end{align}
\noindent
For more detail on Maximum likelihood and Newton–Raphson Iteration procedure, see \cite{giudici2013wiley},\cite{chesneau2017estimateur},\cite{willis2020maximum},\cite{edwards1984likelihood},\cite{held2014applied}.
\section{Fitting Variance Gamma (VG) Model}
 \subsection {VG Model and Asset/Index price}
\noindent
The VG model was introduced by Madan et al\cite{madan1990variance,seneta2004fitting}. The idea \cite{o2010variance} was to model the asset price on bussiness time rather than on calender time. The asset price is modelled on bussiness time $(k)$ as follows:
\begin{align}
 Y_{k}=\mu + \delta V_{k} +\sigma \sqrt{V_{k}}Z  \quad \quad  Z \sim N(0,1) \quad \quad V_{k}\sim \Gamma(\alpha,\theta)  \label {eq:l20a}
 \end{align} 
\begin{align}
 S_{K}=S_{k-1}e^{\sum_{j=k}^{K}Y_{j}}  \quad  \quad  T_{K}=\sum_{k=1}^{K}V_{k} \label {eq:l20b} 
 \end{align} 
\noindent 
Where $\mu, \delta \in R$ , $\sigma>0$, $\alpha>0$ and $\theta>0$.  $\mu$ is the drift of the physical time scale, $\delta$ is the drift of the activity time process, and $\sigma$ is the volatility. $\{T_{k}\}$ is the activity time process\cite{finlay2009variance}  or intrinsic time process\cite{hurst1997subordinated} and is supposed to be non-negative stationary independent increments. This process is associated  with either the flow of new price-sensitive information or the cumulative trading volume process. $Y_{k} $ is the return variable of the stock or index  price, we have (\ref{eq:l21a}) from (\ref{eq:l20a}) and (\ref{eq:l20b}).
\begin{align}
Y_{k} =log(\frac{S_{k}}{S_{k-1}})  \quad  \quad  E(Y_{k}|V_{k}) \sim N(\mu +\delta V_{k}, V_{k}\sigma^{2}) \quad \quad V_{k}\sim \Gamma(\alpha,\theta) \label{eq:l21a} 
 \end{align} 

\noindent 
The density of $Y_{j}$  in (\ref{eq:l20a}) and (\ref{eq:l21a}) has the following expression
\begin{align}
f(y) &=\frac {1} {\sigma\Gamma(\alpha) \theta^{\alpha}}\int_{0}^{+\infty} \frac{1}{\sqrt{2\pi v}}e^{-\frac{(y-\mu-\delta v)^2}{2v\sigma^2}}v^{\alpha -1}e^{-\frac{v}{\theta}} \,dv \label{eq:l23}
 \end{align} 
\noindent
The Fourier transform of (\ref{eq:l23})  can be shown to be . 
  \begin{align}
 F[f](x) =E[e^{-ixY_{j}}]=\frac{e^{-i\mu x}}{\left(1+\frac{1}{2}\theta \sigma^{2}x^{2} + i\delta \theta x\right)^{\alpha}}  \label {eq:l24a}
 \end{align}
 See A.1 for proof (\ref{eq:l24a}).\\
 \noindent
  For $\alpha=\frac{1}{\theta}$  and $0<\theta<<1$, $Y_{k}$ in (\ref{eq:l20a}) becomes
   \begin{align}
Y_{k} \sim N(\mu + \delta,\sigma^2)  \label {eq:l24b}
 \end{align}
See A.1 for proof (\ref{eq:l24b})\\ 

\noindent   
The relationship between means $(E(Y_{j}))$, variance $(Var(Y_{j}))$, Skewness $(Skew(Y_{j}))$ and Kurtosis $(Kurt(Y_{j}))$ indicators  and the VG model with five parameters can be shown by Cumulant-generating function\cite{kendall1946advanced}.
\begin{align}
E(Y_{j}) &=\mu + \alpha \theta\delta \label{eq:l25b}\\
Var(Y_{j}) &=\alpha (\delta^{2} \theta^{2} + \theta \sigma^2) \label{eq:l26}\\
Skew(Y_{j}) &=\frac{\delta (2\delta^2 + 3\frac{\sigma^2}{\theta})}{\alpha^{\frac{1}{2}}(\delta^2 + \frac{\sigma^2}{\theta})^\frac{3}{2}} \label{eq:l27}\\
Kurt(Y_{j}) &=3\left(1 +\frac{2\delta^4 + \frac{\sigma^4}{\theta^2} + \frac{4\sigma^2 \delta^2}{\theta}}{\alpha(\delta^2 + \frac{\sigma^2}{\theta})^2}\right)\label{eq:l28}
 \end{align} 
\noindent
By applying the method of Moments, the system of equations has more parameters than equations. The issue can be overcome by imposing\cite{Mercuri2010OptionPI} $E(V)=1$ or $\sigma=1$. $E(V)=1$ was chosen in \cite{seneta2004fitting,goncu2013fitting} and the implication was that the gamma shape and scale parameters should be equal $(\alpha=\theta)$. In the next section, we will assume $\sigma=1$ in order to solve the system of equation (\ref{eq:l25b})\dots(\ref{eq:l28}). However, the constraints will be relaxed in the Maximum Likelihood method where all five VG parameters will be estimated. 
\subsection{data}
\begin{figure}[ht]
\vspace{-0.5cm}
     \centering
         \includegraphics[scale=0.44] {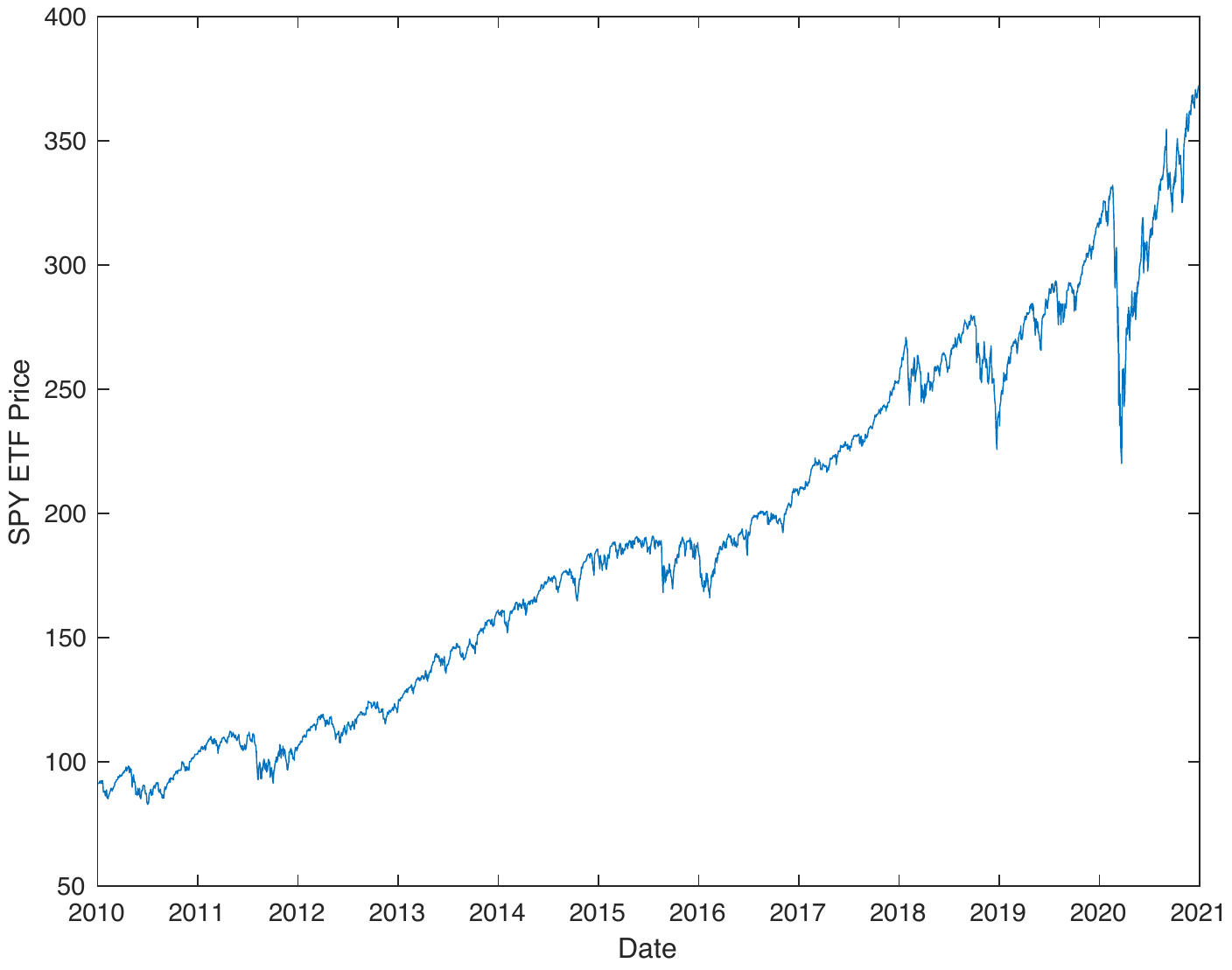}
\vspace{-0.5cm}
        \caption {Daily SPY ETF Price}
        \label{fig61}
\vspace{-0.5cm}
\end{figure}

 \noindent 
The data comes from the SPY ETF, called SPDR S\&P 500   ETF (SPY). The SPY is an Exchange-Traded Fund (ETF) managed by State Street Global Advisors that tracks the Standard \& Poor's 500 index (S\&P 500 ), which comprises 500 large and mid-cap US stocks. The SPY is a well-diversified basket of assets, listed on the New York Stock Exchange (NYSE). Like other ETFs, SPY provides the diversification of a mutual fund and the flexibility of a stock.\\
    \noindent 
The SPY data was extracted from Yahoo finance. The daily data was adjusted for splits and dividends. The period spans from January 4, 2010 to December 30, 2020. 2768 daily SPY ETF price was collected, around 252 observations per year, over 10 years. The dynamic of daily adjusted SPY price is provided in Fig \ref{fig61}.\\
\noindent
As a whole, the SPY price has an increasing trend during the period  2010-2020. The increase is temporally disrupted in the first quarter of 2020 by the coronavirus pandemic.\\

\begin{figure}[ht]
    \centering
  \begin{subfigure}[b]{0.45\linewidth}
    \includegraphics[width=\linewidth]{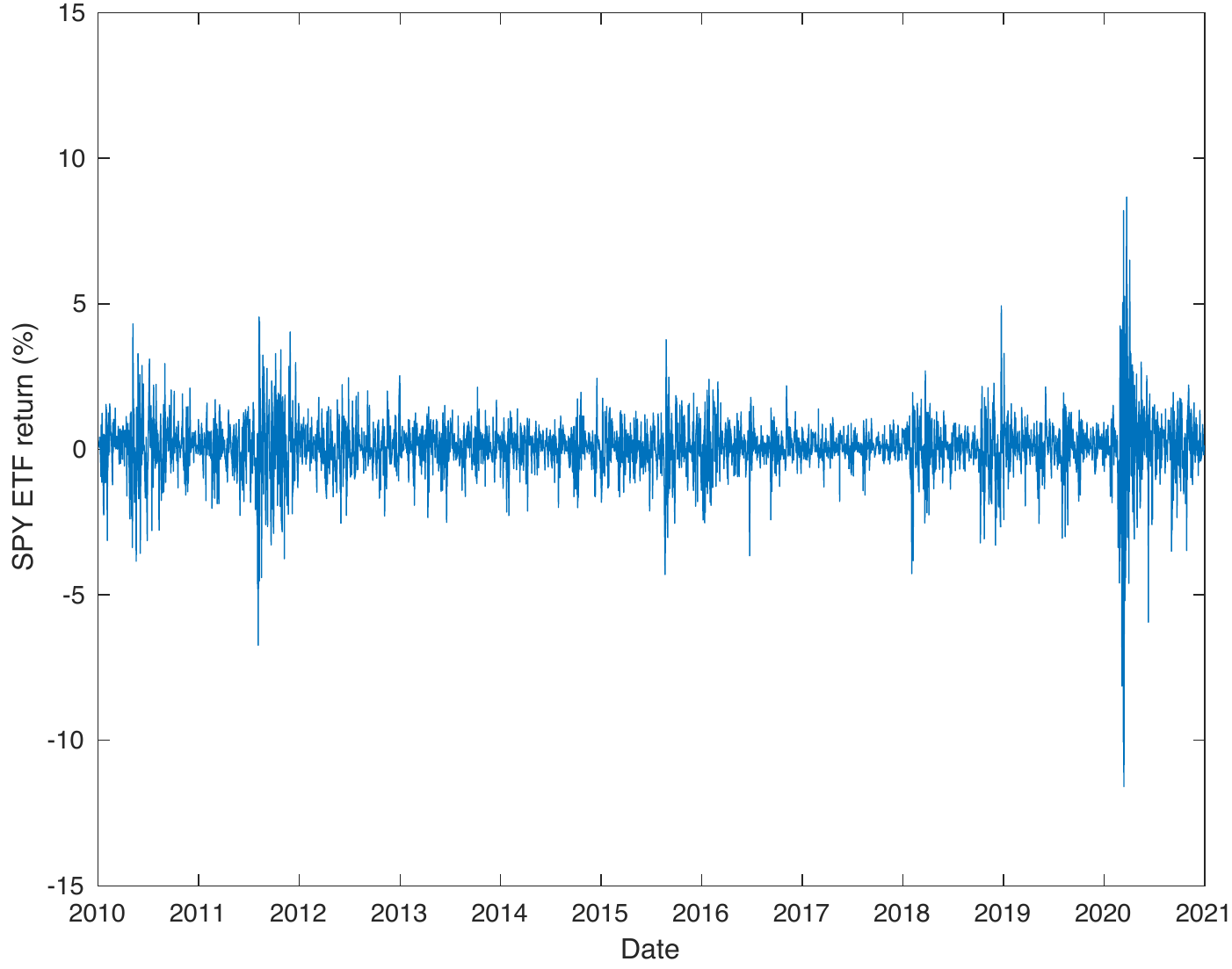}
\vspace{-0.7cm}
     \caption{Daily SPY return  and COVID-19 pandemic}
         \label{fig68}
  \end{subfigure}
  \begin{subfigure}[b]{0.45\linewidth}
    \includegraphics[width=\linewidth]{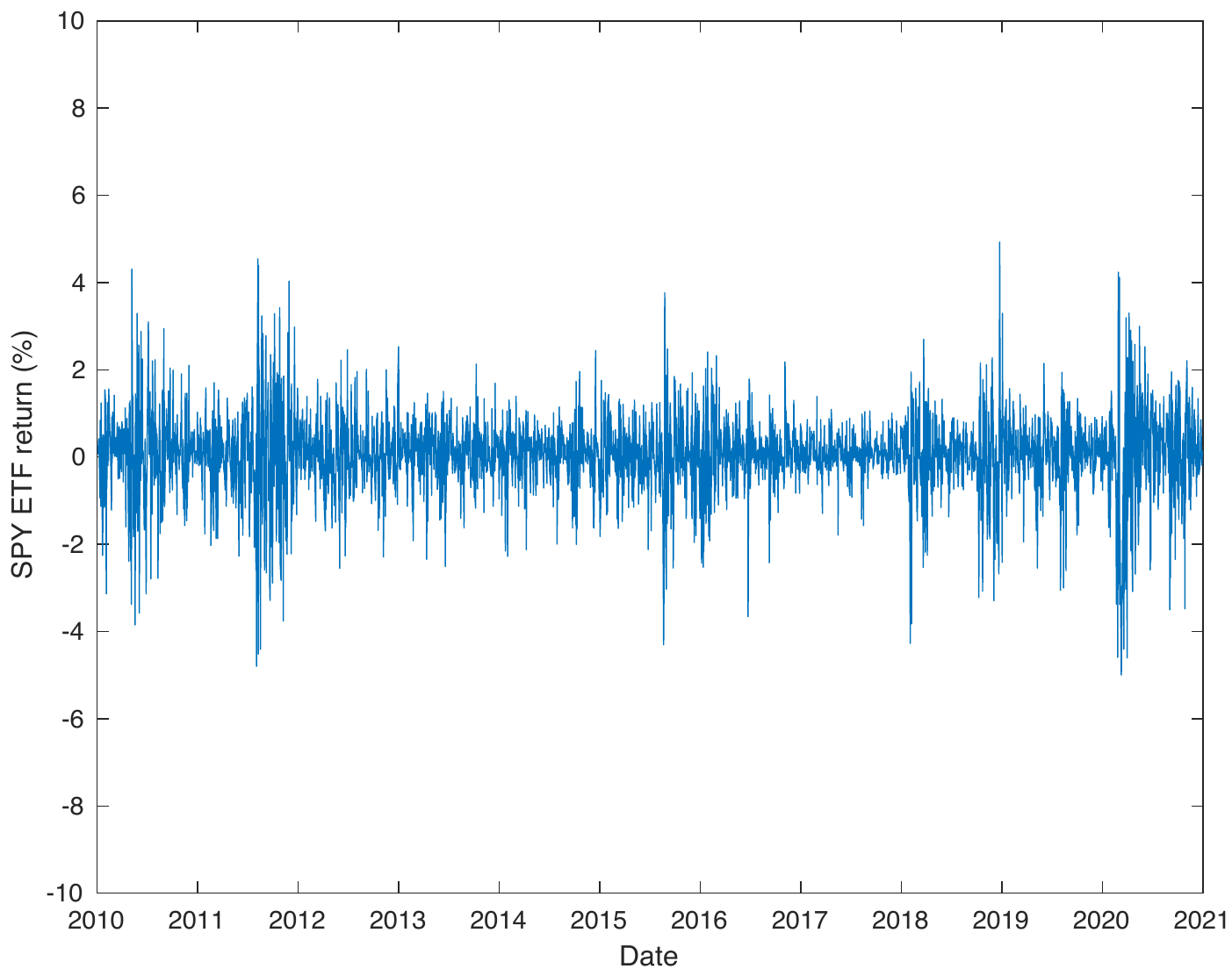}
\vspace{-0.7cm}
     \caption{ Daily SPY return sample without outliers}
         \label{fig67}
          \end{subfigure}
\vspace{-0.5cm}
  \caption{Daily SPY ETF Return ($\%$)}
  \label{fig66}
\vspace{-0.5cm}
\end{figure}

\noindent
Let the number of observations $N=2768$, and the daily observed SPY price $S_{j}$ on day $t_{j}$ with $j=1,\dots,N$; $t_{1}$ is the first observation date (January 4, 2010) and $t_{N}$ is the last observation date (December 30, 2020). The daily SPY ETF log return $(y_{j})$ is computed as in (\ref{eq:l33}).
\begin{align}
y_{j}=\log(S_{j}/S_{j-1}) \hspace{10 mm}  \hbox{ $j=2,\dots,N$}\label{eq:l33}
 \end{align}
\noindent  
One observation will be lost in the computation process. The results of the daily SPY ETF return  are shown in Fig \ref{fig66}.
\noindent 
The volatility of the SPY ETF is higher in the First quarter of $2020$ as displayed in Fig \ref{fig68}. The SPY ETF is an inexpensive way for investors to gain diversified exposure to the U.S. stock market; like others stocks and securities, it was unusually volatile in $2020$ amid the coronavirus pandemic and massive disruptions in the global economy. \\
\noindent 
$13$ daily return observations was identified as outliers and removed from the data set in order to avoid negative impact on the statistics. See Table \ref{tab:1} below. By assuming the independent, identically distribution in the SPY ETF return sample, the method of moments was used to compute  statistics in Table \ref{tab:1} for both samples with outliers and without outliers. The outlier effects are greater on Kurtosis, Skewness, and Variance. The sample  Skewness (0.4687) is between -0.5 and 0.5, which is the skewness interval for symmetric \cite{hurst1997subordinated, burford1968shao}. \\
 \begin{table}[ht]
\vspace{-0.7cm}
 \caption{Impact of outliers on sample Statistics}
 \label{tab:1}
  \centering
\begin{tabular}{@{}lll@{}}
\toprule
\textbf{Statistics}          & \textbf{All}              & \textbf{Without outliers}  \\ \midrule
$N$                   & $2768$              & $2755$              \\
$\hat{E(Y)}$        & $0.0509$   & $0.0541$ \\
$\hat{\sigma^2(Y)}$ & $1.1905$    & $0.9487$  \\
$\hat{Skew(Y)}$     & $-0.8633$            & $-0.4687$            \\
$\hat{Kurt(Y)}$     & $17.3572$           & $6.6853 $           \\ \bottomrule
\end{tabular}
\end{table}

\begin{figure}[ht]
    \centering
  \begin{subfigure}[b]{0.4\linewidth}
    \includegraphics[width=\linewidth]{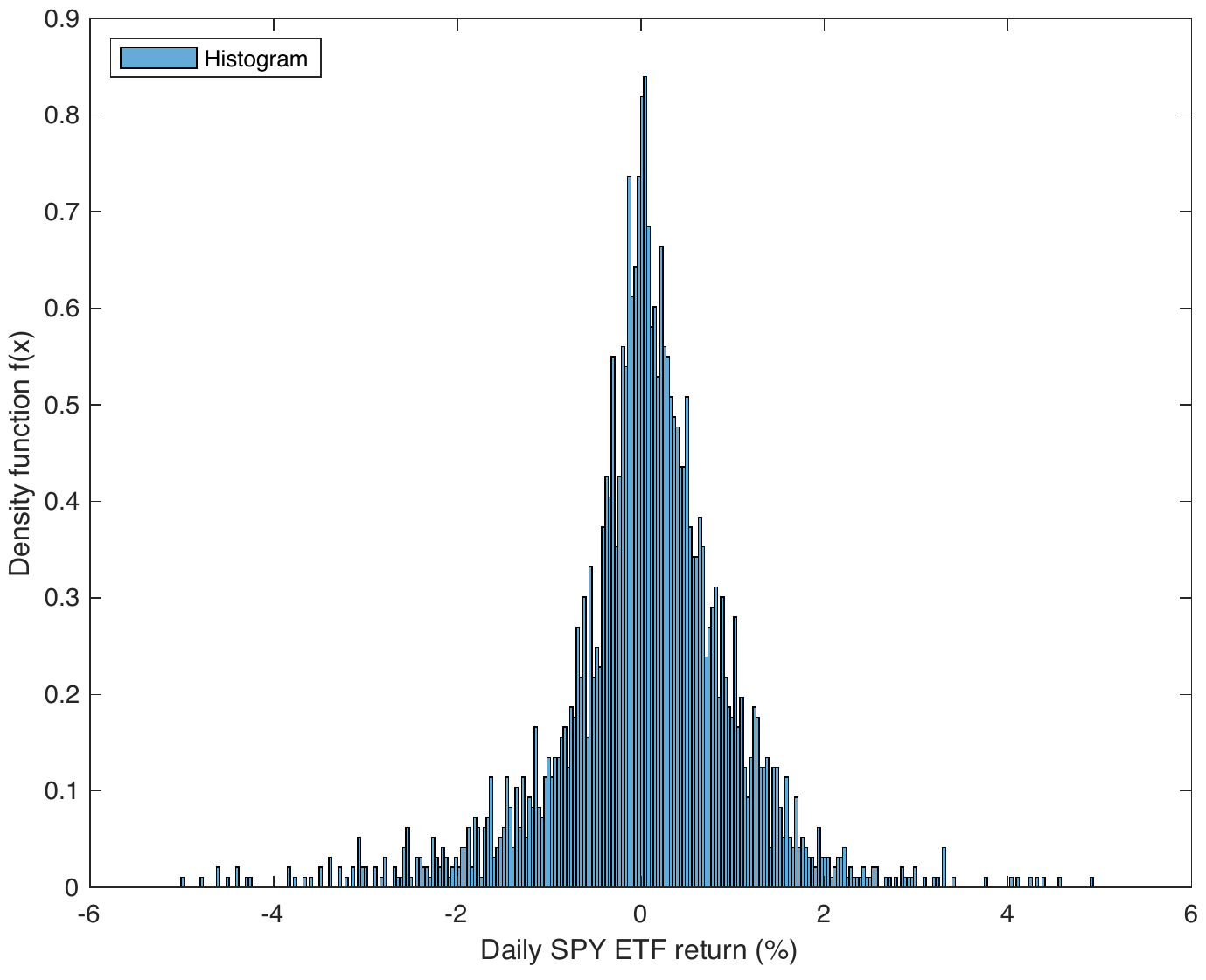}
\vspace{-0.5cm}
     \caption{Daily SPY ETF return Histogram}
         \label{fig91}
  \end{subfigure}
  \begin{subfigure}[b]{0.4\linewidth}
    \includegraphics[width=\linewidth]{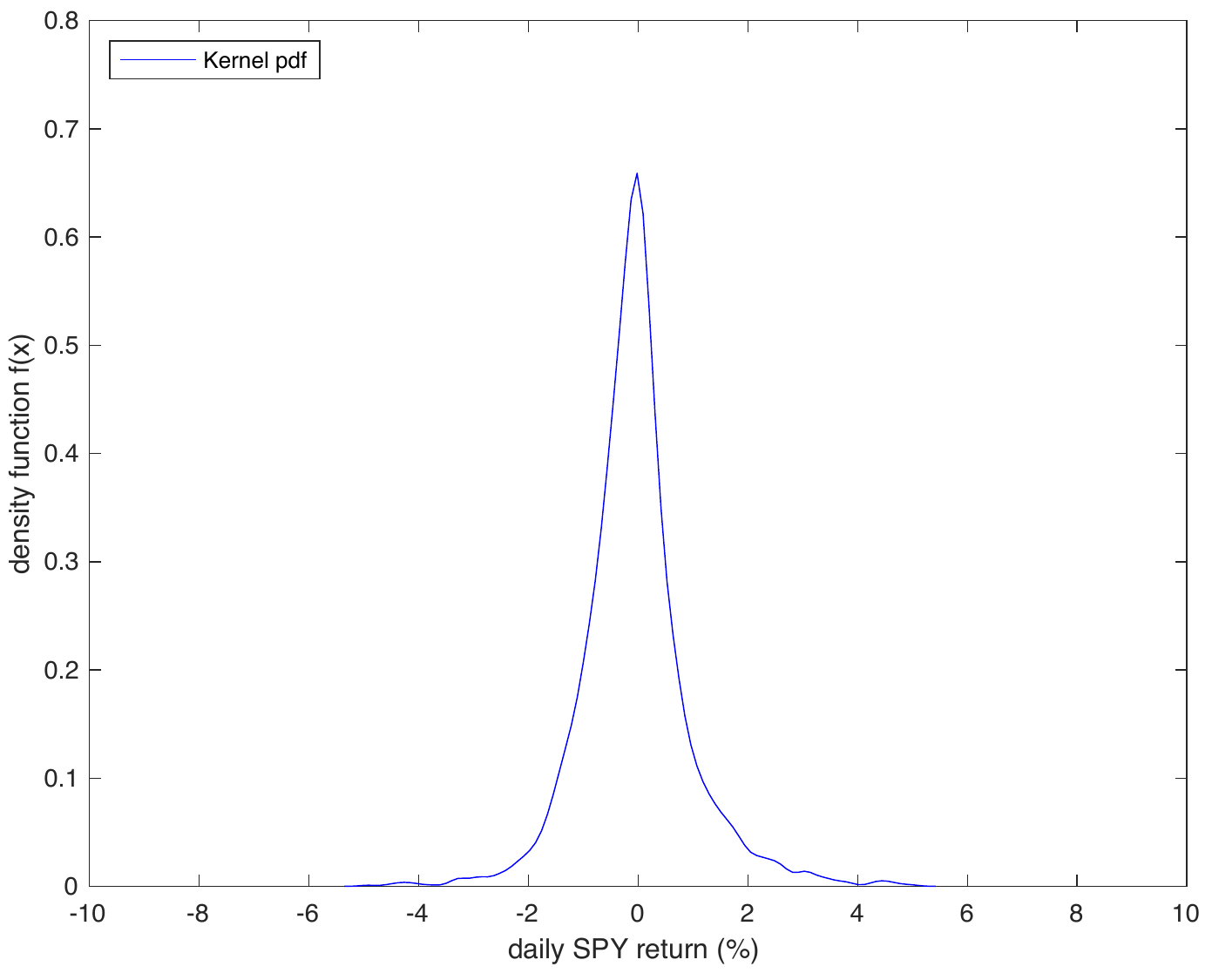}
\vspace{-0.5cm}
     \caption{ Daily SPY ETF return Kernel  density }
         \label{fig92}
          \end{subfigure}
\vspace{-0.7cm}
  \caption{Non-parametric density estimation}
  \label{fig9}
\vspace{-0.5cm}
\end{figure}

\noindent 
The shape of the daily SPY ETF return distribution are estimated from the sample using two non parametric density estimators: Histogram in Fig \ref{fig91} and Kernel  density function in Fig\ref{fig92}. Both graphs show the Leptokurtosis\cite{goncu2013fitting} with sharp peak.
\subsection{VG Model Estimations}
\noindent 
Based on the VG model assumption that $\sigma=1$, the equations (\ref{eq:l25b})-(\ref{eq:l28}) was solved given the statistics provided in Table \ref{tab:1}. The parameter values $\mu$, $\delta$, $\alpha$, and $\theta$ are shown in Table \ref{tab:2}. We have two models, the asymmetric Variance-Gamma model (AVG) with chronometer drift negative ($\delta=-0.1399$); and the symmetric Variance-Gamma model (SVG) without drift.\\
\noindent  
As shown Table \ref{tab:2}, when $\delta=0$, the drift of the physical time scale $\mu$ decreases, but keeps the positive sign. The magnitude of $\alpha$ and $\theta$ remains the same. The Gamma shape parameter ($\alpha$) is less than one. In fact, The larger the sample kurtosis, the smaller the Gamma shape parameter ($\alpha$)  as shown in the  relationship ($Kurtosis=3(1+\frac{1}{\alpha})$) for SVG model.
\newpage
\begin{table}[ht]
\caption{Method of Moments VG Parameters Estimations }
\vspace{-0.3cm}
 \label{tab:2}
\centering
\begin{tabular}{@{}lll@{}}
\toprule
\textbf{Model} &
 \textbf{Parameters} &
  \textbf{Statistics} \\ \midrule
   & $\hat{\mu}=0.1841$   & \multirow{2}{*}{$Skew(Y)=-0.4689$} \\
   & $\hat{\delta}=-0.1399$ &                                   \\
\textbf{AVG} & $\hat{\alpha}=0.8479$         & \multirow{2}{*}{$Kurt(Y)=6.6853$} \\
   & $\hat{\theta}=1.0954$        &                                   \\ 
    & $\hat{\mu}=0.0541$   & \multirow{2}{*}{$Skew(Y)=0$}      \\
\textbf{SVG} & $\hat{\alpha}=0.8140$         &                         \\
\textbf{}                              & $\hat{\theta}=1.1654$         & $\hat{Kurt(Y)}=6.6853$            \\ \bottomrule
\end{tabular}
\vspace{-0.2cm}
\end{table}
\noindent 
Regarding the estimation by Maximum likelihood method, the previous assumption on the volatility ($\sigma=1$) doest not hold and was dropped. The method has a local property that makes the result depending on the initial value. The estimations by Moments method were used as initial values in the maximization procedure.
\begin{table}[ht]
\vspace{-0.3cm}
\caption{Maximum Likelihood VG Parameters Estimations }
\vspace{-0.3cm}
 \label{tab:3}
\centering
\begin{tabular}{@{}lll@{}}
\toprule
\multirow{2}{*}{\textbf{Model}} &
  \multirow{2}{*}{\textbf{Parameters}} &
  \multirow{2}{*}{\textbf{Statistics}}  \\ 
  \multirow{2}{*}{} &
  \multirow{2}{*}{} &
  \multirow{2}{*}{}  \\ \midrule
\multirow{2}{*}{} &
  \multirow{2}{*}{$\hat{\mu}=0.1683$} &
  \multirow{2}{*}{$\hat{Skew(Y)}=  -0.382 $} \\
  \multirow{2}{*}{} &
  \multirow{2}{*}{$\hat{\delta}=-0.1089$} &
  \multirow{2}{*}{$\hat{Kurt(Y)}=6.329$} \\
\multirow{2}{*}{} &
  \multirow{2}{*}{$\hat{\sigma}=0.8987$} &
  \multirow{2}{*}{Log(ML)=-3562} \\
\multirow{2}{*}{\textbf{AVG1}} &
  \multirow{2}{*}{$\hat{\alpha}=0.9284$} &
  \multirow{2}{*}{$P\_{value}=0$} \\
\multirow{2}{*}{} &
  \multirow{2}{*}{$\hat{\theta}=1.0546$} 
  \multirow{2}{*}{} \\
  \multirow{2}{*}{} &
  \multirow{2}{*}{} &
  \multirow{2}{*}{} \\
\multirow{2}{*}{} &
  \multirow{2}{*}{$\hat{\mu}=0.0510$} &
  \multirow{2}{*}{$\hat{Skew(Y)}=0$} \\
  \multirow{2}{*}{} &
  \multirow{2}{*}{$\hat{\sigma}=0.9378$} &
  \multirow{2}{*}{$\hat{Kurt(Y)}=6.534$} \\   
\multirow{2}{*}{\textbf{SVG1}} &
  \multirow{2}{*}{$\hat{\alpha}=0.8490$} &
  \multirow{2}{*}{$Log(ML)=-3559$} \\
  \multirow{2}{*}{} &
  \multirow{2}{*}{$\hat{\theta}=1.0929$} &
  \multirow{2}{*}{$P\_{value}=0$} \\
  \multirow{2}{*}{} &
  \multirow{2}{*}{} &
  \multirow{2}{*}{} \\
 \multirow{2}{*}{} &
  \multirow{2}{*}{$\hat{\mu}=0.0848$} &
  \multirow{2}{*}{$\hat{Skew(Y)}=-0.173 $} \\
 \multirow{2}{*}{} &
  \multirow{2}{*}{$\hat{\delta}=-0.0577$} &
  \multirow{2}{*}{$\hat{Kurt(Y)}=6.412$} \\
  \multirow{2}{*}{\textbf{AVG2}} &
  \multirow{2}{*}{$\hat{\sigma}=1.0295$} &
  \multirow{2}{*}{$Log(ML)=-3548.65$} \\ 
  \multirow{2}{*}{} &
  \multirow{2}{*}{$\hat{\alpha}=0.8845$} &
  \multirow{2}{*}{$P\_{value}= 0$} \\  
  \multirow{2}{*}{} &
  \multirow{2}{*}{$\hat{\theta}=0.9378$} &
  \multirow{2}{*}{} \\  
\multirow{2}{*}{} &
  \multirow{2}{*}{} &
  \multirow{2}{*}{} \\
\multirow{2}{*}{} &
  \multirow{2}{*}{$\hat{\mu}= 0.0652$} &
  \multirow{2}{*}{$\hat{Skew(Y)}=0$} \\
\multirow{2}{*}{} &
  \multirow{2}{*}{$\hat{\sigma}=0.9939$} &
  \multirow{2}{*}{$\hat{Kurt(Y)}=6.421$} \\   
 \multirow{2}{*}{\textbf{SVG2}} &
  \multirow{2}{*}{$\hat{\alpha}=0.8770$} &
  \multirow{2}{*}{$Log(ML)=-3552.29$} \\
\multirow{2}{*}{} &
  \multirow{2}{*}{$\hat{\theta}=0.9937$} &
  \multirow{2}{*}{$P\_{value}= 0$} \\ 
\multirow{2}{*}{} &
  \multirow{2}{*}{} &
  \multirow{2}{*}{} \\
\multirow{2}{*}{\textbf{CLM}} &
  \multirow{2}{*}{$\hat{\mu}=0.0541$} &
  \multirow{2}{*}{$\hat{Kurt(Y)}=3$}\\
\multirow{2}{*}{} &
  \multirow{2}{*}{$\hat{\sigma}=0.9740$} &
  \multirow{2}{*}{Log(ML)=-3836.16}   \\ 
  \multirow{2}{*}{} &
  \multirow{2}{*}{} &
  \multirow{2}{*}{}  \\ \bottomrule
 \end{tabular}
 \vspace{-0.5cm}
\end{table}
\clearpage
\noindent 
 In table \ref{tab:3}, the results are labeled AVG1 for the Asymmetric Variance-Gamma Model and SVG1 for the Symmetric Variance-Gamma Model. 
 Another initial value was chosen with $\sigma=\alpha=\theta=1$, and $\delta=\mu=0$. The results are labeled AVG2 and SVG2 respectively for Asymmetric Variance-Gamma Model and Symmetric Variance-Gamma Model. 
As shown in Appendix \ref{eq:an1}, the maximization procedure convergences after 21 iterations For AVG2, as shown in Table \ref{tab:7}. SVG2 also converges, after 15 iterations. See Table \ref{tab:8} in Appendix \ref{eq:an1}.\\
\noindent
It appears that the drift of the physical time scale $\mu$  is positive, and the drift of the activity time process $\delta$ is negative. The estimation of the Maximum likelihood  underestimates the value of the kurtosis.
\subsection{Comparison of VG Models}
\noindent
For the Classical Lognormal Model (CLM), the Method of Moments and the Maximum likelihood yield the same estimation of the mean ($\hat{\mu}$) and standard deviation ($\hat{\sigma}$), given a high sample size. The results are displayed in Table \ref{tab:3}.\\
It is shown in (\ref{eq:l21a}) that the VG model is a special case of the CLM. It is also known that the parameter values in Table \ref{tab:3} maximized the likelihood function. Therefore, it becomes appropriate to perform the Likelihood Ratio test between the normal distribution and the four other VG models. The Likelihood Ratio estimator (W) is defined as follows.
 \begin{align}
 W &= 2(L_{H_{1}} - L_{H_{0}})  &
   W &\sim \chi^{2}(r)  &
  P\_{value} &= prob(\chi^{2}>\hat{W}|H_{0}) \label{eq:l34}
 \end{align} 
\noindent
 where  $L_{H_{0}}$ is the Maximum of the Log-likelihood under $H_{0}$. Under $H_{0}$: $Y_{k}\sim N(\mu,\sigma^2)$.\\ $L_{H_{1}}$ is the Maximum of the Log-likelihood under $H_{1}$. Under $H_{1}$: $Y_{k} \sim VG(\mu,\delta,\sigma,\alpha,\theta)$.  $P\_values$ is the probability that the real values are even more extreme (more in the tail) than our test statistic ($\hat{W}$). For Likelihood Ratio test review, see section $5.5$ in \cite{held2014applied}, and \cite{willis2020maximum}, \cite{king1989unifying}\\
\noindent
The  value of the Log-likelihood ($Log(ML)$) is displayed in Table \ref{tab:3} for each model. There is a huge discrepancy between the Log-likelihood of the CLM and  VG model, which produces  $P\_{value}$ almost zero. Therefore, the null hypothesis ($H_{0}$) of normal distribution, is rejected based on the sample data. The results provide evidence that the Variance-Gamma model is an improvement compare to the classical Normal model.\\

\begin{figure}[ht]
 \vspace{-0.4cm}
     \centering
         \includegraphics[scale=0.45] {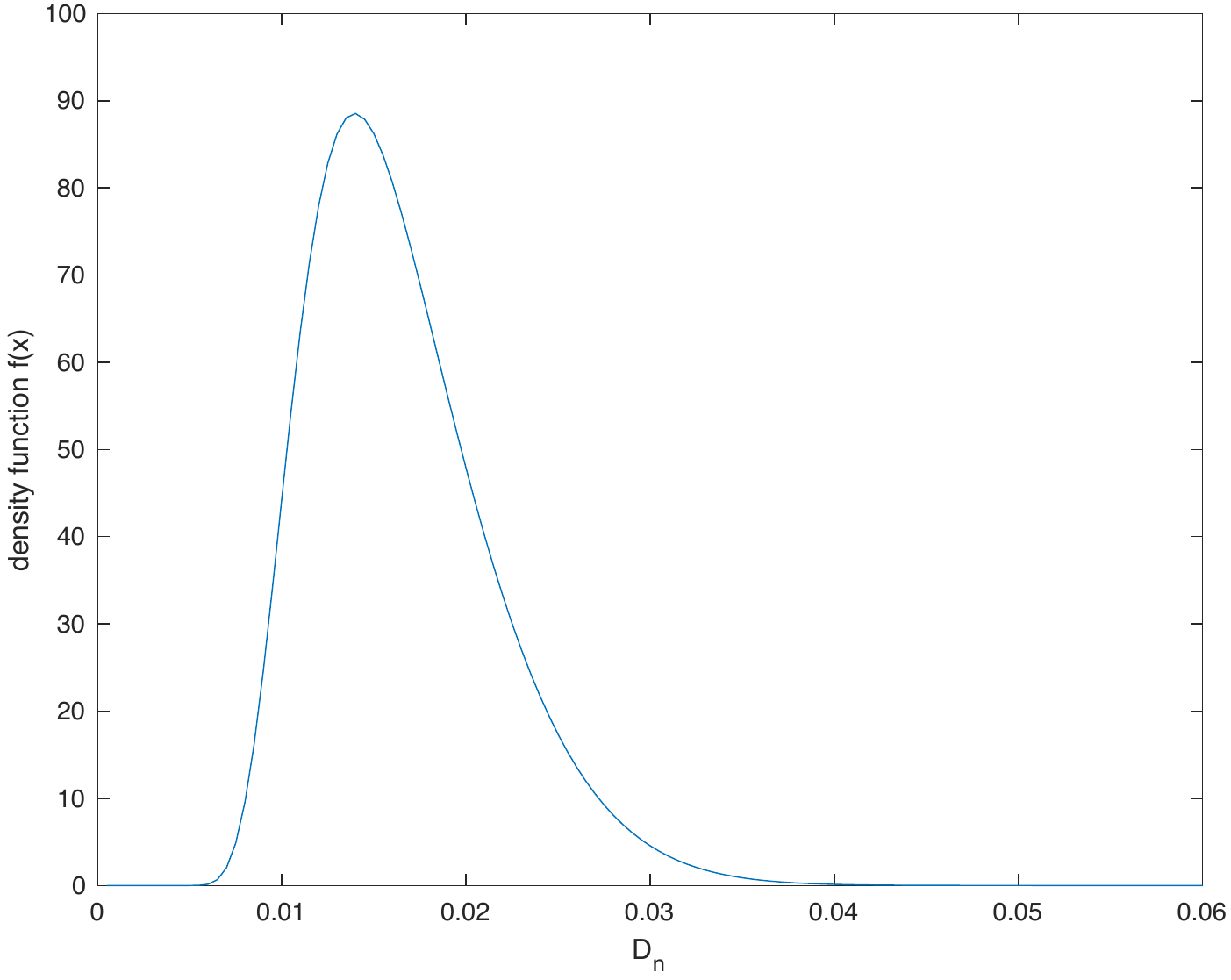}
 \vspace{-0.5cm}
        \caption{Kolmogorov-Smirnov Estimator ($\hat{D_{n}}$) probability density ($n=2755$)  under the null hypothesis $H{0}$}
        \label{fig62}
 \vspace{-0.7cm}
\end{figure}

\noindent
The natural question is to know which VG model fits the empirical distribution of the sample of daily SPY ETF return $\{y_{1}, y_{2}\dots y_{n}\}$. In order to compare the empirical distribution to the VG model theoretical distribution, the Kolmogorov-Smirnov (KS) test was performed under the null hypothesis (H0) that the sample $\{y_{1}, y_{2}\dots y_{n}\}$ comes from VG model.\\
 \noindent
  The cumulative distribution function of the  theoretical distribution needs to be computed.  The density function ($f$) does not have closed-form, the same for the cumulative function($F$) in (\ref{eq:l36}). However, we know the closed form of the Fourier of the density function ($\scrF[f]$) and the relationship in (\ref{eq:l37}) provides the Fourier of the cumulative distribution function ($\scrF[F]$).   
  \begin{align}
 Y &\sim VG(\mu,\delta,\sigma, \alpha,\theta) \label{eq:l35}\\
   F(x)&= \int_{-\infty}^{x} f(t) \mathrm{d}t \hspace{5mm}  \hbox{$f$ is the density function of $Y$} \label{eq:l36}\\
 \scrF[F](x)&=\frac{\scrF[f](x)}{ix} + \pi\scrF[f](0)\delta (x) \label{eq:l37}
 \end{align}
 \begin{align}
 \delta (x)&=
  \begin{cases}
  0     & \quad \text{if } x\neq 0 \\
 \infty   & \quad \text{if } x=0
  \end{cases} \label{eq:l38}
 \end{align}
 \noindent 
  See Appendix \ref{eq:an01} for (\ref{eq:l37}) proof. \\
\noindent
The cumulative distribution function ($F$) was computed from  the inverse of the Fourier of the cumulative distribution ($\scrF[F]$). The results for SVG2 model is shown in Appendix \ref{eq:an2}, Fourth column. The same methodology was followed for the remaining VG models. The Kolmogorov-Smirnov (K-S) estimator ($D_{n}$) is defined as follows.
 \begin{align}
D_{n} &= \sup_{x}{|F(x)-F_{n}(x)|}  &
P_{value} &= prob(D_{n}>d_{n} |H_{0})  \label{eq:l39} 
 \end{align}
\noindent 
 where $n$ is the sample size, $F_{n}(x)$ denotes the empirical cumulative distribution of $\{y_{1}, y_{2}\dots y_{n}\}$ .
\noindent 
The distribution of the Kolmogorov’s goodness-of-fit measure $D_{n}$ has been studied extensively in the litterature \cite{marsaglia2003evaluating},\cite{massey1951kolmogorov},\cite{volition2018}. It was shown  that $D_{n}$ distribution is independent of the theoretical distribution ($F$) under the null hypothesis (H0). More recently in \cite{dimitrova2020computing}, it was determined numerically the exact distribution of $D_{n}$. \cite{dimitrova2020computing} also provides a package KSgeneral\cite{Dimitrina2021} in the R software for computing the  cumulative distribution of $D_{n}$ when F(x) is continuous and $n$ is a  natural number.\\
\noindent
The cumulative distribution of $D_{n}$ under the null hypothesis was computed and the density function was deduced. The computed density function is shown in Fig \ref{fig62}. Under the null hypothesis (H0), $D_{n}$ has a positively skewed distribution with means ($\mu=0.0165$) and standard deviation  ($\sigma=5*10^{-3}$).\\

\noindent
$d_{n}$ is the value of the KS estimator ($D_{n}$) computed from the sample $\{y_{1}, y_{2}\dots y_{n}\}$. Based on \cite{krysicki1999rachunek,kucharska2009nig}, $d_{n}$  can be estimated as  follows.
\begin{align}
d^{+}_{n}&= \sup_{0\leq j\leq P}{|F(x_{j})-F_{n}(x_{j})|}  &
d^{-}_{n}&= \sup_{1\leq j\leq P}{|F(x_{j})-F_{n}(x_{j-1})|} &
d_{n}&= Max(d^{+}_{n}, d^{-}_{n}) \label{eq:l40}
 \end{align} 
The results for SVG2 model is shown in Appendix \ref{eq:an2}, Table  \ref{tab:9}, Fifth and Sixth columns.
From the Appendix \ref{eq:an2}:
$d^{-}_{n}=max((1))=0.023629$, $d^{+}_{n}=max((2))=0.021986$ and $d_{n}=0.023629$.\\
\noindent 
 For each estimation method and VG model, KS-Statistics ($d_{n}$) and P\_values were computed and the results are provided in Table \ref{tab:4}. \\
\begin{table}[ht]
\vspace{-0.3cm}
\centering
\caption{Kolmogorov-Smirnov (KS) test}
\vspace{-0.2cm}
 \label{tab:4}
\begin{tabular}{@{}cccl@{}}
\toprule
\textbf{Method} &
  \textbf{Model} &
  \textbf{KS-Statistics ($d_{n}$)} &
  \textbf{P\_values} \\ \bottomrule
\textbf{Moments}    & \textbf{SVG}  & 0.058145 & 0.00000154\% \\
\textbf{}           & \textbf{AVG}  &0.027211 & 3.3191\%\\
\textbf{}           & \textbf{AVG1} & 0.054290 & 0.00001691\% \\
\textbf{Likelihood} & \textbf{SVG1} & 0.036763 &0.1136\% \\
\textbf{}           & \textbf{AVG2} & 0.028182 & 2.4668\%  \\
\textbf{}           & \textbf{SVG2} &0.023629 & 9.0788\%   \\
\textbf{}           & \textbf{CLM}  & 0.095791 & 0\%            \\ \bottomrule
\end{tabular}
\end{table}

\noindent
As shown in Table \ref{tab:4}, the VG models from the method of moments do not fit the sample data distribution. The $P\_values$ is less than $5\%$  and the null hypothesis $H_{0}$ can not be accepted at some extent. The CLM does not fit the sample data distribution at $5\%$. Regarding the maximum likelihood method, the SVG2 model has $P\_values=9.079\%$, which is high than the classical threshold  $5\%$. Therefore, we can not reject the null hypothesis that SVG2 model fits the empirical distribution.\\
  \begin{figure}[ht]
\vspace{-0.6cm}
  \centering
  \begin{subfigure}[b]{0.38\linewidth}
    \includegraphics[width=\linewidth]{den1}
\vspace{-0.7cm}
     \caption{ Daily SPY ETF return histogram}
         \label{fig71}
  \end{subfigure}
  \begin{subfigure}[b]{0.38\linewidth}
    \includegraphics[width=\linewidth]{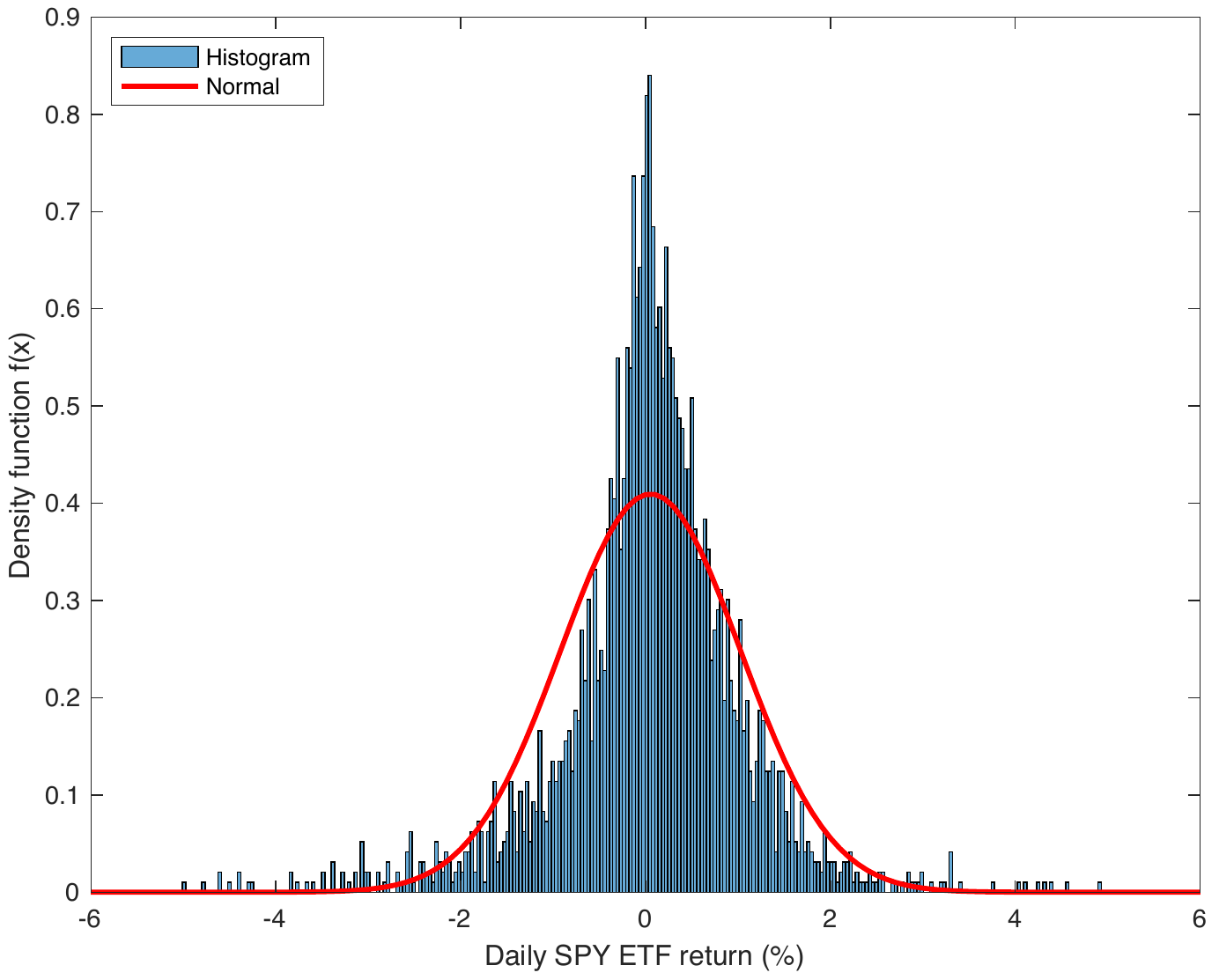}
\vspace{-0.7cm}
      \caption{$\hat{\mu}=0.0541$,$\hat{\sigma}=0.9740$ } \label{fig72}
  \end{subfigure}\\
  \begin{subfigure}[b]{0.38\linewidth}
\vspace{-0.3cm}
    \includegraphics[width=\linewidth]{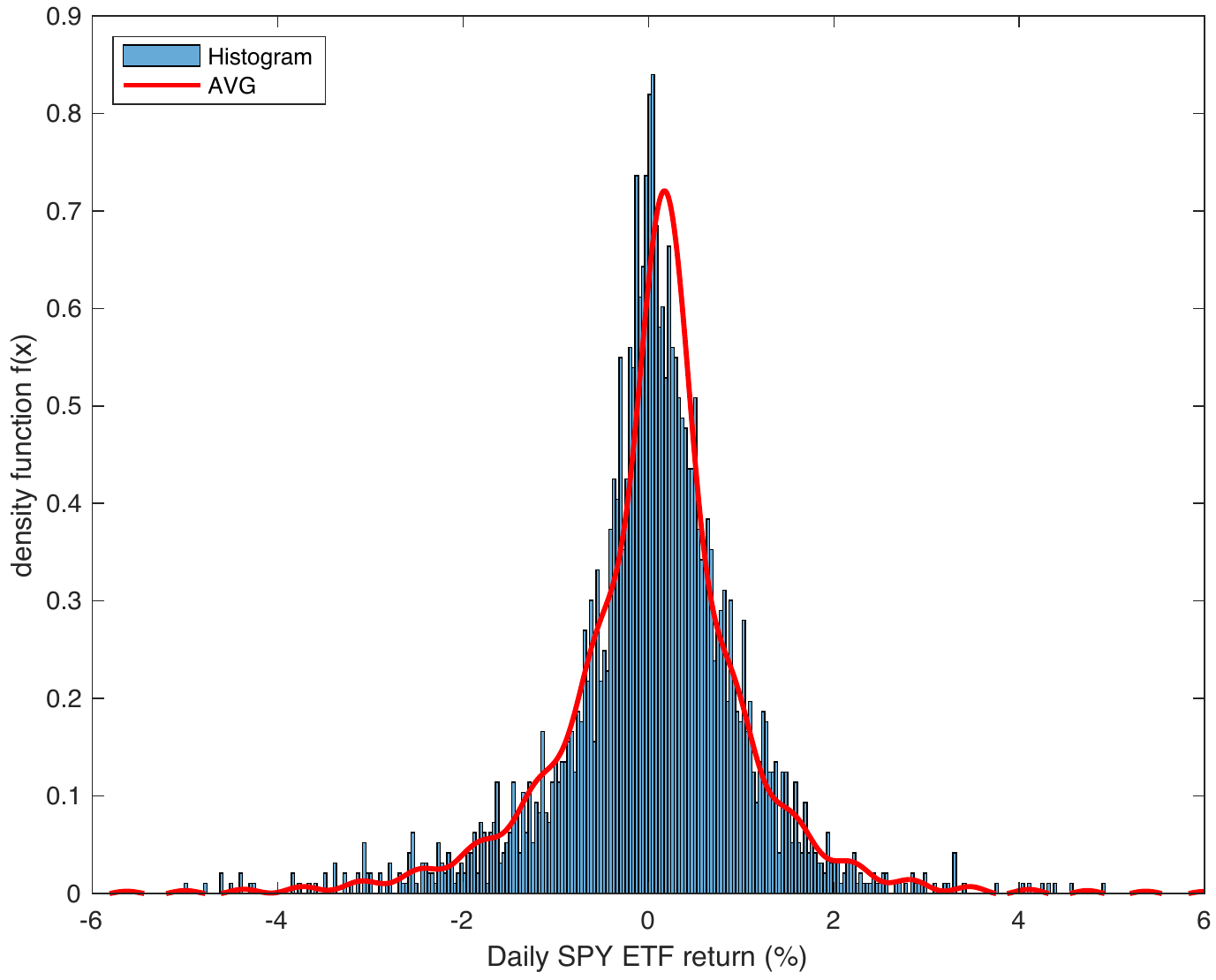}
\vspace{-0.7cm}
    \caption{Moments method ($\hat{f}$): $\hat{\mu}=0.1841$, $\hat{\delta}=-0.1399$, $\sigma=1$, $\hat{\alpha}=0.8479$, $\hat{\theta}= 1.0954$. }
         \label{fig73}
  \end{subfigure}
  \begin{subfigure}[b]{0.38\linewidth}
\vspace{-0.3cm}
    \includegraphics[width=\linewidth]{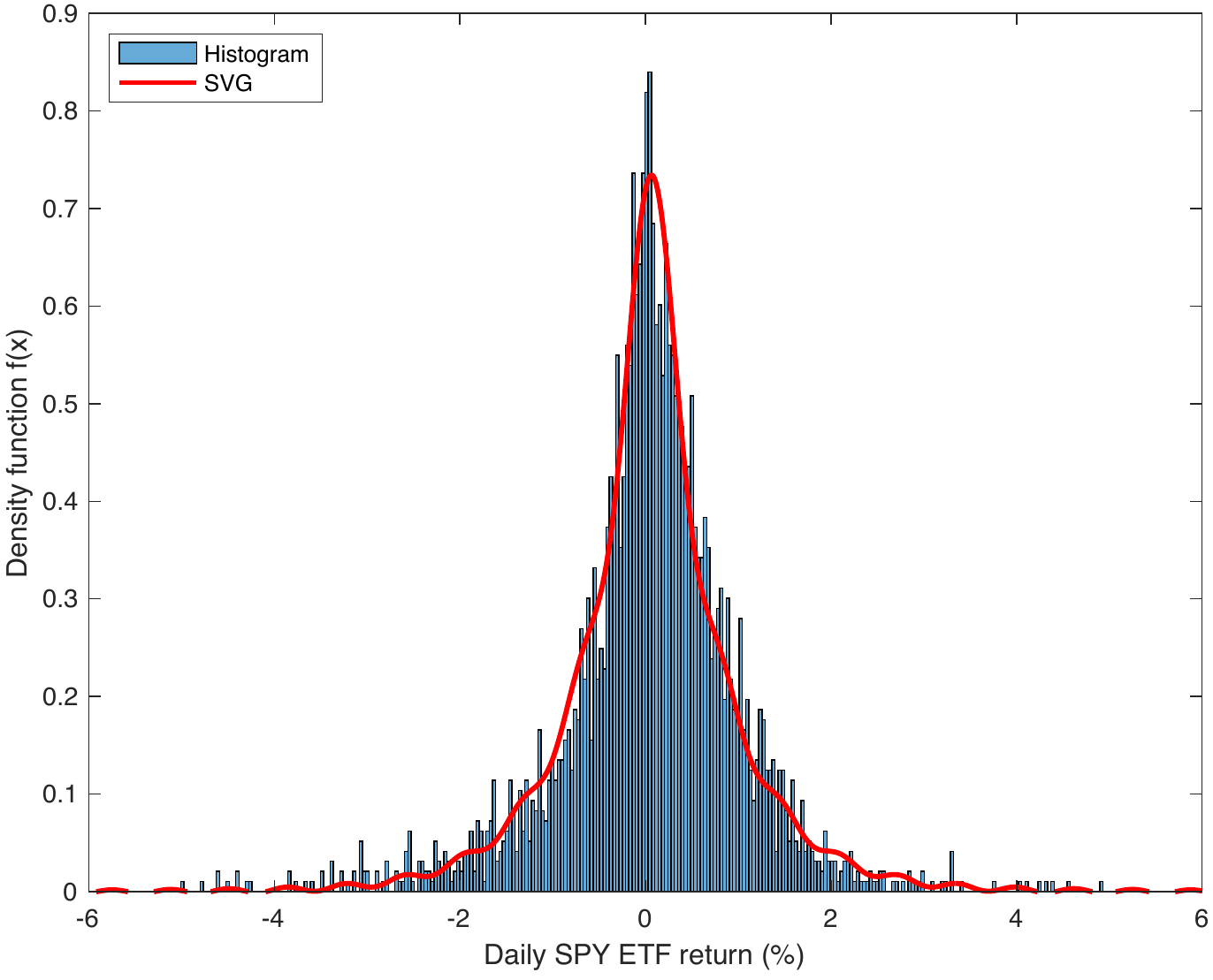}
\vspace{-0.7cm}
     \caption{Maximum likelihood ($\hat{f}$): $\hat{\mu}=0.0652$, $\delta=0$, $\hat{\sigma}=0.9939$, $\hat{\alpha}=0.8770$, $\hat{\theta}=0.9937$. }
         \label{fig74}
  \end{subfigure}
\vspace{-0.7cm}
  \caption{Probability density functions for the daily SPY return}
  \label{fig7}
\vspace{-0.6cm}
\end{figure}

\noindent
The histogram of the daily SPY ETF return was compared to some models in Table \ref{tab:4} as shown in Fig \ref{fig7}. Compare to the CLM in Fig \ref{fig72}, the peakedness of histogram explains the high level of the KS-Statistics ($d_{n}=0.095791$).  But to a lesser extent, the same explanation holds for some  AVG models like SVG, AVG1, SVG1 in table \ref{tab:4}.\\
\noindent
For work related to Normal and exponential distributions , see \cite{nzokem2021sis,aubain2020,Nzokem2020EpidemicDA,aubain2021}

\section {Conclusion} 
\noindent 
The paper explores the use of FRFT based technique as a tool to obtain the probability density function and its derivatives for infinitely divisible distribution.  The probability density functions are computed and the distributional proprieties such as leptokurtosis, peakedness, and asymmetry are reviewed for Variance-Gamma (VG) model and Compound Poisson with Normal Compounding model. The first and second derivatives of the probability density function of the VG model are also computed in order to build the Fisher information matrix for the Maximum likelihood method.\\
\noindent 
We perform the maximum likelihood  to estimate the Variance Gamma (VG) model with five parameters. The sample data comes from the daily SPY ETF  Price data from January 04, 2010 to December 30, 2020. The method of moments and the maximum likelihood were performed on the sample data. Independently of the methods of estimation, the result shows that the VG model provides a better fit than the Classical Lognormal Model (CLM).  The Kolmogorov-Smirnov (KS) goodness-of-fit test shows that the maximum likelihood method with FRFT produces a good estimation for the VG model, which fits the empirical distribution of the sample data. \\
\noindent 
It will be interesting to derive the European option price from the risk-neutral VG model and compare the results with those from the Black-Scholes model.

\pagestyle{plain}
\addcontentsline{toc}{chapter}{References}
\bibliographystyle{unsrt}
\bibliography{fourier.bib}
\newpage
\appendix
\renewcommand{\thesection}{\Alph{section}.\arabic{section}}
\setcounter{section}{0}
\section{Some Proofs}\label{eq:an01}
\textbf{Proof (\ref{eq:l7}):}
 \begin{align*}
   f(x_{k}) = \frac{1}{2\pi}\int_{-\infty}^{+\infty} \scrF[f](y)e^{ix_{k}y} \mathrm{d}y \approx \frac{1}{2\pi}\int_{-a/2}^{a/2} \scrF[f](y)e^{ix_{k}y} \mathrm{d}y &=  \frac{\gamma}{2\pi}\sum_{j=0}^{n-1}\! \scrF[f](y_{j})e^{2\pi i(k-\frac{n}{2})(j-\frac{n}{2})\delta}\\    
       & =  \frac{\gamma}{2\pi}e^{-\pi i(k-\frac{n}{2})n\delta}G_{k}(\scrF[f](y_{j})e^{-\pi i jn\delta}),-\delta)
\end{align*}
\textbf{Proof (\ref{eq:l10}) and (\ref{eq:l24a})}
 \begin{align*}
  \scrF[f](x) &=E[e^{-ixY_{j}}]= E(E[e^{-ixY_{j}}| V_{j}]) \hspace{5mm}   Y_{j}|_{V_{j}=v} \sim N(\mu + \delta v, v \sigma^{2} )  \hspace{5mm}  \hbox{and } \hspace{5mm} V_{j}\sim \Gamma(\alpha,\theta) \\
  &=E\left(e^{-ix(\mu + \delta V_{j}) - \frac{1}{2}\sigma^{2}x^{2}V_{j}}
  \right)=\frac{1}{\Gamma(\alpha)\theta^{\alpha}}\int_{0}^{+\infty} v^{\alpha -1}e^{-i(\mu + \delta v)x}e^{-(\frac{1}{2}\sigma^{2}x^{2} + \frac{1}{\theta})v} \mathrm{d}v \\
    &=\frac{e^{-i\mu x}}{(\frac{1}{2}\theta\sigma^{2}x^{2} + 1)^{\alpha}}E[e^{-i\delta x W}] \hspace{5mm} \hbox{and } \hspace{5mm} W\sim \Gamma(\alpha,\frac{1}{2}\sigma^{2}x^{2} + \frac{1}{\theta}) \\
     &=\frac{e^{-i\mu x}}{(1+\frac{1}{2}\theta\sigma^{2}x^{2} + i\delta \theta x)^{\alpha}}
\end{align*} 

\textbf{Proof (\ref{eq:l12})}
 \begin{align*}
\scrF[f](x) &=E[e^{-ix\sum_{i=1}^{N(\lambda)} X_{j}}]= E\left(E[e^{-ix\sum_{i=1}^{N(\lambda)} X_{j}}| N(\lambda)]\right)\hspace{5mm} \hspace{5mm} N(\lambda) \sim Poisson(\lambda) \hspace{5mm} \hbox{and } \hspace{5mm} X_{j} \sim N(\mu,\sigma^{2}) \\
  &= E\left(E[e^{-ix\sum_{i=1}^{k} X_{j}}| N(\lambda)=k]\right)=E\left(E[e^{-ixX_{j}}]^{N(\lambda)}\right)=e^{\lambda(E[e^{-ixX_{j}}]-1)}=e^{\lambda(e^{-uix -\frac{1}{2}\sigma^{2}x^2}-1)} 
  \end{align*} 
 \textbf{Proof (\ref{eq:l24b}):}
 \begin{align*}
  \scrF[f](x) &=\lim_{\theta \to 0}\frac{e^{-i\mu x}}{(1+\frac{1}{2}\theta\sigma^{2}x^{2} + i\delta \theta x)^{\frac{1}{\theta}}}\\
  &=e^{-i\mu x}\lim_{\theta \to 0} e^{-\frac{1}{\theta} log(1+\frac{1}{2}\theta\sigma^{2}x^{2} + i\delta \theta x)}\\
  &=e^{-i(\mu + \delta) x-\frac{1}{2}\sigma^{2}x^{2} }
  \end{align*} 
\textbf{Proof (\ref{eq:l37}):}\\
Fourier of the step function 
 \begin{align*}
  u (x)&=\begin{cases}
  1    & \quad \text{if } 0\leq x \\
  0   & \quad \text{if }0>x \end{cases}  &
  \scrF[u](y) &=\frac{1}{iy}+\pi\delta(y)
  \end{align*} 
The integral is transformed into convolution of $f$ and $u$ 
  \begin{align*}
 F(x)&= \int_{-\infty}^{x} f(t) \mathrm{d}t=\int_{-\infty}^{+\infty} f(\tau)u(x-\tau)\mathrm{d}\tau =f(x)*u(x)
 \end{align*} 
 Using the convolution theorem
  \begin{align*}
 \scrF[F](x)= \scrF[f](x)\scrF[u](x) \hspace{10mm}  \hbox{and}\hspace{10mm} \scrF[F](x)= \frac{\scrF[f](x)}{ix}+\pi\scrF[f](0)\delta(x)
 \end{align*} 
  
 \newpage
\section{second order derivative of the VG density function}\label{eq:an02}
\begin{figure}[ht]
  \centering
  \begin{subfigure}[b]{0.4\linewidth}
    \includegraphics[width=\linewidth]{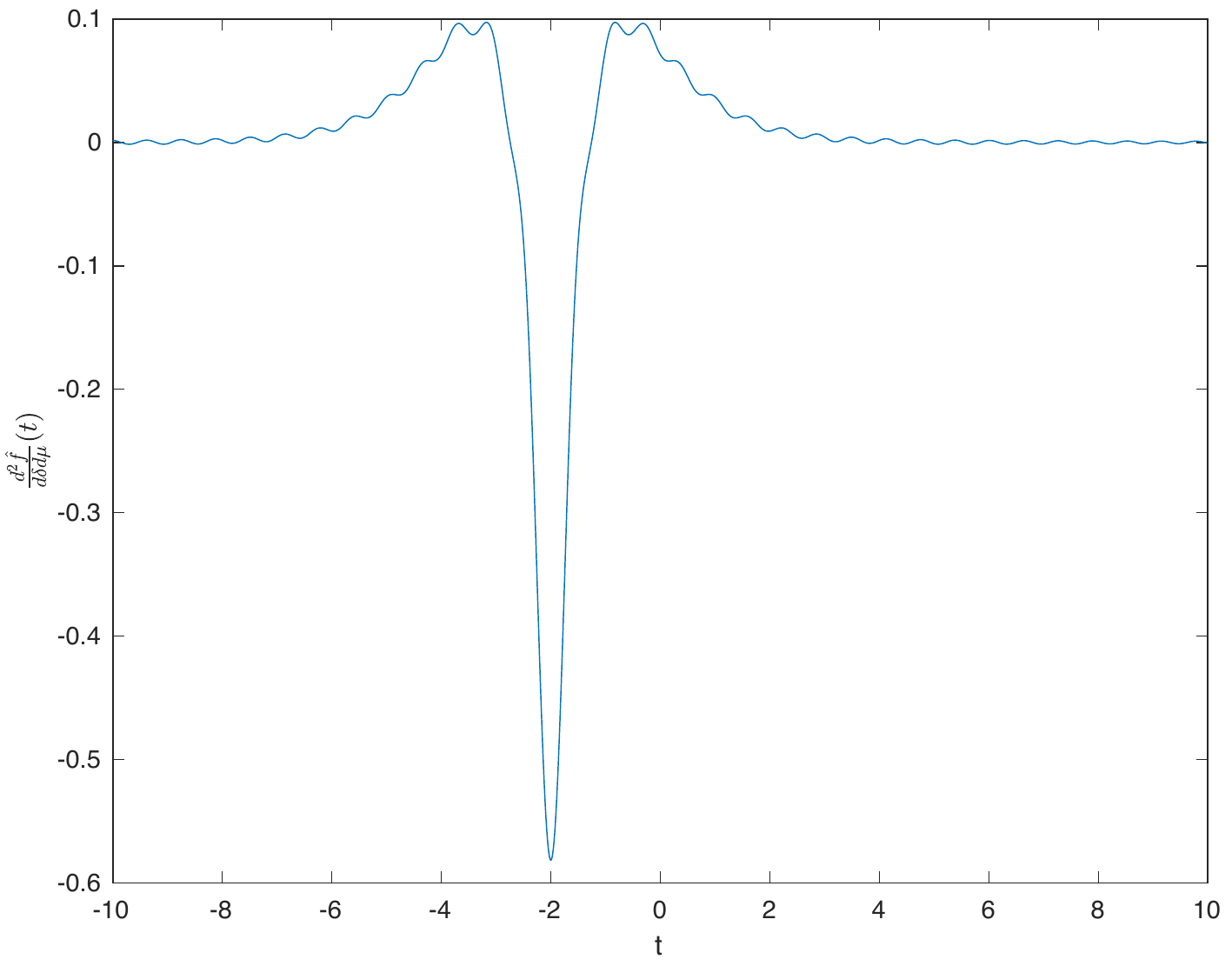}
     \caption{$\frac{d^{2}\hat{f}}{d\delta d\mu}(t)$ }
         \label{fig81}
  \end{subfigure}
  \begin{subfigure}[b]{0.4\linewidth}
    \includegraphics[width=\linewidth]{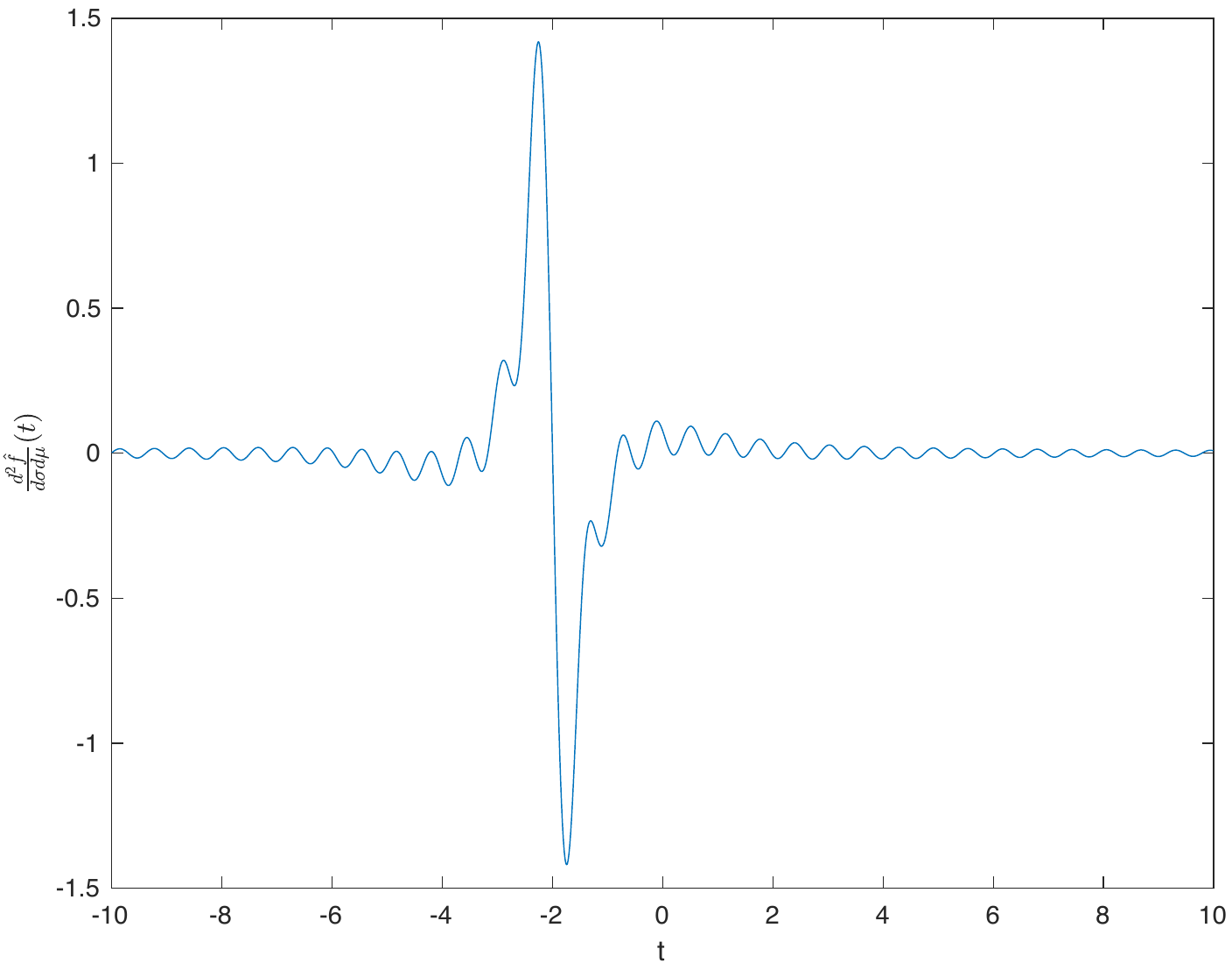}
      \caption{$\frac{d^{2}\hat{f}}{d\sigma d\mu}(t)$}
         \label{fig82}
  \end{subfigure}\\
  \begin{subfigure}[b]{0.4\linewidth}
    \includegraphics[width=\linewidth]{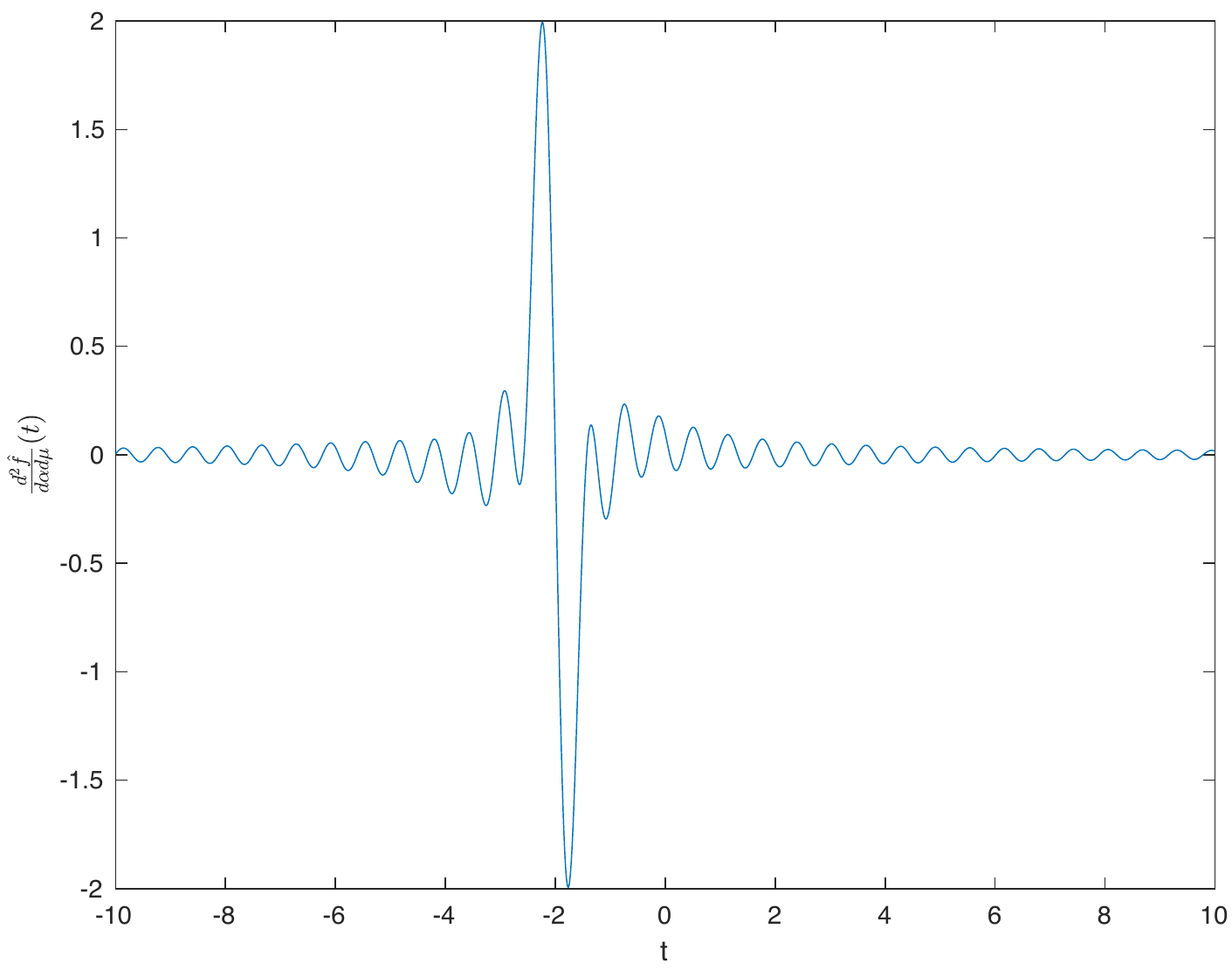}
    \caption{$\frac{d^{2}\hat{f}}{d\alpha d\mu}(t)$}
         \label{fig83}
  \end{subfigure}
  \begin{subfigure}[b]{0.4\linewidth}
    \includegraphics[width=\linewidth]{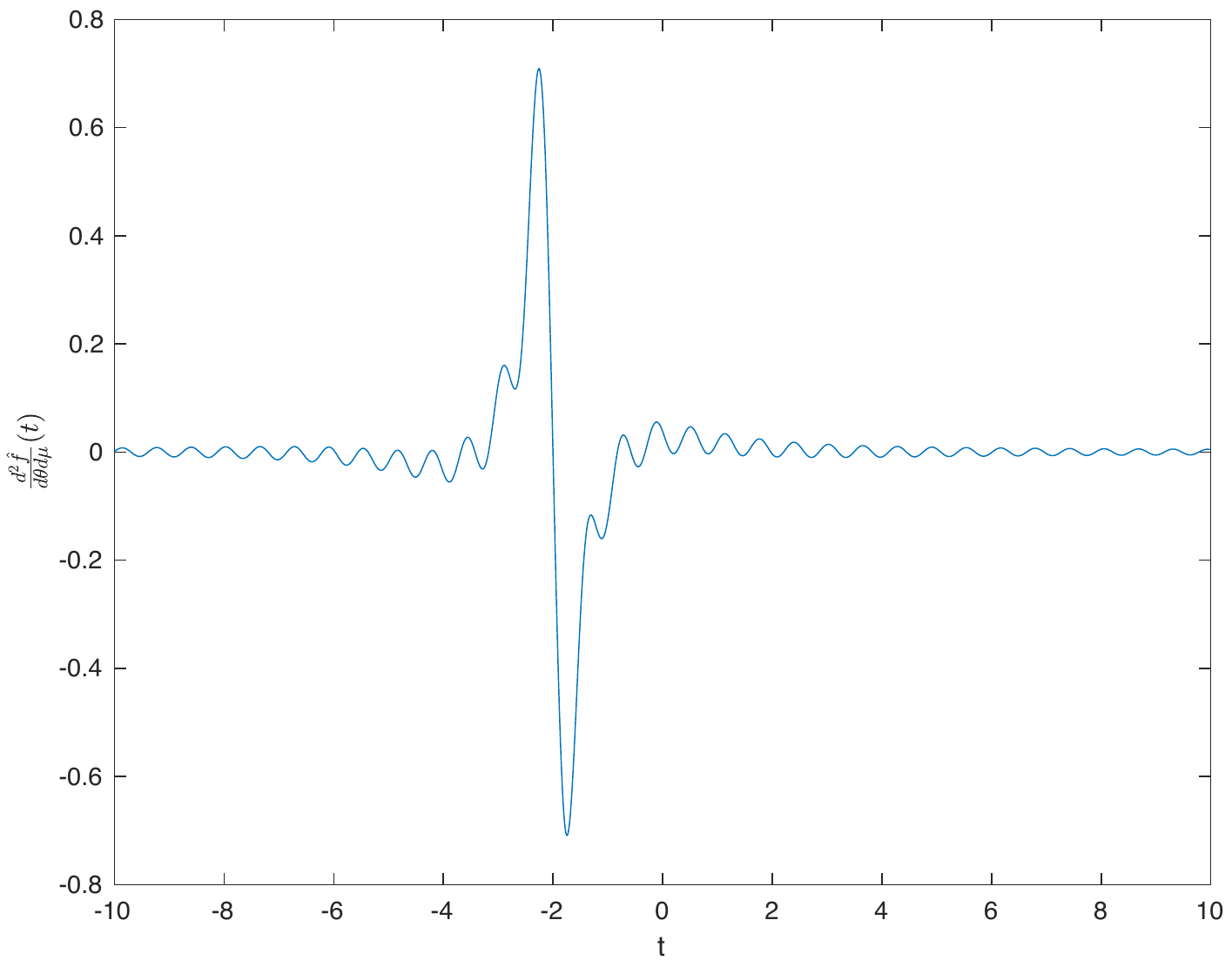}
     \caption{$\frac{d^{2}\hat{f}}{d\theta d\mu}(t)$}
         \label{fig84}
  \end{subfigure}
 
  \begin{subfigure}[b]{0.4\linewidth}
    \includegraphics[width=\linewidth]{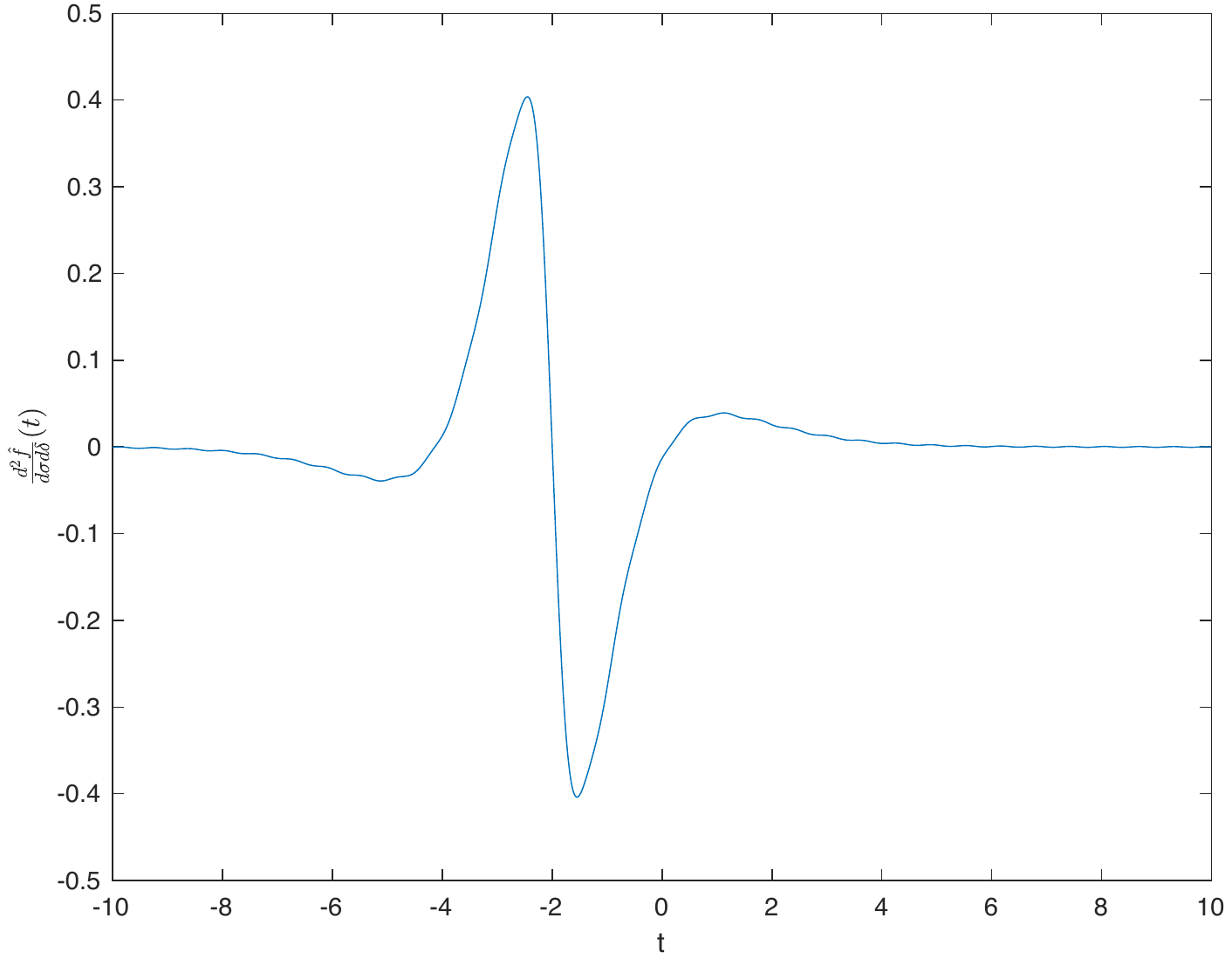}
   \caption{$\frac{d^{2}\hat{f}}{d\sigma d\delta}(t)$}
         \label{fig85}
  \end{subfigure}
  \begin{subfigure}[b]{0.4\linewidth}
    \includegraphics[width=\linewidth]{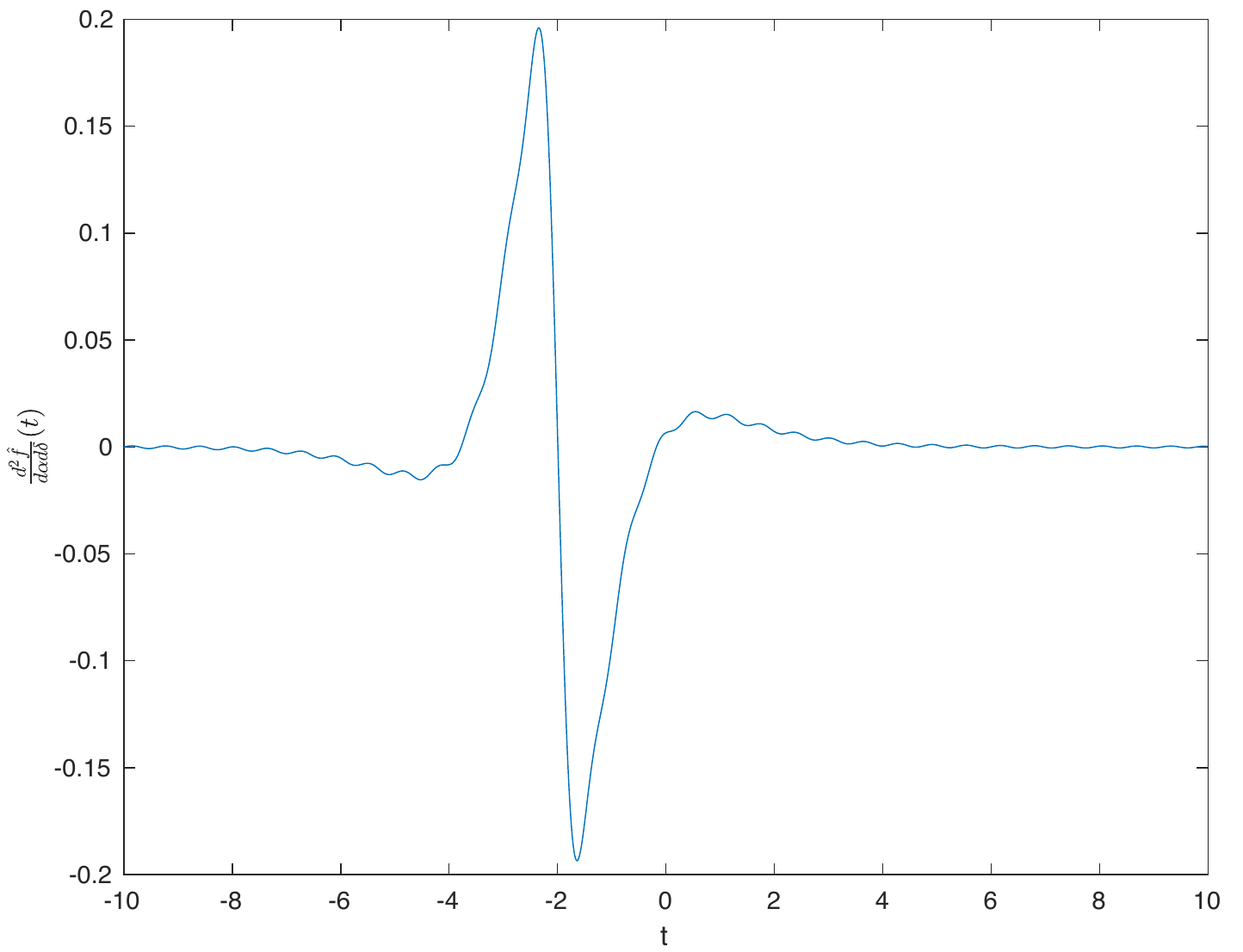}
     \caption{$\frac{d^{2}\hat{f}}{d\alpha d\delta}(t)$}
         \label{fig86}
  \end{subfigure}
  \caption{Second order derivative of the VG density function with $\mu=-2$, $\delta=0$, $\sigma=1$, $\alpha=1$, $\theta=1$}
  \label{fig8}
\end{figure}
\newpage
\section{second order derivative of the VG density function}\label{eq:an03}

\begin{figure}[ht]
  \centering
  \begin{subfigure}[b]{0.4\linewidth}
    \includegraphics[width=\linewidth]{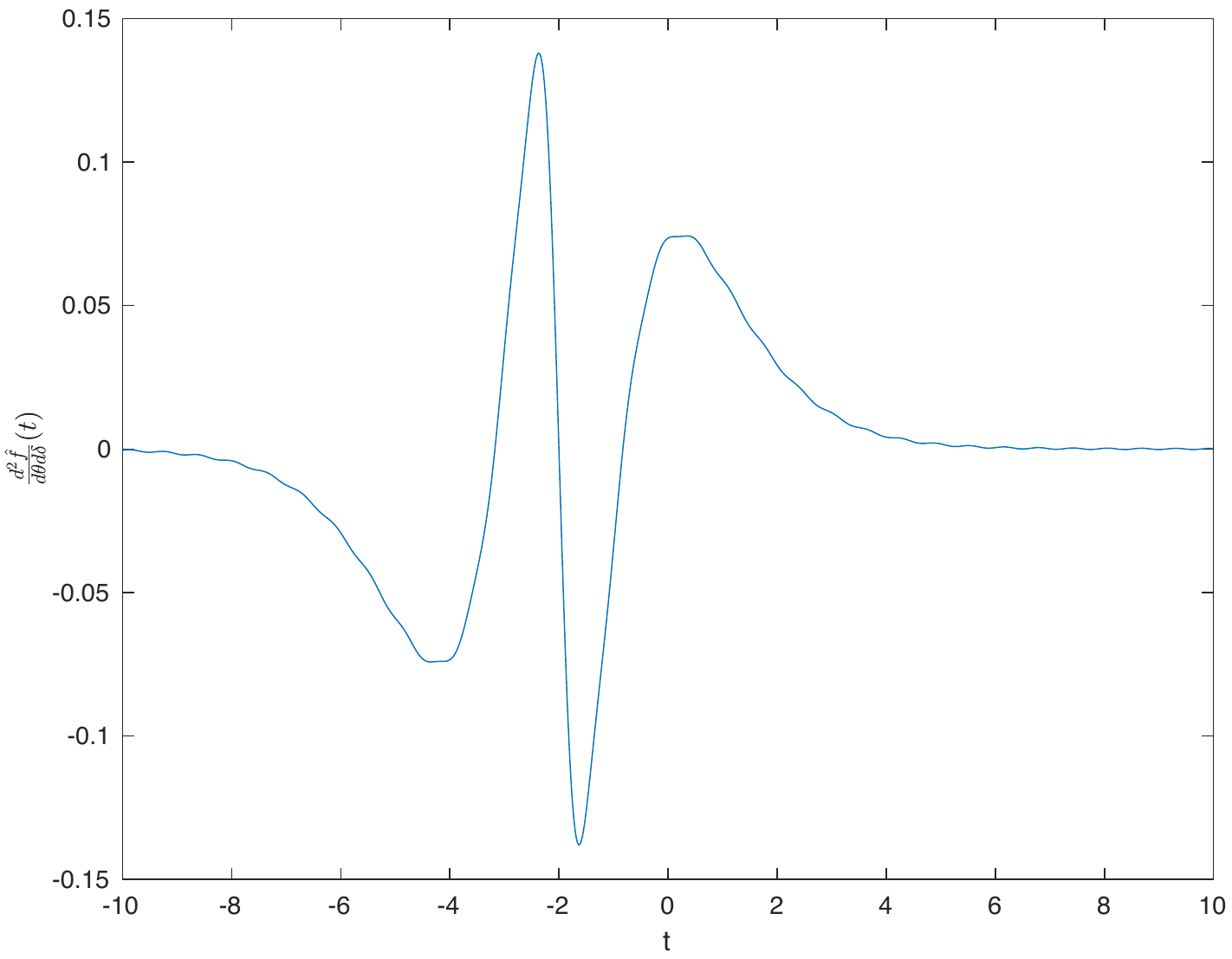}
  \caption{$\frac{d^{2}\hat{f}}{d\theta d\delta}(t)$}
         \label{fig81}
  \end{subfigure}
  \begin{subfigure}[b]{0.4\linewidth}
    \includegraphics[width=\linewidth]{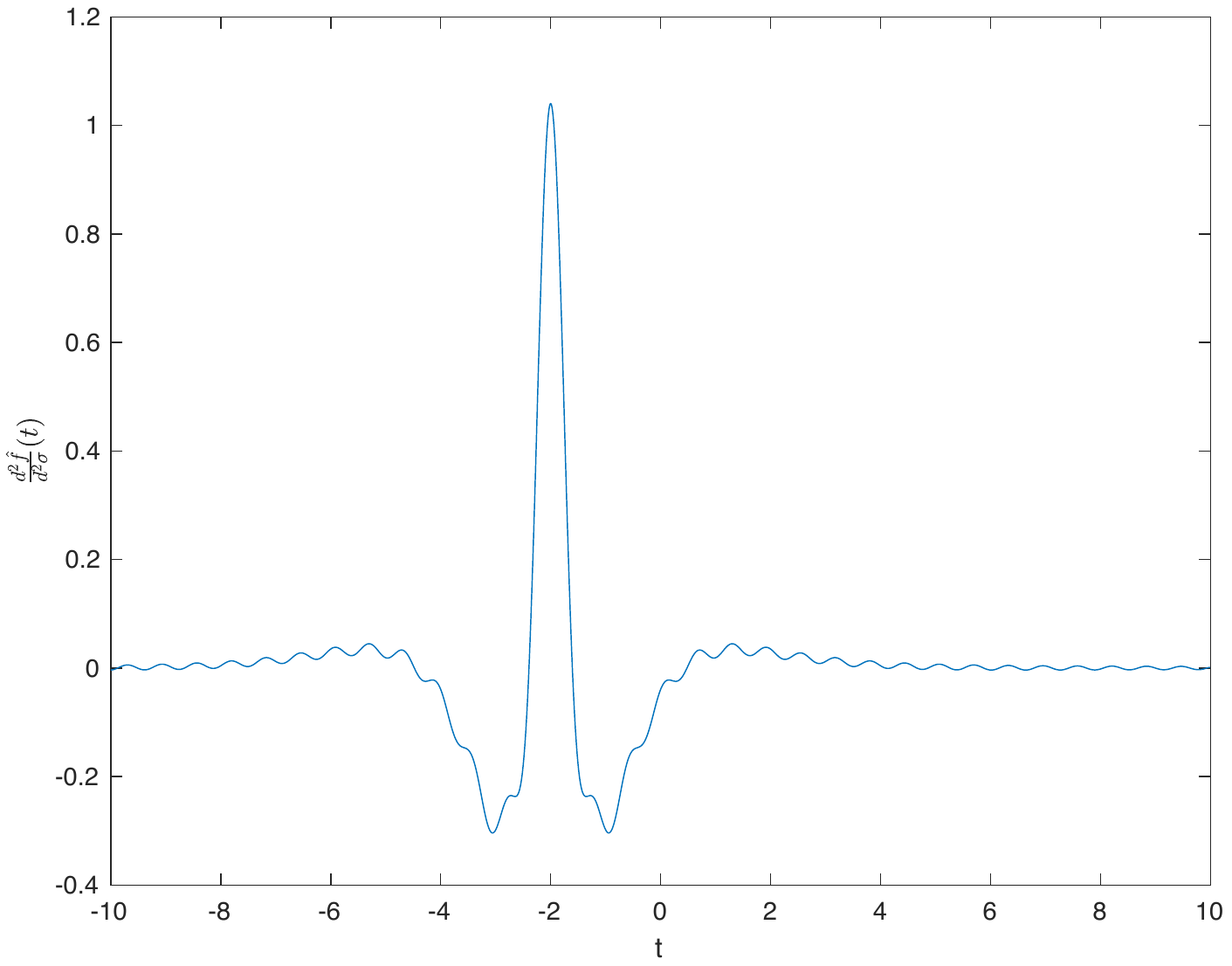}
      \caption{$\frac{d^{2}\hat{f}}{d^{2}\sigma}(t)$}
         \label{fig82}
  \end{subfigure}\\
  \begin{subfigure}[b]{0.4\linewidth}
    \includegraphics[width=\linewidth]{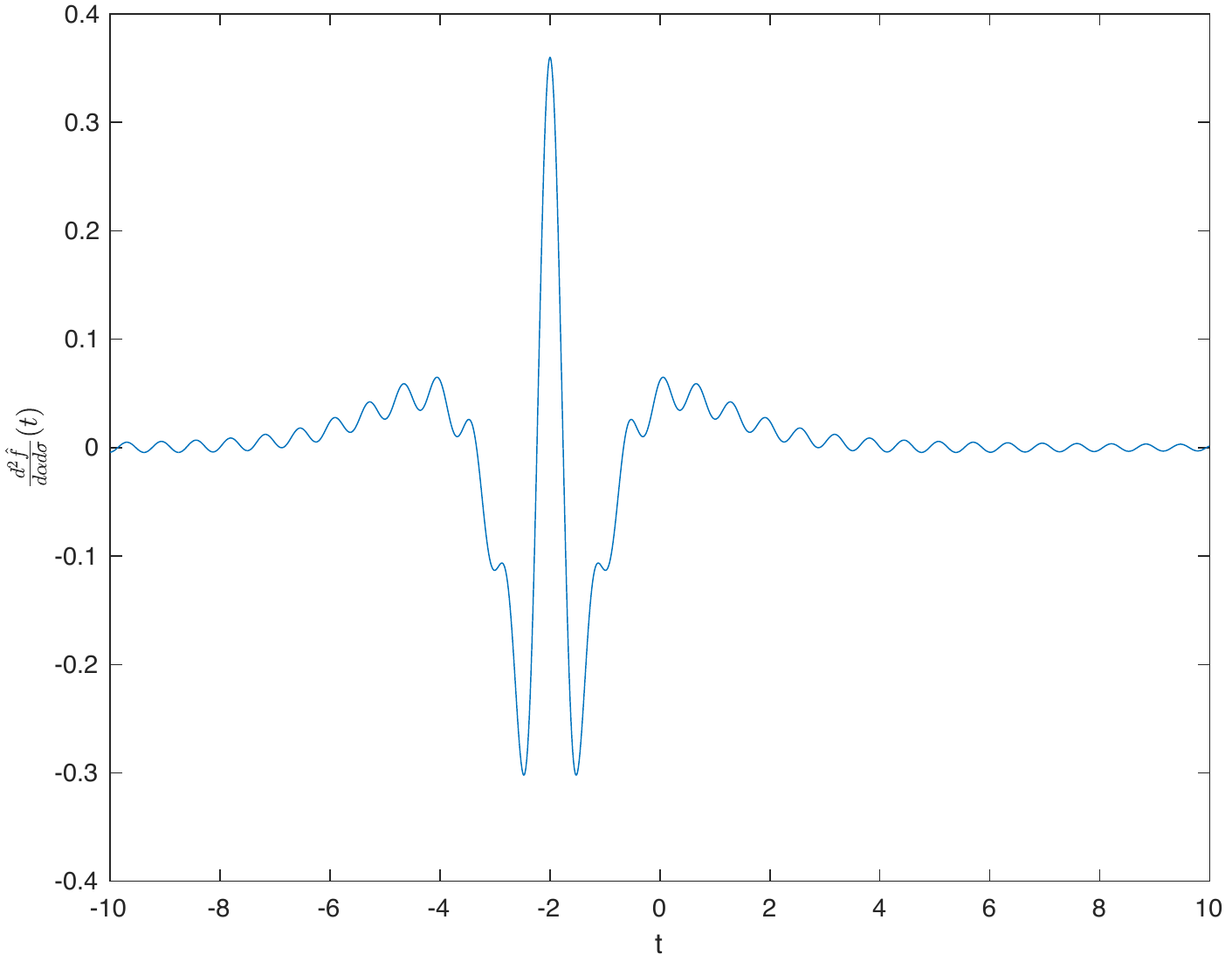}
    \caption{$\frac{d^{2}\hat{f}}{d\alpha d\sigma}(t)$}
         \label{fig83}
  \end{subfigure}
  \begin{subfigure}[b]{0.4\linewidth}
    \includegraphics[width=\linewidth]{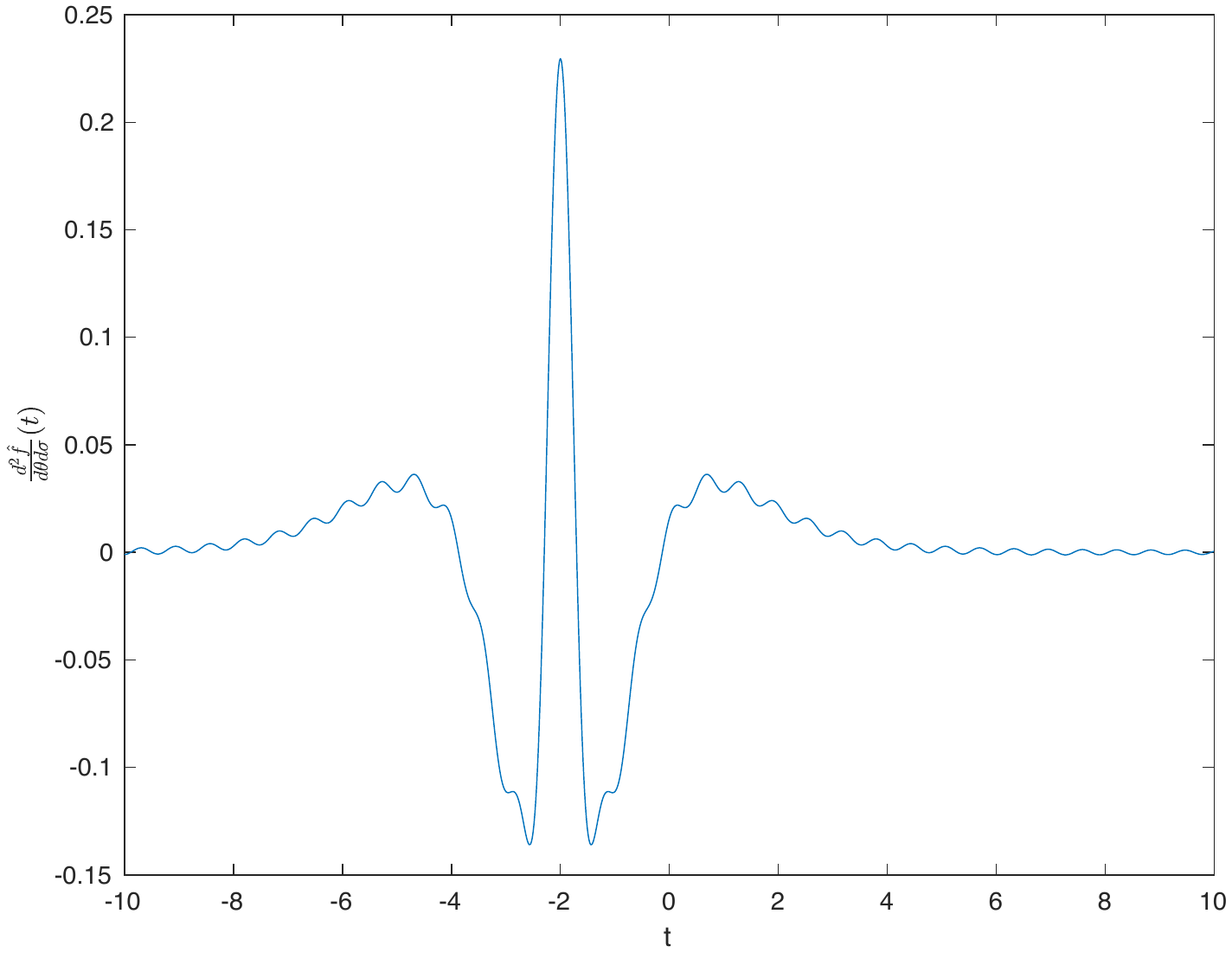}
     \caption{$\frac{d^{2}\hat{f}}{d\theta d\sigma}(t)$}
         \label{fig84}
  \end{subfigure}
 
  \begin{subfigure}[b]{0.4\linewidth}
    \includegraphics[width=\linewidth]{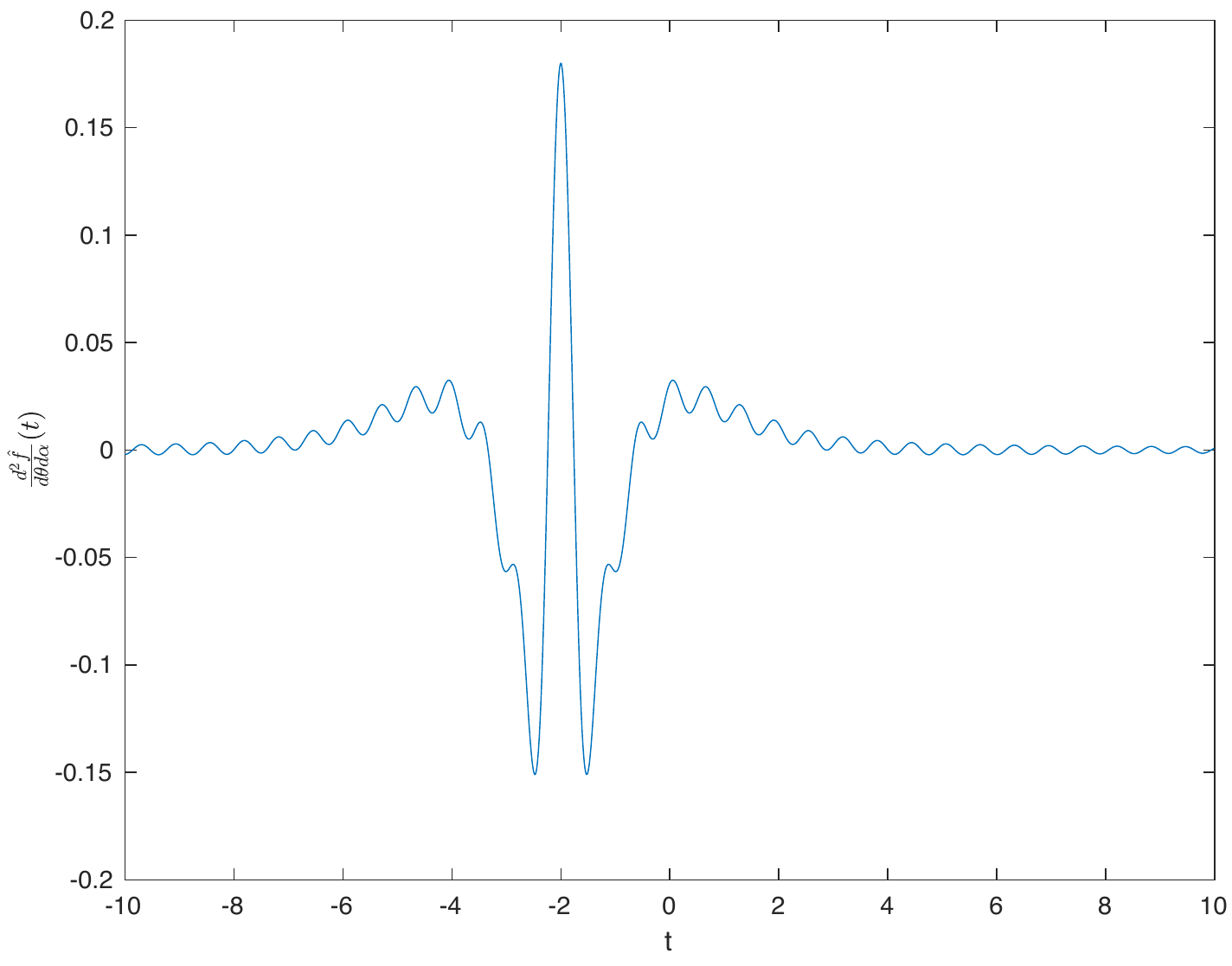}
 \caption{$\frac{d^{2}\hat{f}}{d\theta d\alpha}(t)$}
         \label{fig85}
  \end{subfigure}
 
  \caption{Second order derivative of the VG density function with $\mu=-2$, $\delta=0$, $\sigma=1$, $\alpha=1$, $\theta=1$}
  \label{fig8}
\end{figure}

\newpage
\section{AVG Model parameter estimated \\
by Newton – Raphson Iteration Algorithm (\ref{eq:l21})}\label{eq:an1}

\begin{table}[ht]
\caption{Results of AVG2 Model Parameters Estimations}
 \label{tab:7}
\centering
\resizebox{12cm}{!}{%
\begin{tabular}{cccccccc}
\textbf{Iterations} &
  \textbf{$\mu$} &
  \textbf{$\delta$} &
  \textbf{$\sigma$} &
  \textbf{$\alpha$} &
  \textbf{$\theta$} &
  \textbf{$Log(ML)$} &
  \textbf{$||\frac{d(Log(ML))}{d(\theta)}||$} \\
1  & 0          & 0          & 1          & 1          & 1          & -3582.8388 & 598.743231 \\
2  & 0.05905599 & -0.0009445 & 1.03195903 & 0.9130208  & 1.03208412 & -3561.5099 & 833.530396 \\
3  & 0.06949925 & 0.00400035 & 1.04101444 & 0.88478895 & 1.05131996 & -3559.5656 & 447.807305 \\
4  & 0.07514039 & 0.00055771 & 1.17577397 & 0.67326429 & 1.17778666 & -3569.6221 & 211.365781 \\
5  & 0.08928373 & -0.0263716 & 1.03756321 & 0.83842661 & 0.94304967 & -3554.4434 & 498.289445 \\
6  & 0.08676498 & -0.0521887 & 1.03337015 & 0.85591875 & 0.95066351 & -3550.6419 & 204.467192 \\
7  & 0.086995   & -0.0608517 & 1.02788937 & 0.87382621 & 0.95054954 & -3549.8465 & 66.8039738 \\
8  & 0.08542912 & -0.058547  & 1.02705241 & 0.88258411 & 0.94321299 & -3549.7023 & 15.3209117 \\
9  & 0.08478622 & -0.0576654 & 1.02995166 & 0.88447791 & 0.93670036 & -3549.6921 & 1.14764198 \\
10 & 0.08477798 & -0.0577736 & 1.02922308 & 0.88449072 & 0.93831041 & -3549.692  & 0.17287708 \\
11 & 0.08476475 & -0.0577271 & 1.02960343 & 0.88450434 & 0.93755549 & -3549.692  & 0.07850459 \\
12 & 0.08477094 & -0.0577488 & 1.02942608 & 0.8844984  & 0.93790784 & -3549.692  & 0.03723941 \\
13 & 0.08476804 & -0.0577386 & 1.02950937 & 0.88450117 & 0.93774266 & -3549.692  & 0.01732146 \\
14 & 0.0847694  & -0.0577434 & 1.02947043 & 0.88449987 & 0.93781995 & -3549.692  & 0.00813465 \\
15 & 0.08476876 & -0.0577411 & 1.02948868 & 0.88450048 & 0.93778375 & -3549.692  & 0.00380345 \\
16 & 0.08476906 & -0.0577422 & 1.02948014 & 0.88450019 & 0.9378007  & -3549.692  & 0.00178206 \\
17 & 0.08476892 & -0.0577417 & 1.02948414 & 0.88450033 & 0.93779276 & -3549.692  & 0.00083415 \\
18 & 0.08476898 & -0.0577419 & 1.02948226 & 0.88450026 & 0.93779648 & -3549.692  & 0.00039063 \\
19 & 0.08476895 & -0.0577418 & 1.02948314 & 0.88450029 & 0.93779474 & -3549.692  & 0.00018289 \\
20 & 0.08476897 & -0.0577419 & 1.02948273 & 0.88450028 & 0.93779555 & -3549.692  & 8.56E-05   \\
21 & 0.08476896 & -0.0577418 & 1.02948292 & 0.88450029 & 0.93779517 & -3549.692  & 4.01E-05  
\end{tabular}
}
\end{table}

\begin{table}[ht]
\caption{Results of SVG2 Model Parameters Estimations}
 \label{tab:8}
\centering
\resizebox{12cm}{!}{%
\begin{tabular}{cccccccc}
\textbf{Iterations} &
\textbf{$\mu$} &
\textbf{$\sigma$} &
\textbf{$\alpha$} &
\textbf{$\theta$} &
\textbf{$Log(ML)$} &
\textbf{$||\frac{d(Log(ML))}{d(\theta)}||$} \\
1  & 0          & 1          & 1          & 1          & -3582.8388 & 582.057918 \\
2  & 0.05883419 & 1.03172736 & 0.91367846 & 1.03172736 & -3561.9657 & 1096.4626  \\
3  & 0.06789625 & 1.03903655 & 0.89847415 & 1.03903655 & -3560.2306 & 648.976501 \\
4  & 0.07426127 & 1.06554771 & 0.8507772  & 1.06554771 & -3559.7139 & 270.990173 \\
5  & 0.07141101 & 0.86066094 & 1.14468838 & 0.86066094 & -3567.8236 & 267.821258 \\
6  & 0.06686856 & 0.94781927 & 0.92720936 & 0.94781927 & -3559.7962 & 1338.53688 \\
7  & 0.06739235 & 0.95934348 & 0.90760199 & 0.95934348 & -3558.2696 & 773.42228  \\
8  & 0.0665979  & 0.97263343 & 0.88806105 & 0.97263343 & -3556.6619 & 428.519657 \\
9  & 0.06662118 & 0.98488196 & 0.87499518 & 0.98488196 & -3555.0979 & 203.689715 \\
10 & 0.06614857 & 0.99253289 & 0.87413657 & 0.99253289 & -3554.2092 & 47.3050935 \\
11 & 0.06523245 & 0.99394889 & 0.87669649 & 0.99394889 & -3554.1354 & 1.62304319 \\
12 & 0.06515753 & 0.99386332 & 0.87702703 & 0.99386332 & -3554.1352 & 0.00262982 \\
13 & 0.06515746 & 0.99386267 & 0.87702846 & 0.99386267 & -3554.1352 & 2.69E-08   \\
14 & 0.06515746 & 0.99392282 & 0.87702846 & 0.99374235 & -3554.1352 & 7.64E-05   \\
15 & 0.06515746 & 0.99392282 & 0.87702846 & 0.99374236 & -3554.1352 & 2.27E-12  

\end{tabular}
}
\end{table}
\newpage
\section{SVG2 Cumulative Function ($\hat{F}$)  versus Empirical Function \\ ($\hat{F_{n}}$),  $(1)=|\hat{F(x_{j})}-\hat{F_{n}(x_{j-1})}|$,$(2)=|\hat{F(x_{j})}-\hat{F_{n}(x_{j})}|$, $n=2755$}\label{eq:an2}
 \label{tab:9}
\begin{table}[ht]
\centering
\resizebox{14.25cm}{!}{%
\begin{tabular}{llllllllllllllllll}
\textbf{$x_{j}$} &
  \textbf{$n\_\{j\}$} &
  \textbf{$\hat{F_{n}(x_{j})}$} &
  \textbf{$\hat{F(x_{j})}$} &
  \textbf{(1)} &
  \textbf{(2)} &
 \textbf{$x_{j}$} &
  \textbf{$n\_\{j\}$} &
  \textbf{$\hat{F_{n}(x_{j})}$} &
  \textbf{$\hat{F(x_{j})}$} &
   \textbf{(1)} &
  \textbf{(2)} &
  \textbf{$x_{j}$} &
  \textbf{$n\_\{j\}$} &
  \textbf{$\hat{F_{n}(x_{j})}$} &
  \textbf{$\hat{F(x_{j})}$} &
  \textbf{(1)} &
  \textbf{(2)} \\
 &
   &
   &
   &
   &
   &
   &
   &
   &
   &
   &
   &
   &
   &
   &
   &
   &
   \\
-5.01 &
  1 &
  0.000362976 &
  0.00053753 &
  0.00053753 &
  0.00017456 &
  -1.685 &
  7 &
  0.04827586 &
  0.0350006 &
  0.01073443 &
  0.01327526 &
  1.64 &
  4 &
  0.96442831 &
  0.95534423 &
  0.00763218 &
  0.00908409 \\
-4.975 &
  0 &
  0.000362976 &
  0.00057061 &
  0.00020763 &
  0.00020763 &
  -1.65 &
  11 &
  0.0522686 &
  0.03665193 &
  0.01162393 &
  0.01561667 &
  1.675 &
  9 &
  0.9676951 &
  0.95760224 &
  0.00682607 &
  0.01009286 \\
-4.94 &
  0 &
  0.000362976 &
  0.00057987 &
  0.00021689 &
  0.00021689 &
  -1.615 &
  3 &
  0.05335753 &
  0.03840489 &
  0.01386372 &
  0.01495264 &
  1.71 &
  4 &
  0.96914701 &
  0.959677 &
  0.0080181 &
  0.00947001 \\
-4.905 &
  0 &
  0.000362976 &
  0.00056565 &
  0.00020267 &
  0.00020267 &
  -1.58 &
  4 &
  0.05480944 &
  0.04030396 &
  0.01305358 &
  0.01450548 &
  1.745 &
  5 &
  0.97096189 &
  0.9615774 &
  0.0075696 &
  0.00938449 \\
-4.87 &
  0 &
  0.000362976 &
  0.00053125 &
  0.00016827 &
  0.00016827 &
  -1.545 &
  5 &
  0.05662432 &
  0.042371 &
  0.01243843 &
  0.01425332 &
  1.78 &
  4 &
  0.97241379 &
  0.96333629 &
  0.0076256 &
  0.0090775 \\
-4.835 &
  0 &
  0.000362976 &
  0.00048258 &
  0.0001196 &
  0.0001196 &
  -1.51 &
  6 &
  0.05880218 &
  0.04462677 &
  0.01199755 &
  0.01417541 &
  1.815 &
  3 &
  0.97350272 &
  0.9649836 &
  0.00743019 &
  0.00851912 \\
-4.8 &
  1 &
  0.000725953 &
  0.00042749 &
  6.45E-05 &
  0.00029847 &
  -1.475 &
  11 &
  0.06279492 &
  0.04708394 &
  0.01171824 &
  0.01571098 &
  1.85 &
  3 &
  0.97459165 &
  0.9665451 &
  0.00695762 &
  0.00804655 \\
-4.765 &
  0 &
  0.000725953 &
  0.00037483 &
  0.00035112 &
  0.00035112 &
  -1.44 &
  8 &
  0.06569873 &
  0.04974601 &
  0.01304891 &
  0.01595272 &
  1.885 &
  2 &
  0.9753176 &
  0.96805709 &
  0.00653456 &
  0.00726051 \\
-4.73 &
  0 &
  0.000725953 &
  0.00033343 &
  0.00039252 &
  0.00039252 &
  -1.405 &
  4 &
  0.06715064 &
  0.05260725 &
  0.01309148 &
  0.01454339 &
  1.92 &
  6 &
  0.97749546 &
  0.96953709 &
  0.00578052 &
  0.00795838 \\
-4.695 &
  0 &
  0.000725953 &
  0.00031096 &
  0.00041499 &
  0.00041499 &
  -1.37 &
  10 &
  0.0707804 &
  0.05565387 &
  0.01149677 &
  0.01512653 &
  1.955 &
  3 &
  0.97858439 &
  0.97100001 &
  0.00649546 &
  0.00758439 \\
-4.66 &
  0 &
  0.000725953 &
  0.00031305 &
  0.0004129 &
  0.0004129 &
  -1.335 &
  6 &
  0.07295826 &
  0.05886627 &
  0.01191413 &
  0.01409199 &
  1.99 &
  3 &
  0.97967332 &
  0.97245259 &
  0.00613181 &
  0.00722074 \\
-4.625 &
  2 &
  0.001451906 &
  0.00034253 &
  0.00038342 &
  0.00110938 &
  -1.3 &
  11 &
  0.076951 &
  0.06222216 &
  0.0107361 &
  0.01472884 &
  2.025 &
  3 &
  0.98076225 &
  0.97389748 &
  0.00577584 &
  0.00686477 \\
-4.59 &
  0 &
  0.001451906 &
  0.00039908 &
  0.00105283 &
  0.00105283 &
  -1.265 &
  5 &
  0.07876588 &
  0.06570024 &
  0.01125076 &
  0.01306564 &
  2.06 &
  1 &
  0.98112523 &
  0.97531775 &
  0.00544451 &
  0.00580748 \\
-4.555 &
  0 &
  0.001451906 &
  0.00047921 &
  0.0009727 &
  0.0009727 &
  -1.23 &
  9 &
  0.08203267 &
  0.06929436 &
  0.00947152 &
  0.0127383 &
  2.095 &
  2 &
  0.98185118 &
  0.97669941 &
  0.00442581 &
  0.00515177 \\
-4.52 &
  1 &
  0.001814882 &
  0.00057667 &
  0.00087523 &
  0.00123821 &
  -1.195 &
  8 &
  0.08493648 &
  0.07297576 &
  0.00905691 &
  0.01196072 &
  2.13 &
  3 &
  0.98294011 &
  0.97803303 &
  0.00381815 &
  0.00490708 \\
-4.485 &
  0 &
  0.001814882 &
  0.00068316 &
  0.00113172 &
  0.00113172 &
  -1.16 &
  16 &
  0.0907441 &
  0.07674627 &
  0.00819021 &
  0.01399783 &
  2.165 &
  3 &
  0.98402904 &
  0.97929655 &
  0.00364356 &
  0.00473249 \\
-4.45 &
  0 &
  0.001814882 &
  0.00078932 &
  0.00102556 &
  0.00102556 &
  -1.125 &
  8 &
  0.09364791 &
  0.08064272 &
  0.01010138 &
  0.01300519 &
  2.2 &
  4 &
  0.98548094 &
  0.98047385 &
  0.00355519 &
  0.00500709 \\
-4.415 &
  2 &
  0.002540835 &
  0.00088588 &
  0.000929 &
  0.00165495 &
  -1.09 &
  7 &
  0.09618875 &
  0.08468243 &
  0.00896548 &
  0.01150631 &
  2.235 &
  1 &
  0.98584392 &
  0.98155588 &
  0.00392506 &
  0.00428804 \\
-4.38 &
  0 &
  0.002540835 &
  0.0009648 &
  0.00157603 &
  0.00157603 &
  -1.055 &
  11 &
  0.10018149 &
  0.08890488 &
  0.00728386 &
  0.0112766 &
  2.27 &
  2 &
  0.98656987 &
  0.98253026 &
  0.00331366 &
  0.00403961 \\
-4.345 &
  0 &
  0.002540835 &
  0.00102027 &
  0.00152057 &
  0.00152057 &
  -1.02 &
  13 &
  0.10490018 &
  0.09335839 &
  0.0068231 &
  0.01154179 &
  2.305 &
  1 &
  0.98693285 &
  0.98339832 &
  0.00317155 &
  0.00353453 \\
-4.31 &
  1 &
  0.002903811 &
  0.00104947 &
  0.00149136 &
  0.00185434 &
  -0.985 &
  11 &
  0.10889292 &
  0.09809628 &
  0.00680391 &
  0.01079665 &
  2.34 &
  1 &
  0.98729583 &
  0.98417215 &
  0.0027607 &
  0.00312368 \\
-4.275 &
  1 &
  0.003266788 &
  0.00105301 &
  0.0018508 &
  0.00221378 &
  -0.95 &
  13 &
  0.11361162 &
  0.10317228 &
  0.00572064 &
  0.01043934 &
  2.375 &
  1 &
  0.9876588 &
  0.98486216 &
  0.00243367 &
  0.00279664 \\
-4.24 &
  0 &
  0.003266788 &
  0.00103485 &
  0.00223193 &
  0.00223193 &
  -0.915 &
  13 &
  0.11833031 &
  0.10863572 &
  0.00497589 &
  0.00969459 &
  2.41 &
  2 &
  0.98838475 &
  0.98548587 &
  0.00217293 &
  0.00289889 \\
-4.205 &
  0 &
  0.003266788 &
  0.00100196 &
  0.00226483 &
  0.00226483 &
  -0.88 &
  15 &
  0.12377495 &
  0.11452697 &
  0.00380334 &
  0.00924798 &
  2.445 &
  1 &
  0.98874773 &
  0.98606329 &
  0.00232146 &
  0.00268444 \\
-4.17 &
  0 &
  0.003266788 &
  0.00096346 &
  0.00230332 &
  0.00230332 &
  -0.845 &
  16 &
  0.12958258 &
  0.1208738 &
  0.00290115 &
  0.00870877 &
  2.48 &
  1 &
  0.98911071 &
  0.98661629 &
  0.00213144 &
  0.00249441 \\
-4.135 &
  0 &
  0.003266788 &
  0.00092963 &
  0.00233716 &
  0.00233716 &
  -0.81 &
  12 &
  0.13393829 &
  0.12768918 &
  0.0018934 &
  0.00624912 &
  2.515 &
  2 &
  0.98983666 &
  0.9871586 &
  0.00195211 &
  0.00267806 \\
-4.1 &
  0 &
  0.003266788 &
  0.00091061 &
  0.00235617 &
  0.00235617 &
  -0.775 &
  18 &
  0.14047187 &
  0.13497076 &
  0.00103247 &
  0.00550111 &
  2.55 &
  2 &
  0.99056261 &
  0.98770945 &
  0.00212721 &
  0.00285317 \\
-4.065 &
  0 &
  0.003266788 &
  0.00091524 &
  0.00235155 &
  0.00235155 &
  -0.74 &
  17 &
  0.14664247 &
  0.14270246 &
  0.00223059 &
  0.00394001 &
  2.585 &
  0 &
  0.99056261 &
  0.98827661 &
  0.002286 &
  0.002286 \\
-4.03 &
  0 &
  0.003266788 &
  0.00094992 &
  0.00231686 &
  0.00231686 &
  -0.705 &
  26 &
  0.15607985 &
  0.15085789 &
  0.00421542 &
  0.00522196 &
  2.62 &
  0 &
  0.99056261 &
  0.9888634 &
  0.00169921 &
  0.00169921 \\
-3.995 &
  0 &
  0.003266788 &
  0.00101785 &
  0.00224893 &
  0.00224893 &
  -0.67 &
  21 &
  0.16370236 &
  0.15940559 &
  0.00332573 &
  0.00429677 &
  2.655 &
  1 &
  0.99092559 &
  0.98946709 &
  0.00109552 &
  0.0014585 \\
-3.96 &
  0 &
  0.003266788 &
  0.00111857 &
  0.00214821 &
  0.00214821 &
  -0.635 &
  29 &
  0.17422868 &
  0.16831548 &
  0.00461312 &
  0.0059132 &
  2.69 &
  1 &
  0.99128857 &
  0.99008131 &
  0.00084428 &
  0.00120725 \\
-3.925 &
  0 &
  0.003266788 &
  0.00124799 &
  0.0020188 &
  0.0020188 &
  -0.6 &
  15 &
  0.17967332 &
  0.17756601 &
  0.00333734 &
  0.00210731 &
  2.725 &
  0 &
  0.99128857 &
  0.99069034 &
  0.00059822 &
  0.00059822 \\
-3.89 &
  0 &
  0.003266788 &
  0.00139882 &
  0.00186797 &
  0.00186797 &
  -0.565 &
  32 &
  0.19128857 &
  0.18715116 &
  0.00747784 &
  0.0041374 &
  2.76 &
  1 &
  0.99165154 &
  0.99127971 &
  8.86E-06 &
  0.00037184 \\
-3.855 &
  2 &
  0.00399274 &
  0.00156146 &
  0.00170533 &
  0.00243128 &
  -0.53 &
  21 &
  0.19891107 &
  0.19708652 &
  0.00579795 &
  0.00182455 &
  2.795 &
  0 &
  0.99165154 &
  0.99183786 &
  0.00018632 &
  0.00018632 \\
-3.82 &
  0 &
  0.00399274 &
  0.00172515 &
  0.00226759 &
  0.00226759 &
  -0.495 &
  24 &
  0.2076225 &
  0.2074137 &
  0.00850263 &
  0.0002088 &
  2.83 &
  2 &
  0.9923775 &
  0.99234968 &
  0.00069813 &
  2.78E-05 \\
-3.785 &
  1 &
  0.004355717 &
  0.00187926 &
  0.00211349 &
  0.00247646 &
  -0.46 &
  22 &
  0.21560799 &
  0.21820243 &
  0.01057993 &
  0.00259445 &
  2.865 &
  1 &
  0.99274047 &
  0.99280499 &
  0.00042749 &
  6.45E-05 \\
-3.75 &
  0 &
  0.004355717 &
  0.00201464 &
  0.00234108 &
  0.00234108 &
  -0.425 &
  36 &
  0.22867514 &
  0.22954979 &
  0.01394181 &
  0.00087466 &
  2.9 &
  1 &
  0.99310345 &
  0.99319919 &
  0.00045871 &
  9.57E-05 \\
-3.715 &
  0 &
  0.004355717 &
  0.00212481 &
  0.00223091 &
  0.00223091 &
  -0.39 &
  41 &
  0.24355717 &
  0.24157648 &
  0.01290134 &
  0.00198069 &
  2.935 &
  1 &
  0.99346642 &
  0.9935296 &
  0.00042615 &
  6.32E-05 \\
-3.68 &
  1 &
  0.004718693 &
  0.00220677 &
  0.00214894 &
  0.00251192 &
  -0.355 &
  39 &
  0.25771325 &
  0.25442008 &
  0.01086292 &
  0.00329316 &
  2.97 &
  2 &
  0.99419238 &
  0.99380157 &
  0.00033515 &
  0.00039081 \\
-3.645 &
  0 &
  0.004718693 &
  0.00226154 &
  0.00245715 &
  0.00245715 &
  -0.32 &
  53 &
  0.276951 &
  0.26822579 &
  0.01051254 &
  0.00872521 &
  3.005 &
  0 &
  0.99419238 &
  0.99402612 &
  0.00016626 &
  0.00016626 \\
-3.61 &
  1 &
  0.00508167 &
  0.00229408 &
  0.00242461 &
  0.00278759 &
  -0.285 &
  34 &
  0.2892922 &
  0.28313512 &
  0.00618412 &
  0.00615708 &
  3.04 &
  0 &
  0.99419238 &
  0.99421539 &
  2.30E-05 &
  2.30E-05 \\
-3.575 &
  0 &
  0.00508167 &
  0.0023128 &
  0.00276887 &
  0.00276887 &
  -0.25 &
  41 &
  0.30417423 &
  0.29927354 &
  0.00998135 &
  0.00490069 &
  3.075 &
  1 &
  0.99455535 &
  0.99438437 &
  0.00019199 &
  0.00017099 \\
-3.54 &
  0 &
  0.00508167 &
  0.00232865 &
  0.00275302 &
  0.00275302 &
  -0.215 &
  54 &
  0.32377495 &
  0.31673807 &
  0.01256384 &
  0.00703689 &
  3.11 &
  0 &
  0.99455535 &
  0.99454829 &
  7.07E-06 &
  7.07E-06 \\
-3.505 &
  2 &
  0.005807623 &
  0.00235387 &
  0.0027278 &
  0.00345375 &
  -0.18 &
  52 &
  0.34264973 &
  0.33558577 &
  0.01181081 &
  0.00706396 &
  3.145 &
  0 &
  0.99455535 &
  0.99472139 &
  0.00016604 &
  0.00016604 \\
-3.47 &
  0 &
  0.005807623 &
  0.00240051 &
  0.00340711 &
  0.00340711 &
  -0.145 &
  71 &
  0.36842105 &
  0.35582438 &
  0.01317465 &
  0.01259667 &
  3.18 &
  1 &
  0.99491833 &
  0.99491285 &
  0.0003575 &
  5.48E-06 \\
-3.435 &
  0 &
  0.005807623 &
  0.00247905 &
  0.00332858 &
  0.00332858 &
  -0.11 &
  59 &
  0.38983666 &
  0.37740585 &
  0.0089848 &
  0.01243081 &
  3.215 &
  1 &
  0.99528131 &
  0.99513078 &
  0.00021245 &
  0.00015053 \\
-3.4 &
  3 &
  0.006896552 &
  0.00259707 &
  0.00321056 &
  0.00429949 &
  -0.075 &
  62 &
  0.4123412 &
  0.40022352 &
  0.01038686 &
  0.01211768 &
  3.25 &
  0 &
  0.99528131 &
  0.99537645 &
  9.51E-05 &
  9.51E-05 \\
-3.365 &
  0 &
  0.006896552 &
  0.00275835 &
  0.00413821 &
  0.00413821 &
  -0.04 &
  71 &
  0.43811252 &
  0.42411332 &
  0.01177213 &
  0.0139992 &
  3.285 &
  4 &
  0.99673321 &
  0.99564687 &
  0.00036556 &
  0.00108634 \\
-3.33 &
  0 &
  0.006896552 &
  0.00296235 &
  0.0039342 &
  0.0039342 &
  -0.005 &
  79 &
  0.46678766 &
  0.44885915 &
  0.01074663 &
  0.01792851 &
  3.32 &
  0 &
  0.99673321 &
  0.99593476 &
  0.00079845 &
  0.00079845 \\
-3.295 &
  2 &
  0.007622505 &
  0.00320427 &
  0.00369228 &
  0.00441823 &
  0.03 &
  81 &
  0.49618875 &
  0.47420195 &
  0.00741429 &
  0.0219868 &
  3.355 &
  0 &
  0.99673321 &
  0.99623048 &
  0.00050274 &
  0.00050274 \\
-3.26 &
  0 &
  0.007622505 &
  0.00347555 &
  0.00414695 &
  0.00414695 &
  0.065 &
  66 &
  0.52014519 &
  0.49985209 &
  0.00366334 &
  0.0202931 &
  3.39 &
  1 &
  0.99709619 &
  0.99651915 &
  0.00021407 &
  0.00057704 \\
-3.225 &
  1 &
  0.007985481 &
  0.00376493 &
  0.00385758 &
  0.00422055 &
  0.1 &
  56 &
  0.54047187 &
  0.52550399 &
  0.0053588 &
  0.01496788 &
  3.425 &
  0 &
  0.99709619 &
  0.9967907 &
  0.00030549 &
  0.00030549 \\
-3.19 &
  0 &
  0.007985481 &
  0.00405981 &
  0.00392567 &
  0.00392567 &
  0.135 &
  58 &
  0.5615245 &
  0.55085199 &
  0.01038012 &
  0.01067251 &
  3.46 &
  0 &
  0.99709619 &
  0.99703301 &
  6.32E-05 &
  6.32E-05 \\
-3.155 &
  2 &
  0.008711434 &
  0.00434785 &
  0.00363763 &
  0.00436359 &
  0.17 &
  51 &
  0.5800363 &
  0.57560617 &
  0.01408167 &
  0.00443013 &
  3.495 &
  0 &
  0.99709619 &
  0.99723748 &
  0.00014129 &
  0.00014129 \\
-3.12 &
  0 &
  0.008711434 &
  0.00461851 &
  0.00409292 &
  0.00409292 &
  0.205 &
  64 &
  0.60326679 &
  0.59950707 &
  0.01947077 &
  0.00375972 &
  3.53 &
  0 &
  0.99709619 &
  0.99739926 &
  0.00030307 &
  0.00030307 \\
-3.085 &
  5 &
  0.010526316 &
  0.00486449 &
  0.00384694 &
  0.00566182 &
  0.24 &
  54 &
  0.62286751 &
  0.62233802 &
  0.01907123 &
  0.00052949 &
  3.565 &
  0 &
  0.99709619 &
  0.99751804 &
  0.00042185 &
  0.00042185 \\
-3.05 &
  2 &
  0.011252269 &
  0.00508274 &
  0.00544358 &
  0.00616953 &
  0.275 &
  53 &
  0.64210526 &
  0.64393434 &
  0.02106682 &
  0.00182907 &
  3.6 &
  0 &
  0.99709619 &
  0.99759689 &
  0.0005007 &
  0.0005007 \\
-3.015 &
  2 &
  0.011978221 &
  0.00527498 &
  0.00597729 &
  0.00670324 &
  0.31 &
  49 &
  0.65989111 &
  0.66418868 &
  0.02208342 &
  0.00429757 &
  3.635 &
  0 &
  0.99709619 &
  0.99764367 &
  0.00054748 &
  0.00054748 \\
-2.98 &
  0 &
  0.011978221 &
  0.00544774 &
  0.00653048 &
  0.00653048 &
  0.345 &
  47 &
  0.676951 &
  0.68305233 &
  0.02316123 &
  0.00610134 &
  3.67 &
  0 &
  0.99709619 &
  0.99766906 &
  0.00057287 &
  0.00057287 \\
-2.945 &
  0 &
  0.011978221 &
  0.00561168 &
  0.00636654 &
  0.00636654 &
  0.38 &
  46 &
  0.69364791 &
  0.70058043 &
  0.02362944 &
  0.00693252 &
  3.705 &
  0 &
  0.99709619 &
  0.99768495 &
  0.00058876 &
  0.00058876 \\
-2.91 &
  2 &
  0.012704174 &
  0.00578052 &
  0.0061977 &
  0.00692365 &
  0.415 &
  42 &
  0.70889292 &
  0.71672986 &
  0.02308195 &
  0.00783694 &
  3.74 &
  1 &
  0.99745917 &
  0.99770357 &
  0.00060738 &
  0.00024441 \\
-2.875 &
  0 &
  0.012704174 &
  0.00596947 &
  0.0067347 &
  0.0067347 &
  0.45 &
  42 &
  0.72413793 &
  0.73160823 &
  0.02271531 &
  0.0074703 &
  3.775 &
  0 &
  0.99745917 &
  0.9977359 &
  0.00027673 &
  0.00027673 \\
-2.84 &
  1 &
  0.013067151 &
  0.00619354 &
  0.00651064 &
  0.00687361 &
  0.485 &
  49 &
  0.74192377 &
  0.74542577 &
  0.02128784 &
  0.003502 &
  3.81 &
  0 &
  0.99745917 &
  0.99779057 &
  0.0003314 &
  0.0003314 \\
-2.805 &
  3 &
  0.01415608 &
  0.00646575 &
  0.0066014 &
  0.00769033 &
  0.52 &
  36 &
  0.75499093 &
  0.75827955 &
  0.01635577 &
  0.00328862 &
  3.845 &
  0 &
  0.99745917 &
  0.99787202 &
  0.00041286 &
  0.00041286 \\
-2.77 &
  0 &
  0.01415608 &
  0.00679564 &
  0.00736044 &
  0.00736044 &
  0.555 &
  33 &
  0.76696915 &
  0.77031477 &
  0.01532384 &
  0.00334562 &
  3.88 &
  0 &
  0.99745917 &
  0.99798187 &
  0.00052271 &
  0.00052271 \\
-2.735 &
  0 &
  0.01415608 &
  0.00718809 &
  0.00696799 &
  0.00696799 &
  0.59 &
  33 &
  0.77894737 &
  0.78166917 &
  0.01470003 &
  0.0027218 &
  3.915 &
  0 &
  0.99745917 &
  0.998117 &
  0.00065784 &
  0.00065784 \\
-2.7 &
  2 &
  0.014882033 &
  0.00764271 &
  0.00651337 &
  0.00723932 &
  0.625 &
  37 &
  0.7923775 &
  0.79246369 &
  0.01351632 &
  8.62E-05 &
  3.95 &
  0 &
  0.99745917 &
  0.99827094 &
  0.00081178 &
  0.00081178 \\
-2.665 &
  1 &
  0.015245009 &
  0.00815393 &
  0.00672811 &
  0.00709108 &
  0.66 &
  34 &
  0.80471869 &
  0.8027957 &
  0.0104182 &
  0.00192299 &
  3.985 &
  0 &
  0.99745917 &
  0.99843458 &
  0.00097541 &
  0.00097541 \\
-2.63 &
  1 &
  0.015607985 &
  0.00871162 &
  0.00653338 &
  0.00689636 &
  0.695 &
  23 &
  0.81306715 &
  0.81273525 &
  0.00801655 &
  0.0003319 &
  4.02 &
  1 &
  0.99782214 &
  0.9985973 &
  0.00113813 &
  0.00077515 \\
-2.595 &
  4 &
  0.017059891 &
  0.00930242 &
  0.00630557 &
  0.00775747 &
  0.73 &
  26 &
  0.82250454 &
  0.82232427 &
  0.00925711 &
  0.00018027 &
  4.055 &
  0 &
  0.99782214 &
  0.99874832 &
  0.00092618 &
  0.00092618 \\
-2.56 &
  6 &
  0.01923775 &
  0.00991136 &
  0.00714853 &
  0.00932639 &
  0.765 &
  28 &
  0.83266788 &
  0.83157862 &
  0.00907409 &
  0.00108925 &
  4.09 &
  1 &
  0.99818512 &
  0.99887802 &
  0.00105588 &
  0.0006929 \\
-2.525 &
  1 &
  0.019600726 &
  0.01052387 &
  0.00871388 &
  0.00907686 &
  0.8 &
  30 &
  0.84355717 &
  0.84051747 &
  0.00784959 &
  0.0030397 &
  4.125 &
  0 &
  0.99818512 &
  0.9989791 &
  0.00079398 &
  0.00079398 \\
-2.49 &
  0 &
  0.019600726 &
  0.01112771 &
  0.00847302 &
  0.00847302 &
  0.835 &
  19 &
  0.85045372 &
  0.84906839 &
  0.00551122 &
  0.00138533 &
  4.16 &
  0 &
  0.99818512 &
  0.99904741 &
  0.0008623 &
  0.0008623 \\
-2.455 &
  3 &
  0.020689655 &
  0.01171472 &
  0.00788601 &
  0.00897494 &
  0.87 &
  29 &
  0.86098004 &
  0.85722732 &
  0.0067736 &
  0.00375272 &
  4.195 &
  0 &
  0.99818512 &
  0.99908247 &
  0.00089735 &
  0.00089735 \\
-2.42 &
  3 &
  0.021778584 &
  0.0122821 &
  0.00840755 &
  0.00949648 &
  0.905 &
  21 &
  0.86860254 &
  0.8649413 &
  0.00396127 &
  0.00366124 &
  4.23 &
  1 &
  0.99854809 &
  0.99908738 &
  0.00090226 &
  0.00053929 \\
-2.385 &
  2 &
  0.022504537 &
  0.01283309 &
  0.00894549 &
  0.00967144 &
  0.94 &
  18 &
  0.87513612 &
  0.87222815 &
  0.0036256 &
  0.00290797 &
  4.265 &
  0 &
  0.99854809 &
  0.99906853 &
  0.00052044 &
  0.00052044 \\
-2.35 &
  2 &
  0.02323049 &
  0.01337696 &
  0.00912758 &
  0.00985353 &
  0.975 &
  17 &
  0.88130672 &
  0.87904889 &
  0.00391277 &
  0.00225783 &
  4.3 &
  1 &
  0.99891107 &
  0.99903489 &
  0.0004868 &
  0.00012382 \\
-2.315 &
  1 &
  0.023593466 &
  0.01392822 &
  0.00930227 &
  0.00966525 &
  1.01 &
  27 &
  0.89110708 &
  0.88540103 &
  0.00409432 &
  0.00570605 &
  4.335 &
  0 &
  0.99891107 &
  0.99899639 &
  8.53E-05 &
  8.53E-05 \\
-2.28 &
  5 &
  0.025408348 &
  0.01450523 &
  0.00908823 &
  0.01090311 &
  1.045 &
  16 &
  0.8969147 &
  0.89129736 &
  0.00019028 &
  0.00561735 &
  4.37 &
  1 &
  0.99927405 &
  0.99896338 &
  5.23E-05 &
  0.00031067 \\
-2.245 &
  3 &
  0.026497278 &
  0.0151283 &
  0.01028005 &
  0.01136898 &
  1.08 &
  19 &
  0.90381125 &
  0.89676548 &
  0.00014922 &
  0.00704577 &
  4.405 &
  0 &
  0.99927405 &
  0.99894501 &
  0.00032904 &
  0.00032904 \\
-2.21 &
  2 &
  0.02722323 &
  0.01581745 &
  0.01067983 &
  0.01140578 &
  1.115 &
  12 &
  0.90816697 &
  0.90184565 &
  0.0019656 &
  0.00632132 &
  4.44 &
  0 &
  0.99927405 &
  0.99894827 &
  0.00032578 &
  0.00032578 \\
-2.175 &
  4 &
  0.028675136 &
  0.01659023 &
  0.010633 &
  0.01208491 &
  1.15 &
  9 &
  0.91143376 &
  0.9065871 &
  0.00157987 &
  0.00484666 &
  4.475 &
  0 &
  0.99927405 &
  0.99897717 &
  0.00029687 &
  0.00029687 \\
-2.14 &
  3 &
  0.029764065 &
  0.01745976 &
  0.01121537 &
  0.0123043 &
  1.185 &
  13 &
  0.91615245 &
  0.91104355 &
  0.00039021 &
  0.0051089 &
  4.51 &
  0 &
  0.99927405 &
  0.99903235 &
  0.0002417 &
  0.0002417 \\
-2.105 &
  1 &
  0.030127042 &
  0.01843324 &
  0.01133082 &
  0.0116938 &
  1.22 &
  18 &
  0.92268603 &
  0.91526835 &
  0.0008841 &
  0.00741767 &
  4.545 &
  1 &
  0.99963702 &
  0.99911103 &
  0.00016302 &
  0.000526 \\
-2.07 &
  2 &
  0.030852995 &
  0.01951117 &
  0.01061587 &
  0.01134182 &
  1.255 &
  17 &
  0.92885662 &
  0.91932123 &
  0.00336479 &
  0.00953539 &
  4.58 &
  0 &
  0.99963702 &
  0.99920743 &
  0.0004296 &
  0.0004296 \\
-2.035 &
  3 &
  0.031941924 &
  0.02068741 &
  0.01016559 &
  0.01125452 &
  1.29 &
  12 &
  0.93321234 &
  0.92321878 &
  0.00563785 &
  0.00999357 &
  4.615 &
  0 &
  0.99963702 &
  0.99931353 &
  0.00032349 &
  0.00032349 \\
-2 &
  2 &
  0.032667877 &
  0.02195002 &
  0.0099919 &
  0.01071785 &
  1.325 &
  12 &
  0.93756806 &
  0.92700085 &
  0.00621149 &
  0.01056721 &
  4.65 &
  0 &
  0.99963702 &
  0.99942007 &
  0.00021695 &
  0.00021695 \\
-1.965 &
  4 &
  0.034119782 &
  0.02328294 &
  0.00938493 &
  0.01083684 &
  1.36 &
  13 &
  0.94228675 &
  0.9306828 &
  0.00688526 &
  0.01160395 &
  4.685 &
  0 &
  0.99963702 &
  0.99951794 &
  0.00011908 &
  0.00011908 \\
-1.93 &
  4 &
  0.035571688 &
  0.02466813 &
  0.00945165 &
  0.01090356 &
  1.395 &
  4 &
  0.94373866 &
  0.9342573 &
  0.00802945 &
  0.00948136 &
  4.72 &
  0 &
  0.99963702 &
  0.99959806 &
  3.90E-05 &
  3.90E-05 \\
-1.895 &
  6 &
  0.037749546 &
  0.02608817 &
  0.00948352 &
  0.01166138 &
  1.43 &
  12 &
  0.94809437 &
  0.93773667 &
  0.00600199 &
  0.01035771 &
  4.755 &
  0 &
  0.99963702 &
  0.9996549 &
  1.79E-05 &
  1.79E-05 \\
-1.86 &
  2 &
  0.038475499 &
  0.0275288 &
  0.01022075 &
  0.0109467 &
  1.465 &
  12 &
  0.95245009 &
  0.94109407 &
  0.0070003 &
  0.01135602 &
  4.79 &
  0 &
  0.99963702 &
  0.99968469 &
  4.77E-05 &
  4.77E-05 \\
-1.825 &
  7 &
  0.041016334 &
  0.02898126 &
  0.00949423 &
  0.01203507 &
  1.5 &
  8 &
  0.9553539 &
  0.94430824 &
  0.00814185 &
  0.01104566 &
  4.825 &
  0 &
  0.99963702 &
  0.99968708 &
  5.01E-05 &
  5.01E-05 \\
-1.79 &
  6 &
  0.043194192 &
  0.03044404 &
  0.01057229 &
  0.01275015 &
  1.535 &
  5 &
  0.95716878 &
  0.94735687 &
  0.00799703 &
  0.00981191 &
  4.86 &
  0 &
  0.99963702 &
  0.99966487 &
  2.78E-05 &
  2.78E-05 \\
-1.755 &
  1 &
  0.043557169 &
  0.03192378 &
  0.01127042 &
  0.01163339 &
  1.57 &
  11 &
  0.96116152 &
  0.95022032 &
  0.00694846 &
  0.0109412 &
  4.895 &
  1 &
  1 &
  0.9996235 &
  1.35E-05 &
  0.0003765
\end{tabular}%
}
\end{table}

\end{document}